# Structural Chemistry of Layered Lead Halide Perovskites Containing Single Octahedral Layers

Jason A. McNulty and Philip Lightfoot*

School of Chemistry and EaStChem, University of St Andrews, St Andrews, Fife, KY16 9ST, U.K.

*e-mail: pl@st-andrews.ac.uk

**Abstract:**

We present a comprehensive review of the structural chemistry of hybrid lead halides of stoichiometry $APbX_4$, $A_2PbX_4$ or $AA'PbX_4$, where A and A′ are organic ammonium cations and X = Cl, Br or I. These compounds may be considered as layered perovskites, containing isolated, infinite layers of corner-sharing $PbX_4$ octahedra separated by the organic species. We first extract over 250 crystal structures from the CCDC and classify them in terms of unit cell metrics and crystal symmetry. Symmetry mode analysis is then used to identify the nature of key structural distortions of the $[PbX_4]_\infty$ layers. Two generic types of distortion are prevalent in this family: tilting of the octahedral units and shifts of the inorganic layers relative to each other. Although the octahedral tilting modes are well-known in the crystallography of purely inorganic perovskites, the additional layer shift modes are shown to enrich enormously the structural options available in layered hybrid perovskites. Some examples and trends are discussed in more detail in order to show how the nature of the interlayer organic species can influence the overall structural architecture, although the main aim of the paper is to encourage workers in the field to make use of the systematic crystallographic methods used here to further understand and rationalise their own compounds, and perhaps to be able to design-in particular structural features in future work.

## 1. Introduction

Lead halide perovskites (LHPs) have recently revolutionised the field of solar cells, in addition to showing novel and promising properties in several other areas, such as luminescence, ferroelectricity etc.[1–3] The diversity of chemical composition and structural architecture in this enormous and rapidly expanding family of materials not only creates great opportunities for the synthetic and structural solid-state chemist, but also makes the field somewhat overwhelming for the newcomer. For those of us with long memories, the excitement and opportunities available for the solid-state chemist are somewhat reminiscent of the explosion in work on layered cuprate perovskites in the late 80's and early 90's, at the peak of the High-Tc superconductor revolution. Indeed, with the advent of 'hybrid' inorganic-organic systems in LHPs, the diversity of the field is clearly much greater than in more traditional inorganic-only systems. There have been many excellent reviews of the field of LHPs over the past few years,[4–6] which have focussed on various aspects from the underlying chemistry and crystal structure to optimisation of electronic and optical properties and further towards material processing and device manufacture. The purpose of the present review is to take a more crystallographically-oriented view of the state-of-the-art in the area of *layered hybrid perovskites,* LHPs, specifically those containing a single 'perovskite-like' octahedral layer of stoichiometry $[PbX_4]_\infty$ (X = Cl, Br, I), in which the layers are separated by cationic organic moieties (A, A′) to give overall compositions $APbX_4$, $A_2PbX_4$ or $AA'PbX_4$. Even within this sub-field there are well over 250 crystal structures reported in the Cambridge Crystallographic Database (census date 11/11/10). Hence, we shall not refer to the copious body of work on three-dimensional (3D) perovskite structures, such $(CH_3NH_3)PbI_3$ or the variety of '0-D' or '1-D' perovskite-related materials, based on chain-like structural fragments or isolated $PbX_6$ octahedra. Moreover, in order to keep the review of a manageable size and digestible to the less-expert reader, we also limit our analysis to so-called (001)-cut layered perovskites: hence neither the related (110) or (111)-cut families nor the (001)-cut families with double, triple or higher-order perovskite-like layer thicknesses will be covered here.

We first briefly introduce the various families of perovskites, before proceeding to describe the types of structural distortion that exist in layered perovskites, then using these as a means of classification of the currently known examples. We shall primarily focus on the detailed structural nature of the inorganic layers $[PbX_4]_\infty$ themselves, and then consider how this is influenced by the variety of A-site molecular cations which might determine the detailed architecture of these layers and their interactions: these features are ultimately the main driver influencing the physical properties of the resulting materials.

## 2. A brief introduction to perovskite crystallography

### 2.1 What is a perovskite?

The name ‘perovskite’ originated in the discovery of the mineral Perovskite, $CaTiO_3$, in 1839.[7] This mineralogical curiosity later blossomed into arguably the most diverse and important class of compounds in solid state chemistry. The generic composition of perovskite may be regarded as $ABX_3$, where A and B are ‘large’ and ‘small’ cations, respectively, and X is an anion. The aristotype crystal structure has cubic symmetry, space group $Pm\bar{3}m$, and consists of a cubic-close-packed array of A and X, with B occupying ¼ of the octahedral interstices, in an ordered manner (Fig. 1a). Note that the word ‘cubic’ here is used in two different senses. A ‘cubic perovskite’ does not necessarily adopt a cubic crystal system, and symmetry-lowering is the norm, due to the well-known tolerance factor and octahedral tilting effects; indeed Perovskite itself is orthorhombic! Moreover, there are many further variants on this basic compositional and crystal chemistry, and there is currently confusion and conflict in the literature regarding ‘what exactly is a perovskite’.[8–10] This is unfortunate, but perhaps inevitable, in such a diverse field, and may require an international committee to propose some clear guidelines and definitions of nomenclature in this area. The use of the phrase ‘layered perovskite’ in the present work corresponds to the personal opinions and preferences of the authors, and it is not intended to impose on other authors.

### 2.2 Octahedral tilting and Symmetry mode analysis

One ubiquitous type of distortion in cubic perovskites is ’tilting’ of the octahedral $BX_6$ units, which occurs due to a size mismatch between the A and B cations, governed by the Goldschmidt tolerance factor, *t*:

$$\boldsymbol{t} = \frac{\boldsymbol{r_A} + \boldsymbol{r_X}}{\sqrt{2}(\boldsymbol{r_B} + \boldsymbol{r_X})}$$

Glazer[11] originally classified all the ‘simple’ tilts in cubic perovskites, and this was later updated by Howard and Stokes, using group-theoretical analysis[12] to give 15 possible simple tilt systems. The Glazer notation uses three lower case letters to specify the relative magnitudes of the tilts along the principal axes of the aristotype (‘parent’) unit cell. Superscripts, + or -, are used to specify whether these tilts occur ‘in-phase’ or ‘out-of-phase’ relative to each other, considering only a 2 × 2 × 2 array of corner-linked rigid octahedra. Thus, for example, Glazer tilt systems $a^0a^0c^+$ and $a^0a^0c^-$ are shown in Figure 1(a,b).

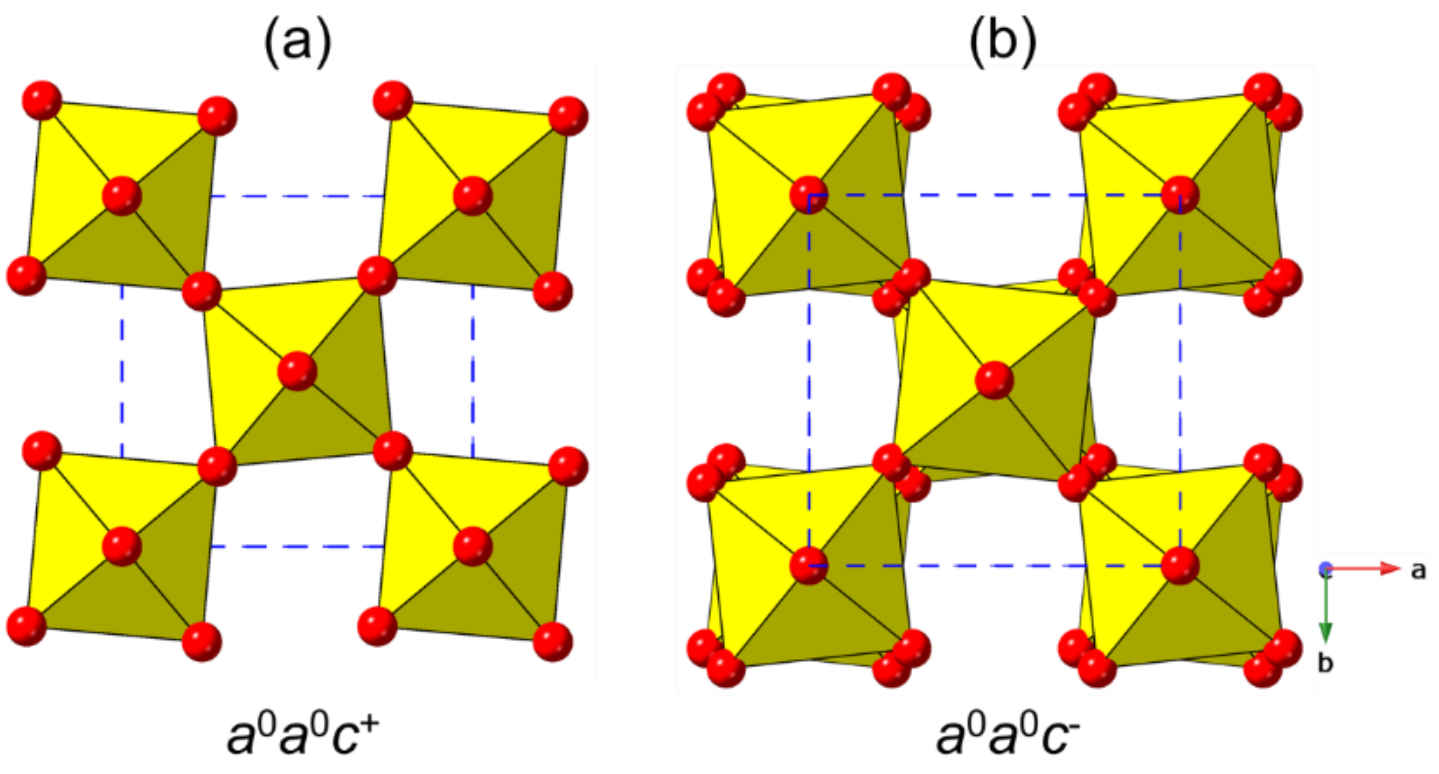

**Figure 1.** Octahedral framework of the ideal cubic perovskite showing the structure with (a) in-phase and (b) out-of-phase octahedral tilting, described by Glazer tilt notations $a^0a^0c^+$ and $a^0a^0c^-$, respectively. B-centred octahedra and X atoms are shown in yellow and red, respectively.

These structural deviations from a high symmetry parent structure can be regarded as 'modes' of distortion (like normal vibrational modes), which are amenable to the application of group-theoretical methods and representational analysis.[12–14] Indeed it is the application of these methods and, in particular, the advent of user-friendly graphical software, such as ISODISTORT[13] and AMPLIMODES[14], amenable to non-expert users, that has allowed solid-state chemists to take a fresh look at structural phenomena of this type, in a much more rigorous and systematic way than was previously available. The $a^0a^0c^+$ and $a^0a^0c^-$ tilts in Glazer notation can be described with irreducible representations (irreps) with labels $M_3^+$ and $R_4^+$, respectively, using the notation of Miller and Love.[15] The distortions associated with these irreps correspond to the 'freezing-out' of phonon modes at specific points of the 1st Brillouin zone of the parent cell. This has two important consequences for solid-state chemists: (i) if a suitable 'parent' model for a particular structure type can be derived, experimentally or otherwise, then structural distortions in 'real' examples of this structure type can be easily and systematically understood in terms of these constituent irreps. (ii) Since the capital letter (e.g. M or R) in the irrep label corresponds to a particular point in reciprocal space, these distortions can fairly easily be identified from a diffraction experiment, as they will give rise to particular types of supercell relative to the parent, higher-symmetry unit cell. Such 'symmetry mode analysis' is therefore an invaluable tool for the solid-state chemist in identifying common structural features across an otherwise apparently diverse range of (related) crystal structures.[12–14,16,17] We shall see that, by using the standard 'parent' phases for either RP (*I4/mmm*) or DJ (*P4/mmm*) structures, we can easily identify tilt modes, and other key types of distortion, unambiguously. Throughout this work we use the on-line tool ISODISTORT to perform this analysis.[13] The symmetry labels for each type of distortion are dependent on the parent phase (and unit cell origin choice) used, but they will be self-consistent for a given sub-family of LHPs. These will be introduced, as required.

### 2.3 Layered perovskites

Here, we shall use the term 'layered perovskite' to mean a compound with a crystal structure that can be easily derived from the cubic perovskite structure by 'slicing' through octahedral apices along a particular crystallographic direction, and inserting additional species between the resultant layers. There are several common types of layered perovskite, of which two are relevant in this work. Ruddlesden-Popper (RP) phases and Dion-Jacobson (DJ) phases were originally observed in mixed metal oxides,[18,19] and these were identified as having generic compositions $A_2A'_{n-1}B_nX_{3n+1}$ and $AA'_{n-1}B_nX_{3n+1}$, respectively. For the $n = 1$ case consider here, the aristotype compounds can be taken as the tetragonal systems $K_2NiF_4$ (space group *I4/mmm*)[20] and $TlAlF_4$ (space group *P4/mmm*),[21] respectively. It should be noted that, in addition to the compositional differences stated above, a key structural distinction between the original two families is in the nature of the relative positioning of adjacent $[BX_4]_\infty$ layers. Thus, we can see (Figure 2) that the adjacent layers in the DJ family are 'eclipsed' relative to each other (a coordinate displacement of (0,0)), whereas those in the RP family are staggered by (½, ½). A further important variant on these two structure types is the intermediate case, with a staggering of (0, ½) or (½, 0): here the parent phase is orthorhombic, with space group *Ammm* (taking *c* as the 'layer stacking' direction)(parent phase, $NaWO_2Cl_2$[22]).

Octahedral tilting is also a recognised and common feature in layered perovskites,[23,24] and we shall describe this in terms of both Glazer-like notation and using irrep labels for the relevant tilt modes. In the case of single-layer layered perovskites, the tolerance factor is clearly of no direct relevance, although the nature of the interaction of the interlayer species with the $[BX_4]_\infty$ framework will certainly influence the nature of tilting. When we use the word 'rotation', rather than tilt this specifically applies to a mode acting around the axis perpendicular to the layer direction.

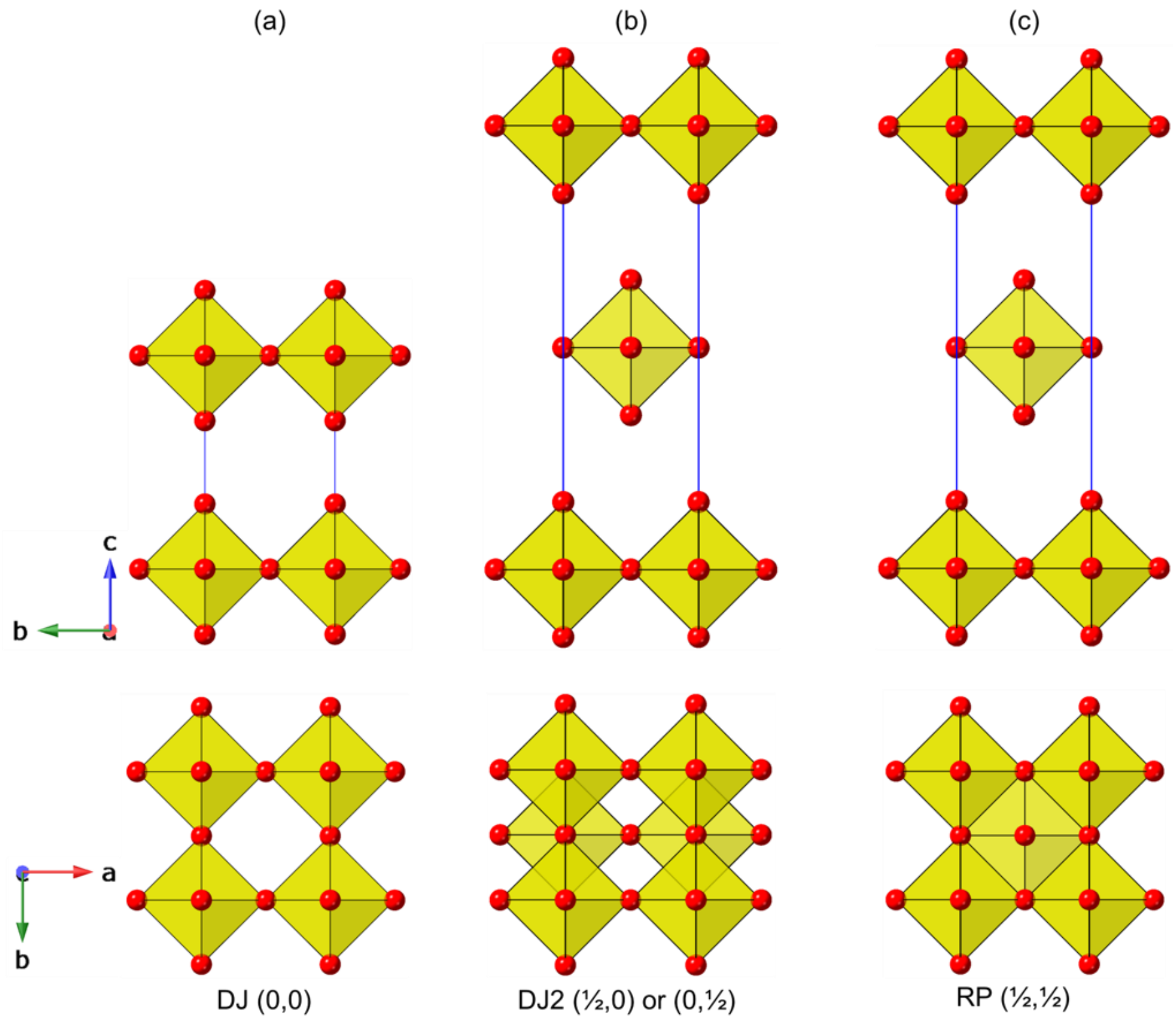

**Figure 2.** Undistorted structures of the simplest *n*=1 layered perovskites viewed between layers (top) and along the layer stacking direction (bottom) highlighting the completely 'eclipsed', partially 'staggered' and completely 'staggered' adjacent layers of the (0,0) and (½,0) Dion-Jacobson (DJ and DJ2, respectively) and (½,½) Ruddlesden-Popper (RP) phases.

### 2.4 Hybrid layered perovskites

With the advent of hybrid perovskites, inorganic solid-state chemistry had already paved the way for a useful description of the structural architectures of these compounds in terms of octahedral tilting and other distortions such as intra-octahedral distortion indices.[12,25–27] However, the inclusion of non-spherical, and often highly anisotropic, molecular species at the inter-layer A sites opens up a new level of complexity in hybrid layered perovskites. One particular feature of note is the much greater tendency for 'slippage', 'shift' or 'staggering' of adjacent perovskite blocks relative to each other, such that the conventional criteria used in recognising RP versus DJ phases can no longer be applied simply. We'll use 'layer shift' from now on, to describe this. In fact, it has already been recognised that a range of degrees of shift of adjacent inorganic layers are observed in LHPs which span the RP-DJ regime.[28] We therefore choose a definition of 'RP' and 'DJ' solely in terms of the degree of layer shift, rather than the original additional differences in A/B stoichiometry. In accord with Tremblay *et al*.[28] we shall refer to any layered LHP, regardless of stoichiometry, having an inter-layer offset close to (½, ½) as RP-like ('near-RP' or nRP), any having an offset near (0, 0) as DJ-like (nDJ) and any having an offset near (0, ½) as DJ2-like (nDJ2). Criteria similar to those of Tremblay will be used to define how close the structures are to one of the ideal types, by use of a layer shift parameter, $\Delta$. We shall also see that layer shift can also be easily understood in terms of symmetry mode analysis, with specific modes occurring commonly, regardless of other simultaneous types of distortion.

The next sections describe a comprehensive survey and classification of all structurally well-characterised (001)-cut LHPs of stoichiometry $APbBr_4$ or $A_2PbBr_4$. These structures are taken from the CCDC[29] up to 11/11/20. In our initial survey, we began by classifying the unit cell metrics of all the known structures in relation to either of the generic structure types RP, DJ or DJ2. It was immediately apparent that, although there is a wide diversity of variants spanning these ideal 'end-members', several common themes in types of distortion and types of supercell emerge.

## 3. A classification of (001)-cut LHPs of stoichiometry $APbBr_4$ or $A_2PbBr_4$

We find it convenient to classify this vast array of structures in terms of the observed unit cell metrics, and their relationship to the parent RP or DJ parents. In particular we shall use the number of octahedral layers per unit cell repeat as a key discriminator. In other words, regardless of whether the resultant layer shift looks 'RP-like', 'DJ-like' or 'DJ2-like', we'll aim to derive all the structure types from either a RP parent (*I*4/*mmm*), for those with two or more layers per unit cell, or a DJ parent (*P*4/*mmm*) for those with one layer per unit cell. In many cases, it is equally feasible to use the alternate parent phase, leading to an equivalent result. For the case of LHPs, the parent tetragonal unit cell metrics are $a_{RP} \sim a_{DJ} \sim 5.5 – 6.5$ Å, for chlorides to iodides, with $c_{RP}$, $c_{DJ}$ obviously being variables, dependent on the nature of the organic moieties.

In all the Tables, we refer to the individual structures using CCDC deposition numbers. We start with the two-layer cases as these more usefully illustrate some of the structural principles observed. The parent phase is the $K_2NiF_4$ type, i.e. with two adjacent, fully staggered layers forming the unit cell repeat perpendicular to the layer direction. Throughout the following survey it is worth noting that the 'layered perovskite' convention of taking the layer-stacking direction as the *c*-axis does not always correspond to the conventional setting of the resultant space group; authors differ on which convention they follow, so occasionally different space group settings appear in the data presented herein. In section 3.1 we describe and classify structures with two octahedral layers and include relatively simple structures with unit cell volumes up to 2 $a_{RP}$ × 2 $a_{RP}$ × $c_{RP}$, i.e. eight formula units per unit cell. In section 3.2 we describe structures with one octahedral layer per unit cell, and in section 3.3 we describe more complex structures with at least one axial metric more than double the parent phase.

## 3.1 Structures with two octahedral layers per unit cell (derived from RP parent)

### 3.1.1 Unit cell metrics equivalent to the parent phase (~ $a_{RP}$ × $a_{RP}$ × $c_{RP}$)

We first note that Balachandran et al.[30] conducted a survey of known inorganic oxides adopting the $A_2BX_4$ RP structure, and found the high-symmetry aristotype phase, in space group *I*4/*mmm* to be the most common variant. In stark contrast, the crystal structures of only three LHPs have been reported in the aristotype space group *I*4/*mmm* (Table 1). These are structures of 'high-temperature' phases, exhibiting disordered organic moieties, and all subsequently undergo phase transitions to ordered polymorphs with lower symmetry supercells on cooling. In addition, there are two simple derivatives which retain the body-centred symmetry. There are three further examples of lower symmetry structures with these cell metrics in our survey. Note that it is impossible for such cell metrics to accommodate octahedral tilting distortions; these require expanded supercells of at least twice the volume of the parent phase. Therefore, the driver for these lower symmetry yet 'parent cell size' structures turns out to be essentially a layer shift mode. This mode is designated $M_5^-$; in fact, it has more flexibility than a 'rigid mode' layer shift, also allowing some intra-octahedral distortions. More specifically, there are often several distinct options for such a mode (and likewise for the octahedral tilt modes). In this case we find one example (1937296) of $M_5^-$(a,a) symmetry and one (1211182) of lower $M_5^-$(a,b) symmetry. The additional notation (a,a) etc. defines the so-called 'order parameter direction' (OPD);[13] for further details see ESI. Examples of these modes are shown schematically in Figure 3, where the mode amplitudes are chosen to keep the octahedra close to regular. It can be seen that each mode, acting individually, causes a specific lowering of the symmetry from the parent symmetry; e.g. the $M_5^-$(a,a) mode naturally results in a space group *Pmmn* with approximate cell metrics of the parent phase. In the case of 1937296 the layers are shifted along the *c*-axis and the magnitude of this mode results in a structure close to (0, ½) staggering, i.e. nDJ2. The second example, 1211182, is a lower symmetry (polar) version of this, also of nDJ2 type. A third case (1186561) was a very early example of a LHP, and was refined with disordered octahedra: the authors stated possible 'unresolved superlattice structure', and we agree that this structure probably does contain octahedral tilting modes, so does not formally belong in this section; in fact, a subsequent re-determination is included later (200737 in section 3.2).

Note there is one further possible symmetry for the $M_5^-$ mode: $M_5^-$(a,0), which we shall see in the next section. The relative directions of the shifts should be clear from Figure 3.

**Table 1.** Summary of experimentally known RP-derived structures with $a_{RP} \times a_{RP} \times c_{RP}$ unit cell metrics. The 'Type' column in all Tables is merely intended to highlight the structures with divalent ([A]) or mixed-cation ([A][A′]) stoichiometries.

| CCDC Number | Amine | Type | Formula | Space group | Key modes | Tilt modes | Ref. |
|---|---|---|---|---|---|---|---|
| 1934895 | 4,4-Difluoropiperidine | | $[C_5H_{10}F_2N]_2PbI_4$ 453 K | *I*4/*mmm* | – | $a^0a^0c^0$ | 31 |
| 1944745 | Benzylamine | | $[C_7H_{10}N]_2PbCl_4$ 493 K | *I*4/*mmm* | – | $a^0a^0c^0$ | 32 |
| 1944744 | 2-Fluorobenzylamine | | $[C_7H_9FN]_2PbCl_4$ 463 K | *I*4/*mmm* | – | $a^0a^0c^0$ | 32 |
| 1944742 | 3-Fluorobenzylamine | | $[C_7H_9FN]_2PbCl_4$ 473 K | *I*4/*mmm* | – | $a^0a^0c^0$ | 32 |
| 1944740 | 4-Fluorobenzylamine | | $[C_7H_9FN]_2PbCl_4$ 493 K | *I*4/*mmm* | – | $a^0a^0c^0$ | 32 |
| 1992694 | 4,4-Difluorohexahydroazepine | | $[C_6H_{12}F_2N]_2PbI_4$ 493 K | *I*-42*m* | – | $a^0a^0c^0$ | 33 |
| 1992693 | 4,4-Difluorohexahydroazepine | | $[C_6H_{12}F_2N]_2PbI_4$ 423 K | *Imm*2 | $\Gamma_5^-$ | $a^0a^0c^0$ | 33 |
| 1937296 | Methylhydrazine | | $[CH_7N_2]_2PbI_4$ RT | *Pmmn* | $M_5^-$ | $a^0a^0c^0$ | 34 |
| 1211182 | 1,4-Dimethylpiperazine | [A] | $[C_6H_{16}N_2]PbBr_4$ | *P*2$_1$ | $M_5^-$ | $a^0a^0c^0$ | 35 |
| 1186561 | 1-Phenylethylamine | | $[C_8H_{12}N]_2PbI_4$ | *C*2/*m* | $M_5^-$ | $a^0a^0c^0$ | 36 |

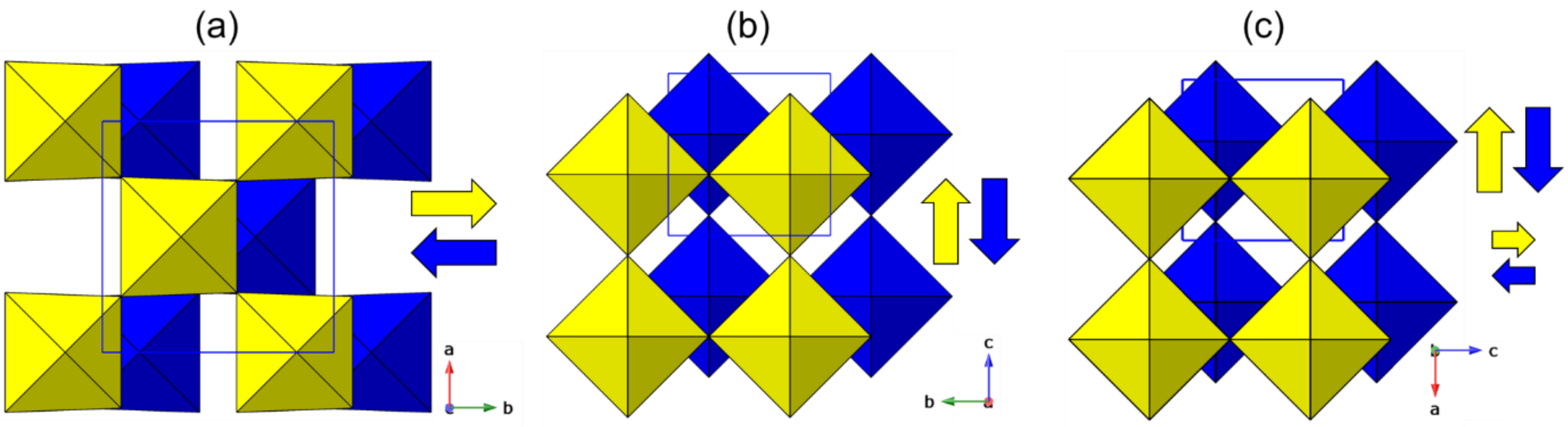

**Figure 3.** The three distinct $M_5^-$ displacive modes, derived from the RP parent. (a) $M_5^-$(a,0) antiparallel displacements of adjacent layers along one in-plane axis of the supercell; (b)$M_5^-$(a,a) showing anti-parallel displacements of adjacent layers along one in-plane axis of the parent cell (c) $M_5^-$(a,b) allows two different amplitudes of displacement along the in-plane axes of the parent cell. It can be seen that the (a,0) mode results in layers shifted towards DJ type, (a,a) towards DJ2 type, and (a,b) provides both degrees of freedom. Note that each mode also allows a distortion of the octahedra, which is illustrated in (a) by the differing lengths of the trans- octahedral edges.

### 3.1.2 Metrics ~ $\sqrt{2}\,a_{RP} \times \sqrt{2}\,a_{RP}$ in the layer plane

These structures are detailed in Table 2. A very common distortion of a cubic perovskite unit cell (unit cell parameter $a_p$) is an approximately $\sqrt{2}\,a_p \times \sqrt{2}\,a_p$ supercell caused by octahedral tilting. Such effects are seen, for example, in both of the Glazer systems in Figure 2. It comes as no surprise that such features are also commonplace in layered perovskites. We note that unit cell volume is effectively doubled in all these derivatives but the *c*-axis remains equivalent to that of the parent phase (i.e. not doubled relative to the RP parent, but still encompassing two adjacent octahedral layers per *c*-axis repeat). In addition, the body-centring is lost. It can be seen that there is a diversity of resultant space groups. As discussed above, we are now anticipating that structures may contain two particular types of distortion of the $[PbX_4]_\infty$ layers (i.e. octahedral tilting and layer shift). Our classification therefore considers those structures with layer shifts and octahedral tilting, either independently or cooperatively, starting from the simplest to the more complex.

**Insert Table 2 here**

**Table 2.** Summary of experimentally known structures with two octahedral layers per unit cell (derived from RP parent) and $\sqrt{2}a_{RP} \times \sqrt{2}a_{RP}$ cell metrics in the layer plane.

***Structures with no octahedral tilting, but with layer shifts***

We include in Table 2, and subsequent Tables, the parameter $\Delta$ which describes the extent of layer shift between neighbouring layers (highlighted in Figure 3). In fact, in a general case this is a two-dimensional parameter ($\Delta_1$, $\Delta_2$). We have included these parameters only for selected series of relatively simple structure types, using manually calculated $\Delta$ values, based on the relative displacements of Pb atoms only, in neighbouring layers. For more complex structures we have chosen to state a visually estimated layer shift outcome (i.e. nRP, nDJ or nDJ2). A $\Delta$ value of (0.25, 0.25) signifies the crossover between 'nDJ' (for $\Delta < 0.25$) and 'nRP' (for $\Delta > 0.25$). nDJ2 structures have $\Delta$ values closer to (½, 0), i.e. $\Delta_1 > 0.25$, $\Delta_2 < 0.25$.

There are two distinct types of layer shift which may be present in layered perovskites. These can be described and classified conveniently in the language of symmetry mode analysis. The first type is represented by a 'gamma mode' (i.e. acts at the Brillouin zone centre), usually designated $\Gamma_5^+$. It acts simply to slide adjacent layers in the same sense relative to each other along one crystallographic axis, and results in a lowering of symmetry to monoclinic (Figure 4). The resultant $\Delta$ values for this type of distortion can be simply calculated from the unit cell metrics (see, for example, the equations in section 3.2.2), although direct graphical measurement (for example using Crystalmaker[102]) also gives a good approximation. It will be seen that this type of monoclinic distortion is a common feature in LHPs, leading to bridging of the RP to DJ regimes.

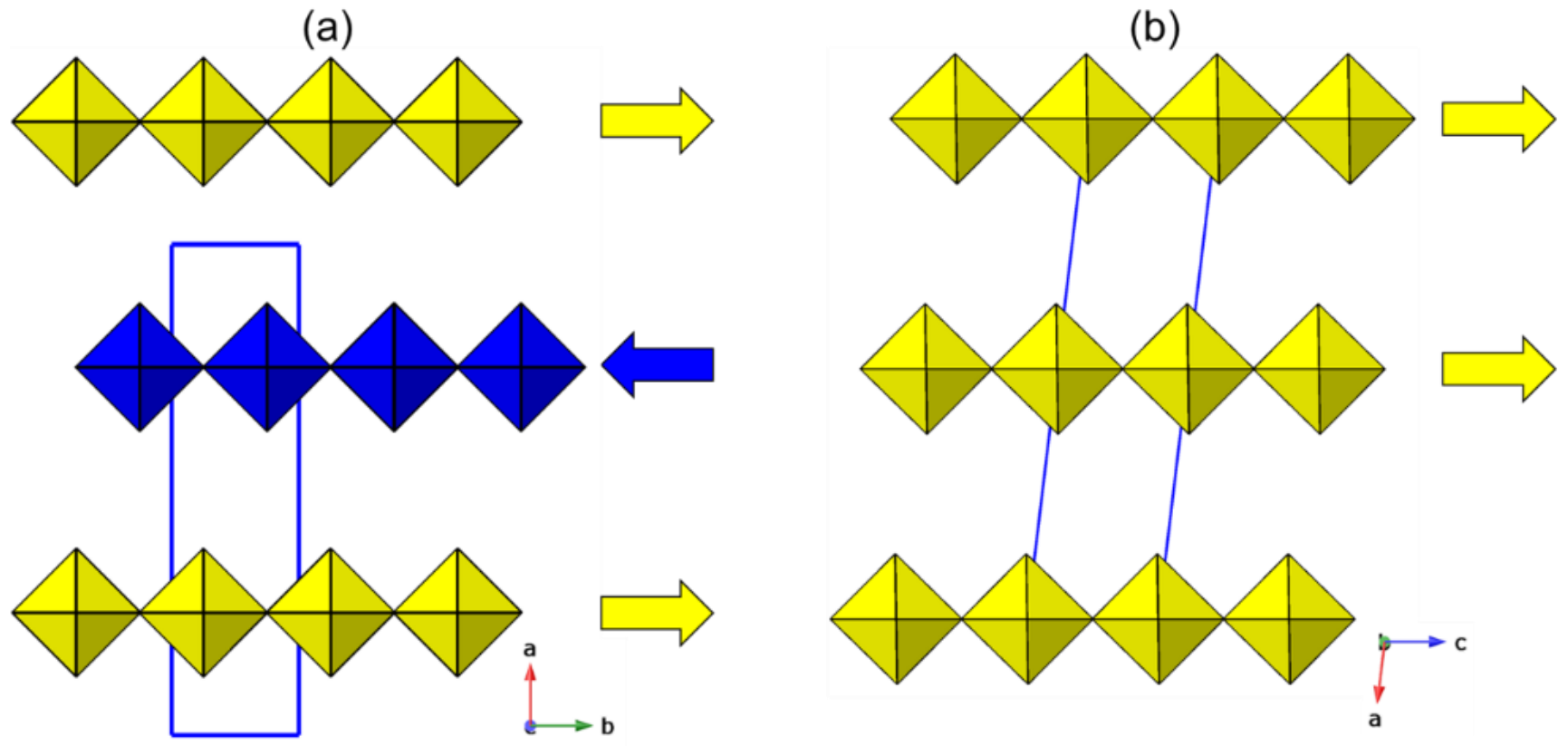

**Figure 4.** Comparative effects of the two generic types of layer shift mode acting on the RP parent (a) $M_5^-$(a,a) mode (see also Figure 3) and (b) the analogous $\Gamma_5^+$ mode. Note that the former leads to orthorhombic symmetry, whereas the latter corresponds to a monoclinic distortion.

The second type of layer shift mode is the antiferrodistortive $M_5^-$ type introduced in section 3.1.1. i.e. a shift of adjacent octahedral layers in *opposite* directions relative to an axis perpendicular to the layer direction (Figures 3, 4). The mode, acting alone, leads to a lowering of symmetry from tetragonal to orthorhombic, and so the corresponding Δ values can be calculated straightforwardly, from the associated difference in *x*, *y*, or *z* coordinates. The first, and unique, example here is 1852626, which occurs in space group *Cmcm*. This example is a high temperature (413 K) polymorph of a phase that appears at ambient temperature in space group *Pbcn* (1845548). It has no octahedral tilting, just the $M_5^-$(a,0) displacive mode (Figure 3), which is distinct from those seen in section 3.1.1. As far as we are aware, there are no previously reported examples of any $M_5^-$ type of distortion in inorganic $A_2BX_4$ structures. Two further examples also exhibit layer shifts, without octahedral tilting: these examples (641642 and 167103) occur in space group *C*2/*c* and again feature no octahedral tilting; however they have the antferrodistortive shift mode $M_5^-$, which describes a displacement along the *b*-axis (as occurs in the first example) and, in addition, a purely displacive $\Gamma_5^+$ mode, which leads to an additional displacement along the *a*-axis, and a monoclinic distortion.

***Structures with a single type of octahedral tilt and no layer shift***

The structure 1863837, in space group $P4_2/ncm$, is a unique example of one of the simplest types of distortion in this family (Balachandran[30] reported five examples of oxides with this structure type). The structure exhibits an out-of-phase tilting of octahedra around the *ab*-plane, but the direction of this tilt alternates in adjacent layers (Figure 5(a)) hence retaining the tetragonal symmetry. The corresponding tilt mode is designated $X_3^+$(a,a) (see ESI for further details of some of these tilt mode descriptions). It is necessary to use an extended Glazer-like notation to describe the tilts in these systems that contain two adjacent octahedral layers which, while they may be symmetry-related, may also contain opposite directions, or signs, of the corresponding tilts. In the adapted Glazer-like notation, e.g. as used by Hayward,[103] the tilt system here is $a^-b^0c^0/b^0a^-c^0$. Aleksandrov[25] undertook a comprehensive group-theoretical analysis of tilting in RP phases, using a different, but equivalent, notation and designated this tilt system $\phi 00/0\phi^-0$; we shall use the Glazer-like notation. A further two examples (1934896 and 1992692) are also based on a single $X_3^+$ tilt mode, but there are two key distinctions from the previous structure: first, the $X_3^+$ mode has a different OPD, and is designated $X_3^+(0,a)$: the octahedral tilt system is $a^-a^-c^0$/-($a^-a^-$)$c^0$. This structure type is the most common tilted type reported amongst the inorganic oxide analogues by Balachandran.[30] Second, there is an additional, purely displacive ($\Gamma_5^-$), mode which leads to a polar space group, *Aba*2 or *Cmc*$2_1$ indeed, the former compound has been demonstrated to exhibit ferroelectricity.[31]

There are nine examples in Table 2 (commencing 2016195) of phases exhibiting a single $X_2^+(0,a)$ rotation mode and no other significant mode. This results in unit cell metrics $c_{RP} \times \sqrt{2}\, a_{RP} \times \sqrt{2}\, a_{RP}$ and space group *Cmca* (alternatively $\sqrt{2}\, a_{RP} \times \sqrt{2}\, a_{RP} \times c_{RP}$ and non-standard space group *Acam*). Note that this, by coincidence only, is the same space group as for the examples discussed, with active mode $X_3^+(0,a)$. The tilt system here can be designated $a^0a^0c/a^0a^0c$, with no tilting relative to the layer-plane but rotations around the axis perpendicular to the layers, with each layer having the same degree of rotation (Figure 5(b)). Note that we do not use the superscript notation for the *c*-axis in the case of layered perovskites with single octahedral layers (as Glazer's original concept explicitly relies on octahedra being linked in the third direction). We use '*c*' to mean 'rotated perpendicular to the layer direction' and $c^0$ to mean 'no rotation'.[104] The symbol (-*c*) is used if the 2nd layer is rotated contrary to the first (which is only really relevant if there is a partial layer shift from the ideal RP parent).

The next subset of structures (commencing 2003637) involves the same single rotation mode ($X_2^+(0,a)$) but the symmetry is lowered to *Cmc*$2_1$. These ten examples are simple derivatives of the corresponding *Cmca* subset above, but they have an additional displacive mode ($\Gamma_5^-$) acting along the *c*-axis, which leads to the polar space group. Nevertheless, they have $\Delta$ = (0.5,0.5) and can be regarded as RP. There is often a considerable distortion of the $PbX_6$ octahedra present in these structures which leads to the observation that the polar axis (*c*) is significantly shorter than the other in-plane axis (*b*), in each case.

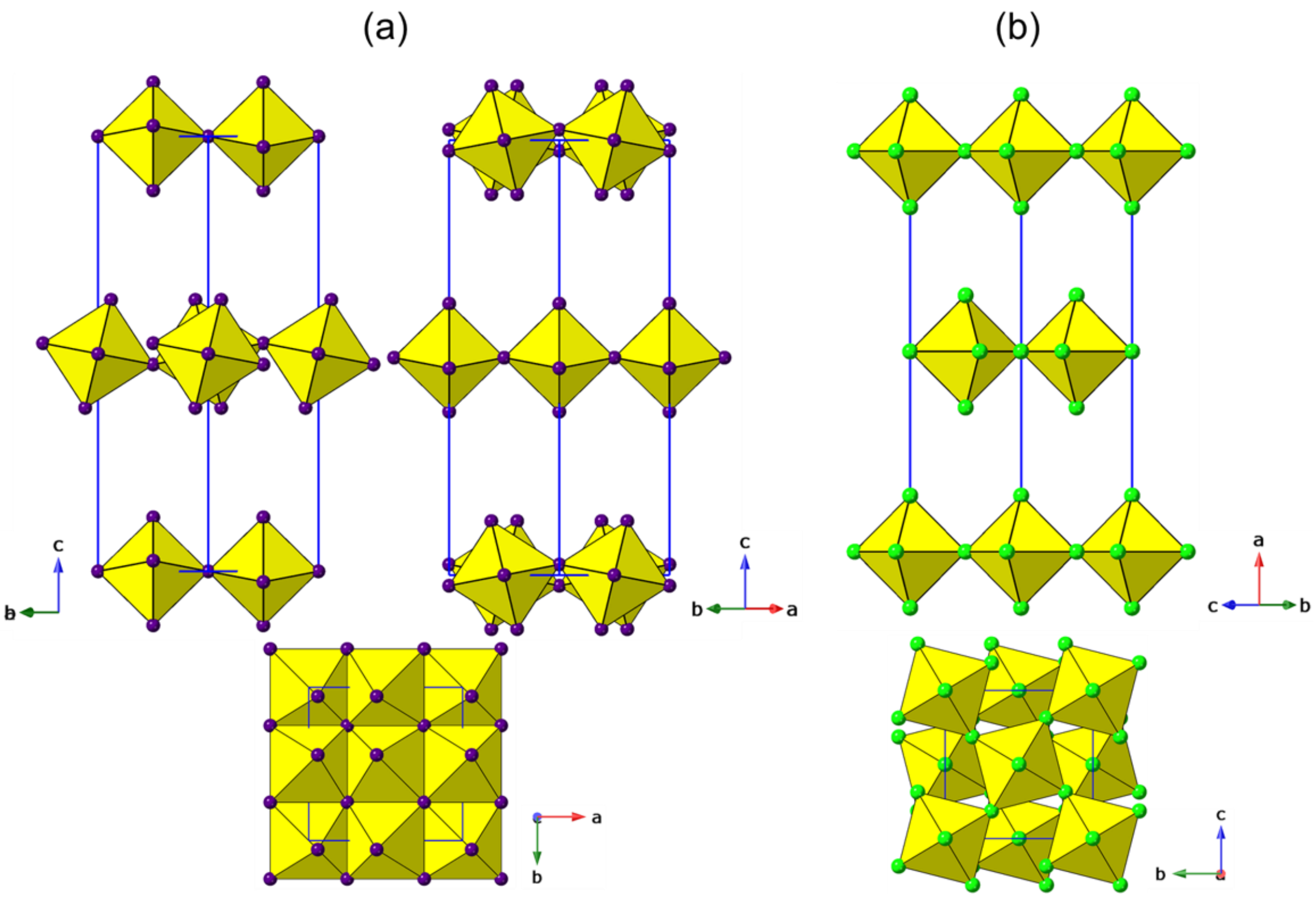

**Figure 5**. Octahedral frameworks of two structures with a single tilt mode (a) $X_3^+$(a,a) tilt mode (1863837) and (b) $X_2^+$(0,a) rotation mode (2016195). Note the out-of-phase tilting around the *ab*-plane and alternating direction of tilts in (a) and the absence of layer-plane tilting but presence of rotations of the same degree around the *c*-axis in (b).

### *Structures with a single type of octahedral tilt and layer shift*

There are several further examples of structures (237190-1975109) containing a single tilt mode, $X_3^+$(0,a); however, this is now supplemented by a further key mode, designated $M_5^-$ (0,-b), which describes a shift of adjacent octahedral layers in *opposite* directions along the *b*-axis (Figure 3). This type of superposition of two modes may be written as $X_3^+ \oplus M_5^-$. The resulting space group is orthorhombic, *Pbcn*. Again, such an 'antiferrodistortive' displacement means that the octahedral layers are no longer in perfectly staggered configuration relative to each other. In other words, such structures to some extent fall between the extremes of 'ideal RP' and 'ideal DJ' types. In these cases, where the unit cell contains two adjacent layers per unit cell repeat, we choose to consider them to be derived from the RP, rather than DJ, parent structure but, of course, the degree of layer shift will dictate whether these examples might be regarded as nDJ or nRP in the classification introduced by Tremblay *et al.*.[28] The parameter Δ therefore comes into play here (defined in these orthorhombic cases as simply the difference in *y*-parameters between two Pb atoms in neighbouring layers, shown in Figure 3). As can be seen, the majority of the examples here are nDJ. Note that it is convenient in simple supercells with "√2 × √2" in-plane metrics to determine the Δ parameters relative to the supercell axes, but $\Delta_1 = \Delta_2$ in these cases.

There are also several further, lower symmetry structures that are derived from a single rotation described by the $X_2^+$ mode, but with additional modes leading to lower symmetry space groups. This set consists of eight examples (1938882 onwards), which adopt orthorhombic space group *Pnma*, with metrics $\sqrt{2}\, a_{RP} \times c_{RP} \times \sqrt{2}\, a_{RP}$. The $X_2^+$(0,a) is a key mode, which could be described as tilt system $a^0a^0c/a^0a^0c$, in the case of a perfect RP structure. However, as in the case of the *Pbcn* structure types above, this is now supplemented by the further key mode, $M_5^-$ (0,b-), which describes a shift of adjacent octahedral layers in *opposite* directions along the *a*-axis (Figure 3). This again means such structures to some extent fall between the extremes of 'ideal RP' and 'ideal DJ' types. The parameter Δ signifies that each of the examples here are nRP. However, for significantly shifted layers, the choice of *c* or (-*c*) symbols is open to definition, and may be taken from the relative rotations of the 'nearest' octahedron in the adjacent layer. For the nDJ structures these should perhaps be described as $a^0a^0c/a^0a^0(-c)$. It should also be noted that the distortive effect of the $M_5^-$ mode is often more dominant than the octahedral rotation mode (perhaps a manifestation of the stereochemically active $Pb^{2+}$ lone pair); an example is 1938883. It can be seen, in even in these relatively 'simple' examples of LHPs, that the unambiguous assignment of Glazer-like tilt systems is not as straightforward as it is in the traditional inorganic layered perovskite families. There is further one example (1914148) of a combination of $X_2^+$(0,a) rotation with a different shift mode, $M_5^-$ (b,0), which naturally leads to space group *Pbcm*; in this case the rotational mode is again near zero.

***Structures with two types of octahedral tilt and no layer shift***

We now consider structures within this family of unit cell metrics which accommodate two distinct types of tilt mode. This subset is very common, with 24 examples (commencing 1938881). It has contributions from the two modes we have seen individually: $X_2^+$(0,a) and $X_3^+$(b,0), and results in unit cell metrics $\sqrt{2}\, a_{RP} \times \sqrt{2}\, a_{RP} \times c_{RP}$ and space group *Pbca*. The tilt system can be regarded as $a^-a^-c/\text{-}(a^-a^-)c$. It is perhaps not surprising that this tilt system is common, as it resembles the most common tilt system in 3D oxide perovskites, $a^-a^-c^+$ (or $GdFeO_3$ type).

### *Structures with two types of octahedral tilt and layer shift*

Finally, we describe several classes of structure having, simultaneously, two tilts and one or two layer shift modes. The first subset has metrics $c_{RP} \times \sqrt{2}\, a_{RP} \times \sqrt{2}\, a_{RP}$ and space group either $C2/c$ (centrosymmetric) or its polar derivative $Cc$. There are nine examples (1826587-956552). These structures have monoclinic, rather than orthorhombic, unit cells which means they have the additional degree of freedom, described by the $\Gamma_5^+$ strain mode, whereby the adjacent layers are permitted to slide 'in-phase' relative to each other leading to shifts intermediate between RP and DJ. The two tilt modes in this case are designated $X_2^+$(0,a) and $X_4^+$(0,a). This leads to the tilt system $a^-a^-c/a^-a^-c$ for unshifted layers; however, the same issue of how to describe this taking into account nRP versus nDJ arises. Due to the additional complexity here, we'll use the ideal 'unshifted' tilt system. At first sight, this may resemble the $a^-a^-c/\text{-}(a^-a^-)c$ system above. However, looking closely at the relationship between directions of tilts in neighbouring layers (Figure 6) the distinction between the $X_4^+$ and $X_3^+$ tilt mode is clear.

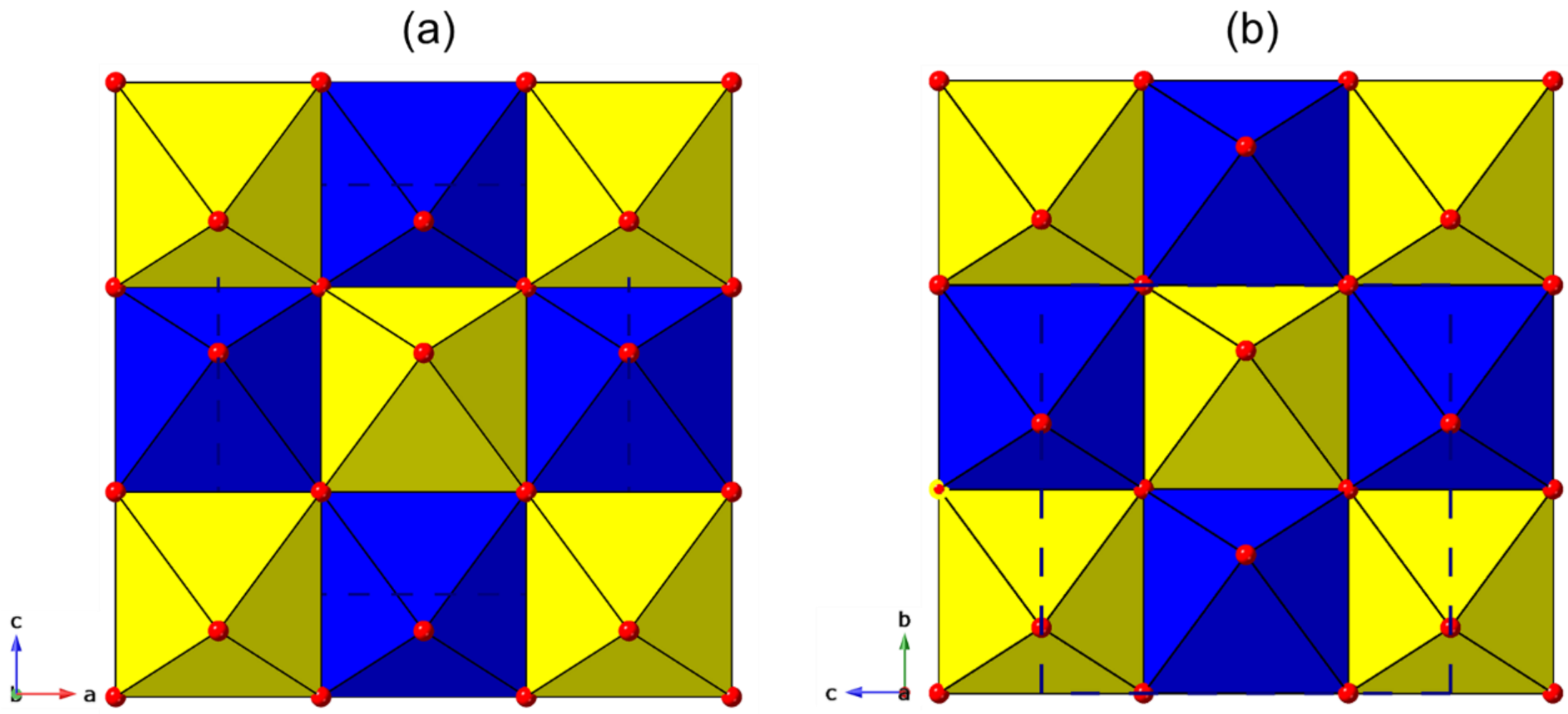

**Figure 6.** Comparison of the (a) $X_3^+$(0,a) and (b) $X_4^+$(0,a) modes, derived from the RP parent. Two layers are plotted, separated by half a unit cell. Notice that the top (yellow) layer tilt pattern is identical for each, but in the bottom (blue) layer tilts change in relative sense. The corresponding Glazer-like notation is $a^-a^-c^0/\text{-}(a^-a^-)c^0$ and $a^-a^-c^0/a^-a^-c^0$, respectively.

The remaining structures in Table 2 fall into several types, which are mostly more complex variants, with different tilt systems, and different degrees of layer shift. We first describe those that are RP-like or nRP. The next two structures (1856671 and 1934873) are simply polar derivatives of the common *Pbca* type, above, with tilt system $a^-a^-c$/-($a^-a^-$)$c$. The following structure (1903531) is again a derivative of the *Pbca* type: the lower symmetry is created, formally, by additional degrees of freedom in both layer shift and tilting (the modes are $X_2^+$(a,b) and $X_3^+$(c,0), leading to a tilt system $a^-a^-b$/-($a^-a^-$)$c$). However, mode decomposition shows that the true symmetry, at least as far as the inorganic network is concerned, is very close to *Pbca*. The next two structures (1119707 and 607740) can be regarded as lower symmetry variants of either the *Pnma* or *Pbca* structures above, having simultaneous $X_2^+$, $X_3^+$ and $M_5^-$ modes; i.e. a formal tilt system $a^-a^-c$/-($a^-a^-$)$c$ together with a layer shift mode. This places the first example close to RP type and the second closer to DJ. The remaining structures are closer to either DJ or DJ2 types. Taking the nDJ types first, three unusual examples are 1942543, 2016669 and 659021. These have a combination of $X_2^+$ and $X_4^+$ rotation/tilt together with the $M_5^-$ and $\Gamma_5^+$ shift modes.

The final examples in this section are most closely related to the DJ2 type, i.e. close to a neighbouring layer offset of (½, 0). The structures (1305732-1521055) in space group $P2_1/c$ exhibit simultaneous $X_2^+$ and $X_3^+$ tilt modes and the antiferrodistortive shift mode $M_5^-$, which, again, leads to a displacement along the *b*-axis. However, in contrast to the two $P2_12_12_1$ examples above, in this case there is also a simultaneous monoclinic distortion ($\Gamma_5^+$ mode) which describes the additional offset of adjacent layers along the *a*-axis, resulting in DJ2 rather than DJ-like behaviour.

The final cases (1525376-1934876) have either tilt modes $X_2^+ \oplus X_4^+$ or $X_3^+ \oplus X_4^+$, each with additional $M_5^-$ and $\Gamma_5^+$ modes.

### 3.1.3 Metrics ~ 2 $a_{RP}$ × $a_{RP}$ or 2 $a_{RP}$ × 2 $a_{RP}$ in the layer plane

There are several structures with two octahedral layers per unit cell which also have one doubled cell axis in the layer plane, and some which have both axes doubled. These are presented in Table 3. We shall save the larger supercell structures for section 3.3.

**Insert Table 3 here**

**Table 3.** Summary of experimentally known structures with two octahedral layers per unit cell (derived from RP parent) and at least one axis in the layer plane doubled.

***Structures with no octahedral tilting, but with layer shifts***

Five structures (1962913-1846391), all iodides, are reported in space group *C*2/*c* with metrics $c_{RP} \times a_{RP} \times 2\ a_{RP}$. They display no octahedral tilting but have $\Gamma_5^+$ layer shifts: this degree of freedom leads to varied structure types between RP and DJ2 type. Unit cell doubling along the *c*-axis arises from an antiferrodistortive displacement of the Pb atoms along the *b*-axis (Figure 7), which is described by $N_1^-$ being the most significant mode. The mode is somewhat reminiscent of the Pb atom shifts in antiferroelectric $PbTiO_3$ (ref), and it is unique among the LHPs we have discussed so far. It should be noted that some of these examples show disorder within the inorganic framework. The next five structures (1938883-641643) have cell metrics $2\ a_{RP} \times c_{RP} \times a_{RP}$ and adopt space group P*nma* (or alternative setting P*bnm*, $a_{RP} \times 2\ a_{RP} \times c_{RP}$). These incorporate the $M_5^-$ displacement mode. In structures with these cells metrics, the $M_5^-$ mode acts along the doubled axis, but simultaneously there is also a $\Sigma_4$ mode [full irrep label in this case is (a,a|0,0;b,-b)] which, rather than being a tilt mode, acts to undulate the layers slightly out of plane. The structures vary between nRP and nDJ2 types.

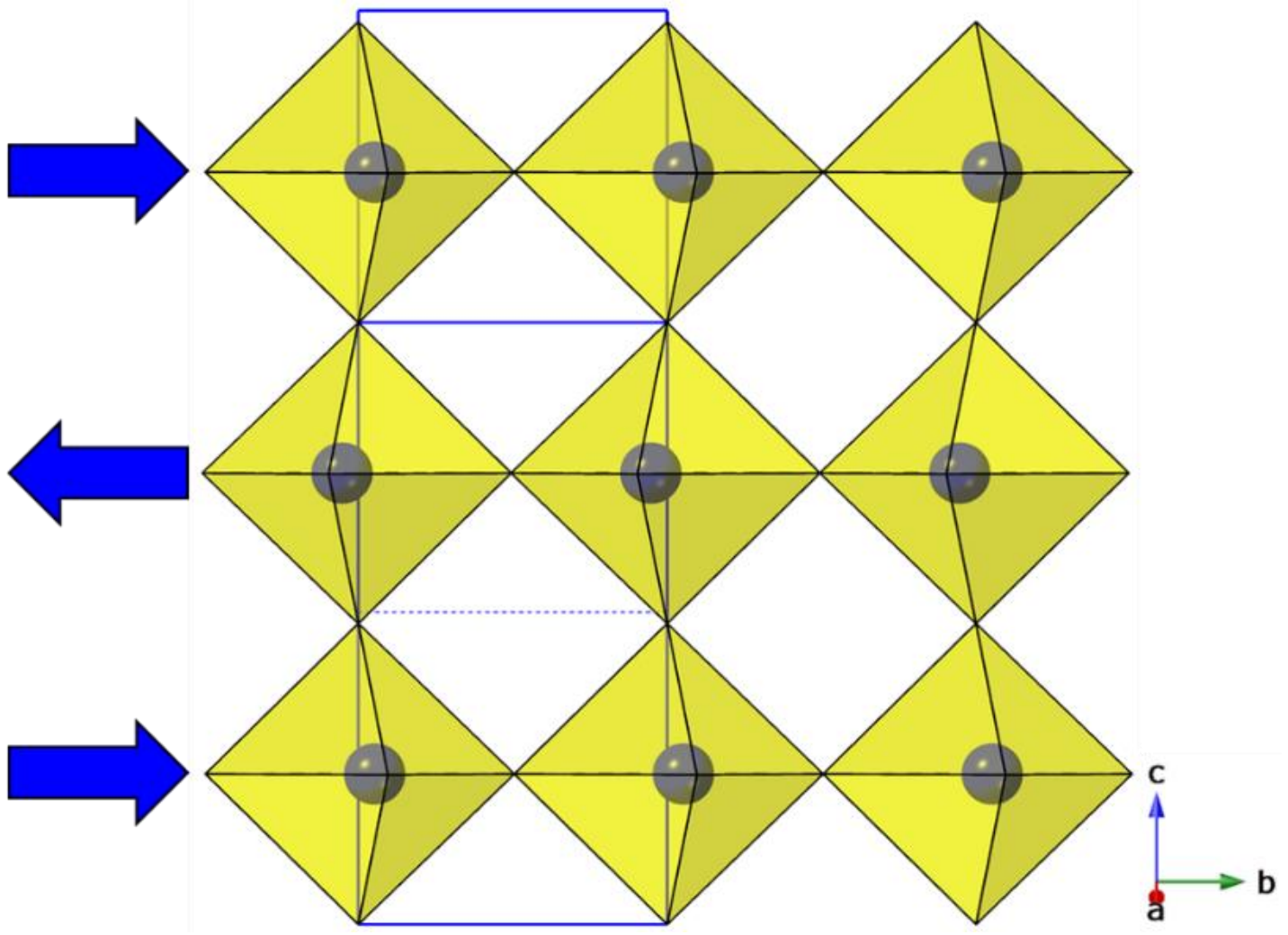

**Figure 7.** Antiferrodistortive displacement of Pb ions causes doubling of the *c*-axis in nRP type $[C_5H_{12}N]_2PbCl_4$ (1962913).

***Structures with a single type of octahedral tilt and no layer shift***

Two quite high symmetry derivatives (1588974 and 1552603, space group *Imma*) have cell metrics $a_{RP} \times 2\ a_{RP} \times c_{RP}$. These are of particular interest as they are rare examples of $AA'PbX_4$ stoichiometries (i.e. having two distinct, ordered interlayer cations). These structures cannot be derived directly from the RP parent phase, so we use the (0, ½)-shifted parent in space group *Ammm*.[22] Indeed, they are perfect examples of DJ2 type, but they also have a single octahedral tilt mode. From the *Ammm* parent, the tilt mode is designated $T_3^+$, and the corresponding tilt system is $a^+b^0c^0$/-($a^+$)$b^0c^0$, although we note that 1552603 was modelled with disorder of the Br ligands. It should also be noted that there is no possibility of an $a^+$ tilt mode in an RP-derived structure (i.e. there is no suitable irrep of the space group *I*4/*mmm*).

***Structures with more complex tilts and layer shifts***

Five further structures (1915486-659016), space group $P2_1/n$, metrics $a_{RP} \times 2\ a_{RP} \times c_{RP}$, display a more complex set of distortions (starting from the RP parent phase these are designated $\Sigma_3$, which is essentially a tilt, and $\Sigma_4$, a octahedral distortion) and there is additional symmetry lowering due two different types of layer shift; $M_5^-$ which acts along the *b*-axis, and $\Gamma_5^+$ (monoclinic distortion) which acts along *a*. The resulting tilts and displacements are shown in Fig. 8; although the symmetry is too low to define a rigorous tilt system, it is reminiscent of the $a^+b^0c^0$/-($a^+$)$b^0c^0$ type above. A further structure (1841680) in $P2_1/c$ ($a_{RP} \times c_{RP} \times 2\ a_{RP}$) has a similar resultant structure.

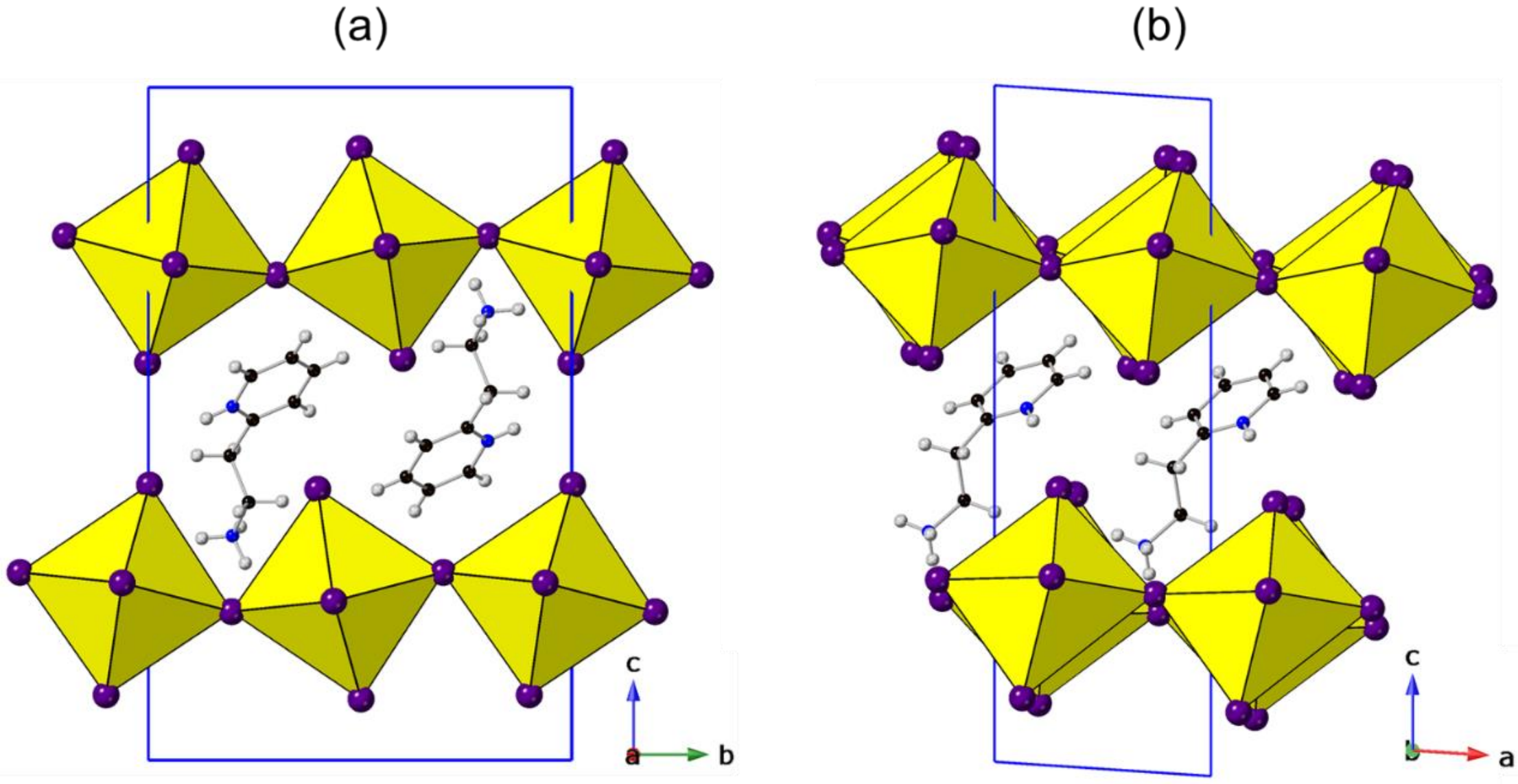

**Figure 8**. $[C_7H_{12}N_2]PbI_4$ (1944786) showing (a) $a^+/a^+$ tilts around *a* and (b) octahedral distortion (not a tilt) relative to *b*.

Finally, for this section, there are a few interesting and complex examples where *both* in-plane axes are doubled. The simplest of these is 628793, which has metrics $2\ a_{RP} \times 2\ a_{RP} \times c_{RP}$, space group *C*2/*c*. A projection of the structure down the *c*-axis is shown in Figure 9a. This combines two key modes: a rotation of octahedra around the *c*-axis, designated $P_4$ (a,-a), together with the now familiar $M_5^-$(a,a) shift mode, leading to an overall description $P_4$ (a,-a|b,b). These can be regarded as the primary order parameters, and acting together lead to the observed space group *C*2/*c* via the transformation matrix [(2,0,0) (0-2,0) (-1,0,-1)]. The $P_4$ mode is new to us, but it is relatively common in purely inorganic RP phases (see Balachandran et al,[30]). Acting alone, this mode would produce a unit cell size $\sqrt{2}\ a_{RP} \times \sqrt{2}\ a_{RP} \times 2\ c_{RP}$, whilst retaining a body-centred tetragonal symmetry and space group, $I4_1/acd$. However, coupled with the $M_5^-$ mode (which in itself produces no expansion of unit cell metrics, and space group *Pmmn*, seen in 1937296 (section 3.1.1)) the unit cell metrics become $2\ a_{RP} \times 2\ a_{RP} \times c_{RP}$. The resulting structure can be regarded as nDJ2, with a Glazer-like system approximately $a^0a^0c/a^0a^0(-c)$. Note that there is also a minor component of a more complex out-of-plane tilt mode ($P_5$(0,0;a,a)) allowed in these structures, in addition to the $\Gamma_5^+$ (monoclinic distortion) mode. This makes the pragmatic assignment of a Glazer-like tilt system difficult. For example, the following structure (1985833) has the same structure type but is nDJ, and contains a more significant $P_5$ tilt mode (Figure 9b). In general, the most useful way to assign a tilt system here will depend on the relative significance of the $P_4$ and $P_5$ modes and the degree of layer shift. The next four examples in Table 3 (1043214-184082) are polar variants (*Cc*) of essentially the same mode combination. The resultant structure type can vary between nDJ, nDJ2 and nRP, depending on the relative degrees of the $M_5^-$ and $\Gamma_5^+$ displacements. For example, 2016668 is nRP and the $P_5$ tilt mode is much more significant than the $P_4$ rotation. The next example (1542463) in polar space group *Pn*, appears to be a lower symmetry variant of the above, with an additional, minor contribution from an antiferrodistortive Pb atom shift perpendicular to the layers, of symmetry $N_2^-$ (compare to 1962913).

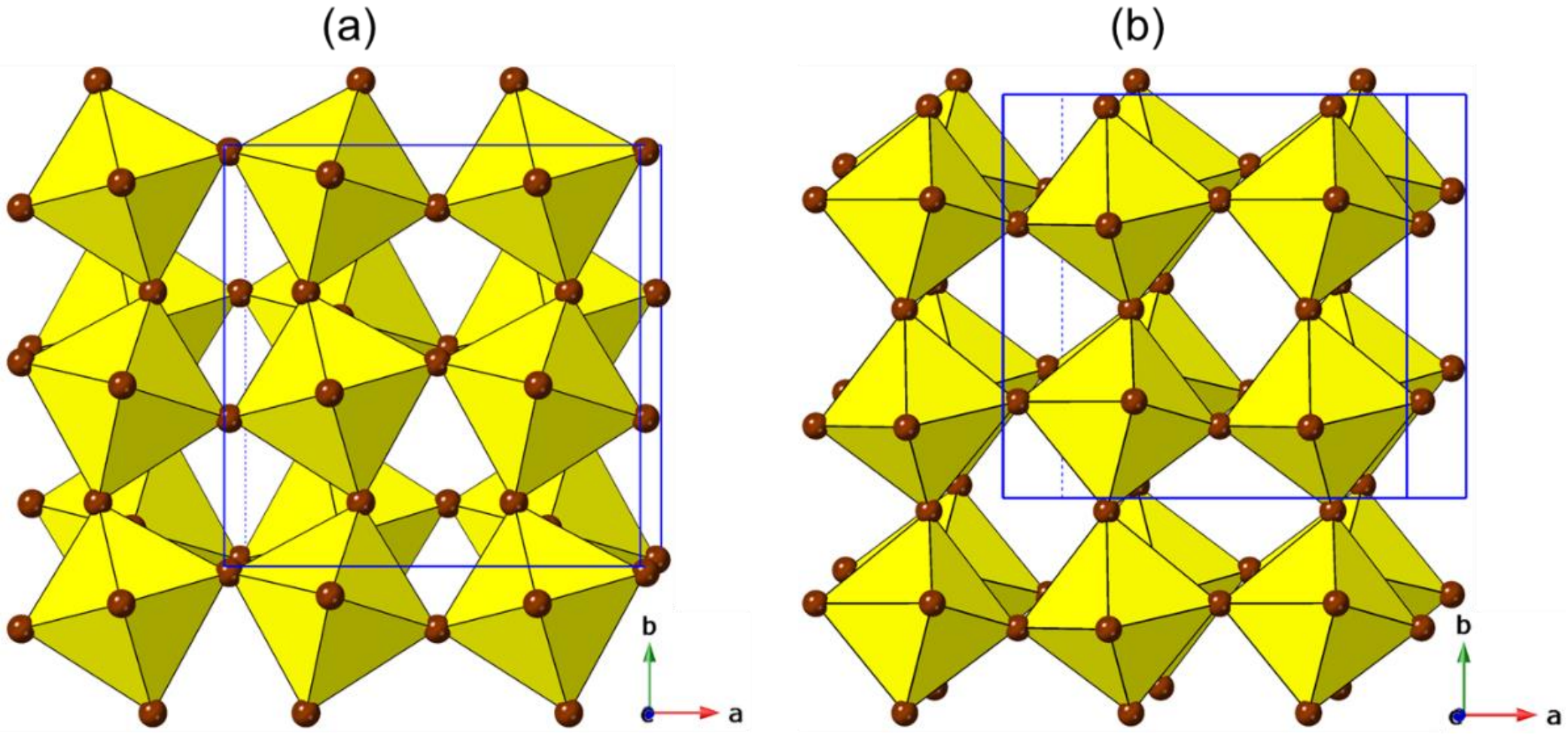

**Figure 9.** Comparison of two structures having the same unit cell metrics ($2\ a_{RP} \times 2\ a_{RP} \times c_{RP}$) and space group (*C*2/*c*), but exhibiting significantly different resultant tilt and shift behaviour (a) $[C_4H_{14}S_2N_2]PbBr_4$ (628793) and (b) $[C_2H_4N_3]_2PbBr_4$ (1985833).

The final three structures (1552604-1841683) in this section have metrics 2 $a_{RP} \times c_{RP} \times$ 2 $a_{RP}$. These larger, low symmetry unit cells have a diversity of allowed distortion modes, but often it is reasonable to pick out the most significant ones. The highest symmetry example (1552604) has space group *Pnnm*. This has the $M_5^-$ displacive mode leading to a nDJ2 structure, and tilt modes leading to a system $a^+b^0c$/ -($a^+b^0c$). The other two structures have four and two unique Pb sites, respectively. The former is close to RP type and has only a rotation mode around the *b*-axis, but there are significant, and differing, distortions of each of the Pb sites. The latter is borderline, RP-DJ with $a^+$-like tilts, but again, other distortion modes are significant.

### 3.2 Structures with one octahedral layer per unit cell (derived from DJ parent)

We chose to discuss structures with two octahedral layers per unit cell rather than a single layer first, not because they are 'simpler' but because they offer a much greater diversity of constituent distortions modes, from single tilt or displacement types to types with much greater degrees of freedom. In fact, the single layer sub-family, discussed in this section, has far fewer degrees of freedom, but nevertheless has surprisingly few examples of 'high symmetry' structures (only one centrosymmetric and orthorhombic, for example). In contrast, despite the structural diversity described in section 3.1, it can be noted that there is only a single example of a triclinic structure there. In this section we shall see that the vast majority of examples have monoclinic symmetry, and several derivatives of these have triclinic symmetry. In fact, 87 out of 108 structures in this section correspond to the same basic structure type! We'll derive these single layer structures (Table 4) from the DJ-type parent (space group *P*4/*mmm*). As we shall see, the common features highlighted in section 3.1, *viz*. octahedral tilting and layer shift modes, also occur here, but the mode labels used to describe them are necessarily different (i.e. different parent Brillouin zone). It is therefore helpful to point out the different labels used to describe the corresponding tilt modes between the two sub-families. These are, for the *I4/mmm* (RP) and *P4/mmm* (DJ) parent, taking *c* as the unique axis:

Rotation around the *c*-axis: $X_2^+$ (RP); $M_3^+$ (DJ)
In-phase tilt around the *ab* plane: not possible for RP; $X_3^+$ (DJ)
Out-of-phase tilt around the ab plane: $X_3^+$ (RP); $M_5^+$ (DJ)

**Insert Table 4 here**

**Table 4.** Summary of experimentally known structures with one octahedral layer per unit cell (derived from DJ parent).

A thorough study of the possible combination of tilt modes in DJ phases has been given by Aleksandrov,[25] and a briefer version, in the context of hybrid systems, by Li *et al.*[104] Layer shift modes in these systems are described by symmetry-lowering to monoclinic or triclinic (strain modes, $\Gamma_5^+$ for example). Note that there is no option of the antiferrodistortive layer shift mode (corresponding to the $M_5^-$ mode prevalent in section 3.1, using the *I*4/*mmm* parent) in this section, although many of the structures exhibiting those modes could equally well be derived from the *P*4/*mmm* parent by a $Z_5^-$ mode, which leads to doubling of the number of layers per unit cell.

### 3.2.1 Structures with no tilts but layer shifts, or tilts but no shifts

There are two examples (993479 and 1871404) with layer shift, but no tilting; both can be regarded as nRP due to the layer shift ($\Gamma_5^+$ mode). The second of these has an unusually large octahedral distortion. There is only one example of a structure type with a single octahedral layer per unit cell, with no layer shift but with octahedral tilting: this is perhaps surprising and contrasts with the common occurrence of such tilted/unshifted structure types in inorganic DJ phases. The example (120686) has the $M_3^+$ rotation mode and resulting tilt system $a^0a^0c$.

### 3.2.2 Structures with metrics $\sqrt{2}\, a_{DJ} \times \sqrt{2}\, a_{DJ} \times c_{DJ}$ or $c_{DJ} \times \sqrt{2}\, a_{DJ} \times \sqrt{2}\, a_{DJ}$

Apart from the examples above, the vast majority of structures (66) in this section (commencing 641641) also have metrics $\sqrt{2}\, a_{DJ} \times \sqrt{2}\, a_{DJ} \times c_{DJ}$ (space group $P2_1/a$) or $c_{DJ} \times \sqrt{2}\, a_{DJ} \times \sqrt{2}\, a_{DJ}$ (space group $P2_1/c$). These are essentially the same structure type, containing the $M_3^+$ rotation and $M_5^+$ (a,a) tilt modes (resulting in tilt system $a^-a^-c$, full model symbol (a|b,b)) and layer shift described, in part, by the $\beta$ angle of the monoclinic unit cell ($\Gamma_5^+$ mode). More precisely, the Δ parameters ($\Delta_1$, $\Delta_2$) can be derived simply by the equations:

for the $P2_1/a$ examples:

$$\Delta 1 = \Delta 2 = \frac{\boldsymbol{c} \times \boldsymbol{sin}(\boldsymbol{\beta} - \mathbf{90})}{\boldsymbol{a}}$$

and, for the $P2_1/c$ examples:

$$\Delta 1 = \Delta 2 = \frac{\boldsymbol{a} \times \boldsymbol{sin}(\boldsymbol{\beta} - \mathbf{90})}{\boldsymbol{c}}$$

We recall (section 3.1.2) that the corresponding tilt system ($a^-a^-c/$-$(a^-a^-)c$) is also the most common in the RP-derived structures, but in that case it is common for these compounds to retain orthorhombic symmetry and have perfect RP-like staggering of adjacent layers. This is symmetry-disallowed in the single layer structures, where combination of these two tilt modes naturally leads to monoclinic symmetry. Nevertheless, it is permissible for these structures to be close to DJ-type, with Δ values close to (0,0). In fact, the full range of Δ values spanning nDJ to nRP is observed (Table 4 and Figure 10). The difference between the $P2_1/a$ and $P2_1/c$ types is simply a choice of crystallographic setting, and has no consequence. The remaining structures in this sub-section fall into three sub-groups of the above ($P2_1$, $P\bar{1}$ and $P1$). They have the same tilt system, $a^-a^-c$, but additional distortions; the additional flexibility in unit cells angles describes the tendency towards structures with layer shifts away from the DJ-RP line.

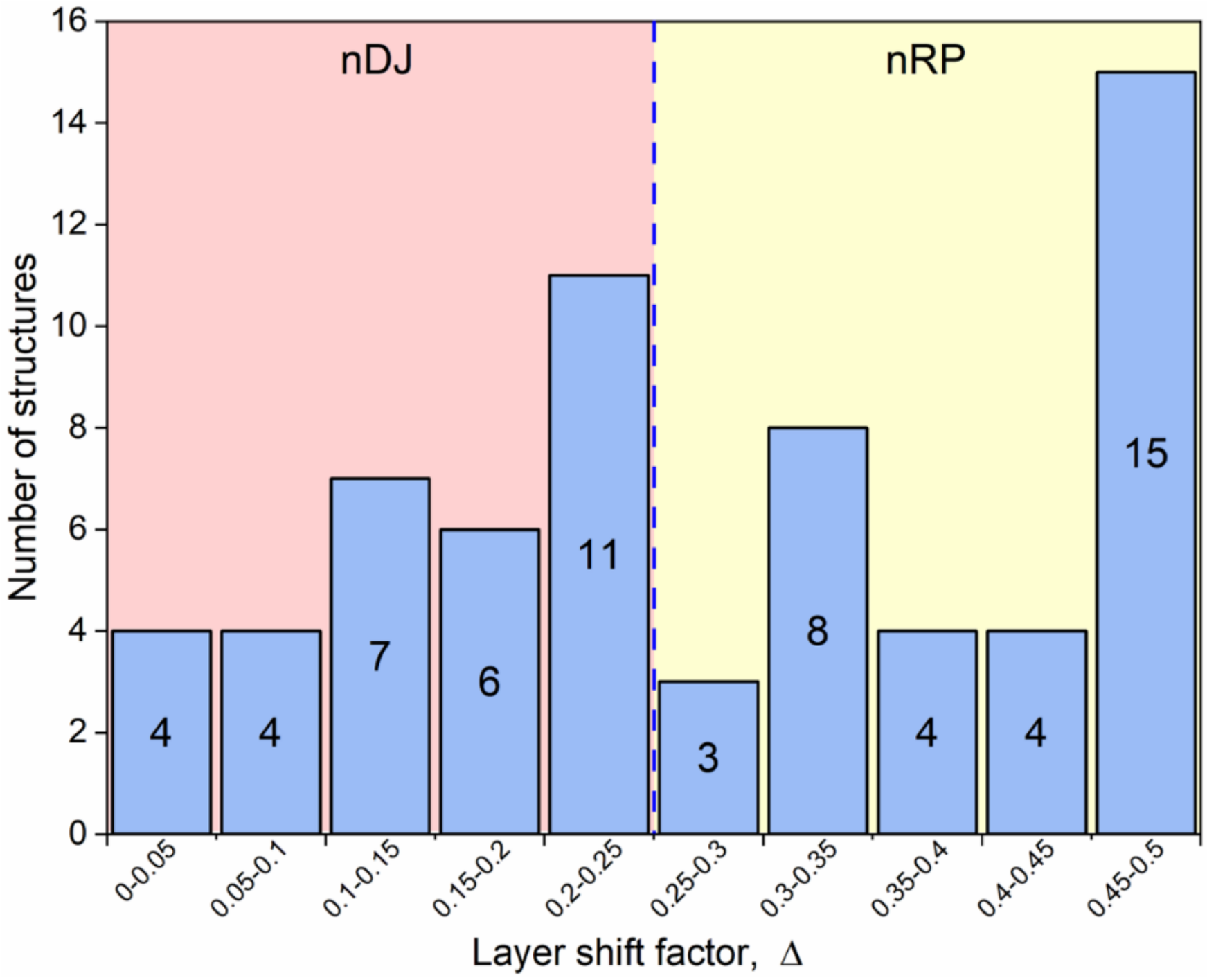

**Figure 10.** Histogram of layer shift factor values (Δ) for $P2_1/a$ and $P2_1/c$ structures from Table 4. Note that Δ consists of two components ($\Delta_1$, $\Delta_2$) but by symmetry $\Delta_1 = \Delta_2$ therefore only one of these is shown.

### 3.2.3 Structures with 2 $a_{DJ}$ or higher order supercells

There are few more complex supercells derived from the single layer DJ parent. The first (1883687) has metrics $a_{DJ} \times 2\, a_{DJ} \times c_{DJ}$ . The key distortion mode is a $X_3^+$(a,0) tilt mode (Figure 11) leading to tilt system $a^+b^0c^0$. This, in itself would lead to a doubled *b*-axis and *Pmma* symmetry, but additional minor distortions lower the symmetry to $P2_1$. The tilt is somewhat reminiscent of 1552603 in section 3.1.2, but in that case the sense of the $a^+$ tilt alternates in adjacent layers ($a^+b^0c^0$/-($a^+$)$b^0c^0$), leading to a further cell doubling. The remaining structures in this section have a unit cell quadrupled in the layer plane; i.e. metrics of the type $2\, a_{DJ} \times 2\, a_{DJ}$ or $2\sqrt{2}\, a_{DJ} \times \sqrt{2}\, a_{DJ}$ . They typically have low symmetry structures, but still exhibit conventional tilts as some of the largest amplitude modes, together with much smaller additional distortions. 1942547 has metrics $2\, a_{DJ} \times c_{DJ} \times 2\, a_{DJ}$, derived from simultaneous rotation ($M_3^+$(a)) and tilt ($X_3^+$(b,c)) modes. This combination leads to an ideal space group *Pmmn* and tilt system $a^+b^+c$, but the space group is lowered further to polar, *Pn*, due to very minor distortions from the ideal DJ type. 295291-1963066 have combinations of rotation/tilt ($M_3^+$ and $M_5^+$(a,a)) plus a shift mode leading to tilt system $a^-b^0c$ and nDJ2 structure type. The metrically-related structures of 1939809 and 1831525 effectively have a simpler tilt system, $a^0a^0c$, and DJ type, with the $M_5^+$ and $\Gamma_5^+$ modes present, but near-zero. Further, minor distortions lower the symmetry slightly from tetragonal to polar monoclinic (see also section 3.4). A series of structures, 1816279-1846391 have metrics $2\, a_{DJ} \times 2\, a_{DJ} \times c_{DJ}$ , the first having space group $P2_12_12$.

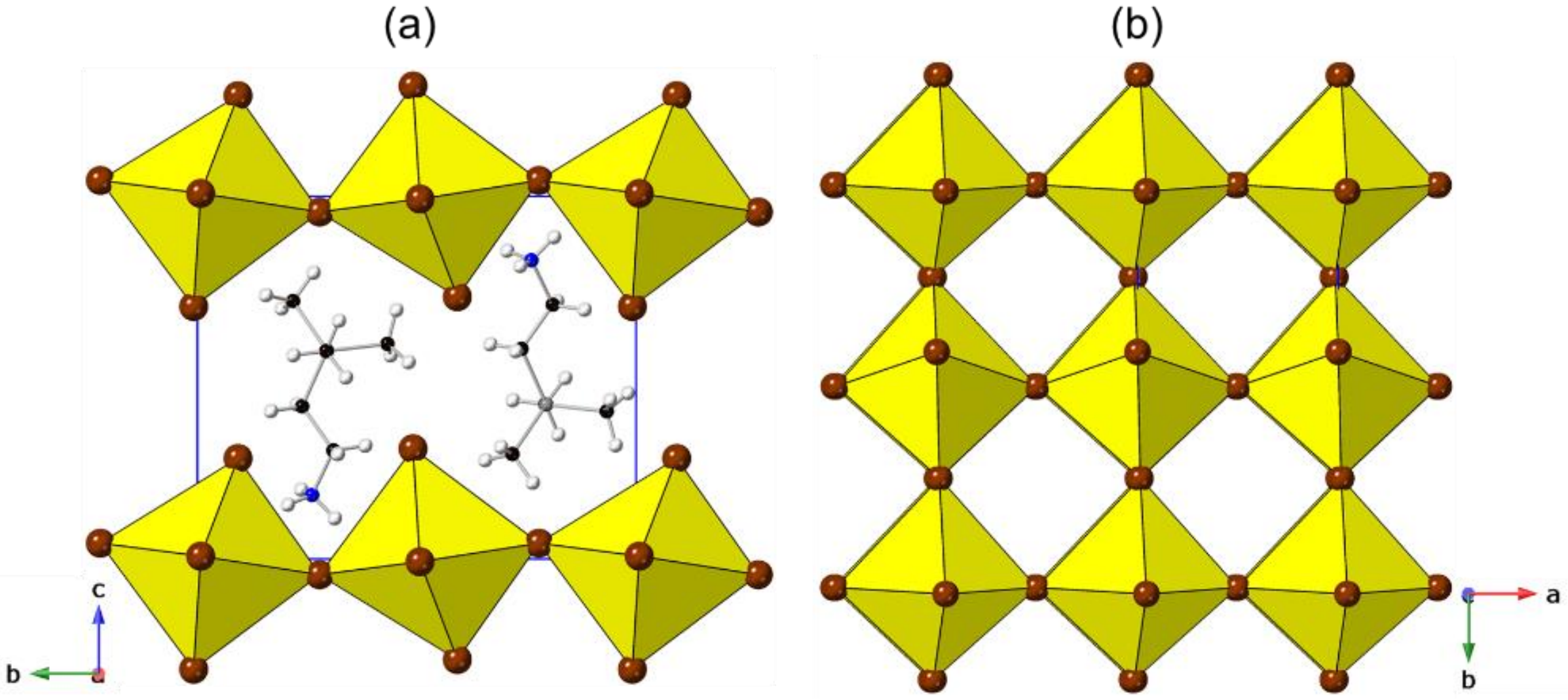

**Figure 11.** $[C_5H_{16}PN]PbBr_4$ (1883687) showing (a) tilting around *a* and (b) the view down the *c*-axis corresponding to the $a^+b^0c^0$ tilt system.

The dominant mode is the $M_3^+$ rotation. There is an additional mode, $X_3^-$, which describes an unusual in-plane antiferrodistortive shift of octahedra within each layer; however, the adjacent layers remain perfectly eclipsed, DJ-style (Figure 12). The metrically-related triclinic structures all have much higher pseudo-symmetry of the inorganic layers than the space group would suggest, typically with only a small number of modes with significant amplitudes. For example, 754084, 1498513 and 616101 have a dominant $A_3^+$ rotation mode: this is unfamiliar, but it is effectively a simple tilt system $a^0a^0c/a^0a^0(-c)$, which couples with a layer shift mode, bringing the structure to nDJ2 type. Without the additional layer shift the $A_3^+$ mode would lead to a doubling of the *c*-axis and space group *I4/mcm* (reminiscent of the situation in the standard Glazer system $a^0a^0c^-$ (Figure 1). 2011085 has the same modes but resulting in an nRP structure. 1861843 and 1846392 have the $M_3^+$ rotation mode, with an additional key mode $X_2^-$, which describes an antiferrodistortive displacement of Pb atoms away from their octahedral centres. The combination of these two modes does lead to a 2 × 2 × 1 supercell, but the highest, ideal symmetry is *Pmna*. The orthorhombic structure, 1875165, has metrics $2\sqrt{2}\, a_{DJ} \times c_{DJ} \times \sqrt{2}\, a_{DJ}$. The $M_3^+$ rotation is again the key mode, and the structure is DJ type, but further complexity arises from intra-octahedral distortions. Finally, 1934874 has a more complex superstructure with a unit cell six times the DJ aristotype ($c_{DJ} \times 3\sqrt{2}\, a_{DJ} \times \sqrt{2}\, a_{DJ}$), involving $M_3^+$ rotations, $M_5^+$ tilts and a rippling or undulation of the $[PbX_4]_\infty$ layers along the *b*-axis. This type of layer undulation/rippling is discussed further below.

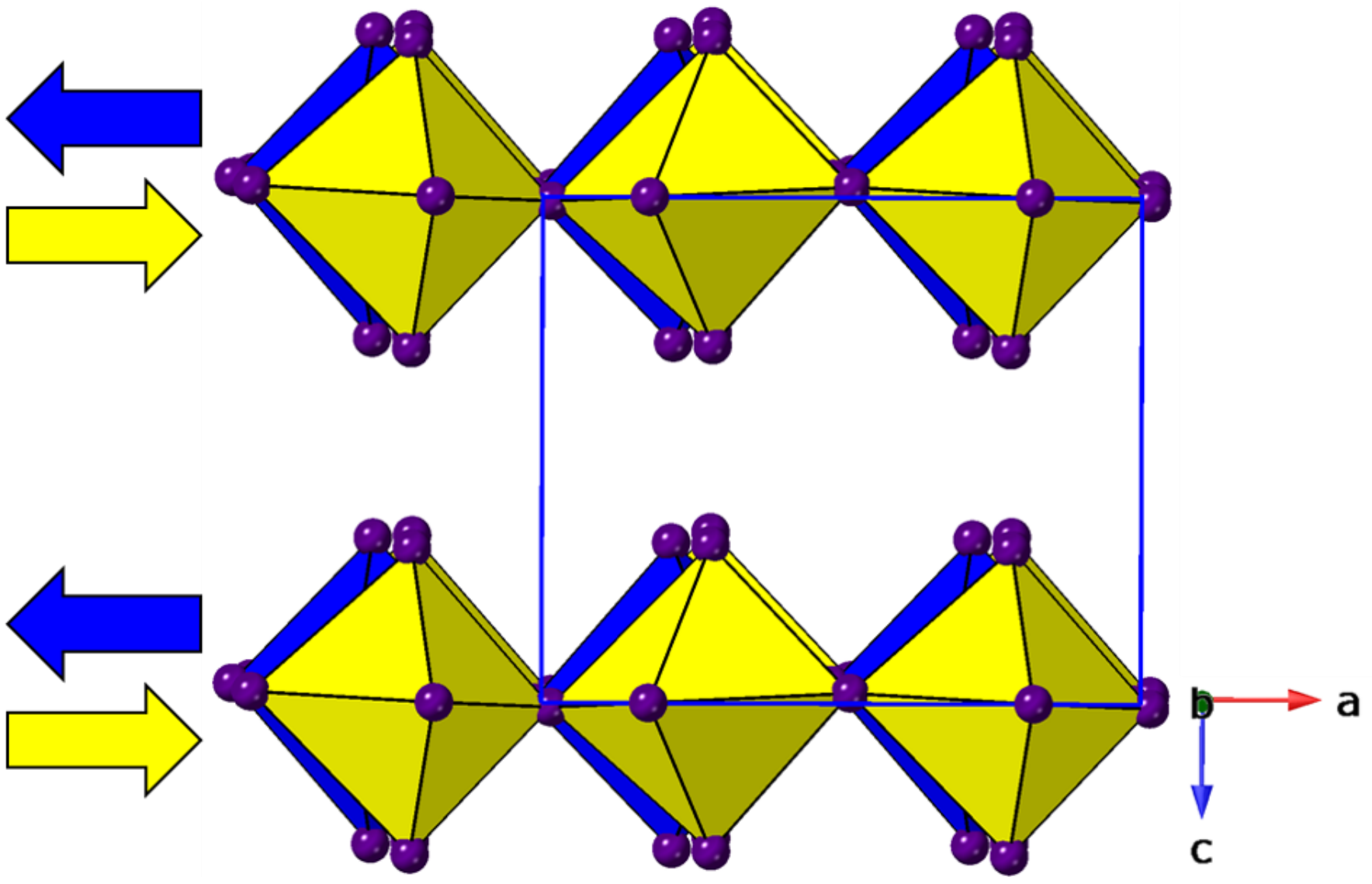

**Figure 12.** $[C_6H_{16}N_2]PbI_4$ (1816279) viewed along the *c*-axis highlighting the fully 'eclipsed' DJ arrangement with an antiferrodistortive shift of the octahedra in the layer plane.

### 3.3 More complex derivatives

In addition to the final example above, a few more complex structures have unit cells where at least one axis has a metric larger than $2a_{RP}$ or $2c_{RP}$. Ultimately, we find the most complex superstructure yet reported, with a unit cell volume 16 times the parent RP phase (i.e. 32 $[PbX_4]$ units per unit cell). These complex structures are discussed here and summarised in Table 5. Unique and unusual features are highlighted. An interesting observation is that the majority of these examples have $APbX_4$ stoichiometries.

**Insert Table 5 here**

**Table 5.** Summary of experimentally known structures with at least one unit cell metric larger than $2a_{RP}$ or $2c_{RP}$.

Several structures have metrics of the type $2\surd2\ a_{RP} \times \surd2\ a_{RP} \times c_{RP}$, the first set adopting space group *Pbca* (995699-1995236). Although the unit cell size here is four times the size of the RP parent, and therefore contains eight $[PbX_4]$ units, the relatively high symmetry of these examples still makes an analysis based on ISODISTORT very informative. In 995699, we immediately recognise the $X_2^+$(0,a) mode (i.e. octahedral rotation around the *c*-axis) and the $M_5^-$(b,0) shift mode, which acts along *b*, leading to an nDJ structure (Figure 13) and tilt system $a^0a^0c/a^0a^0c^-$. In addition, the type of 'rippling' distortion referred to above is also observed here.

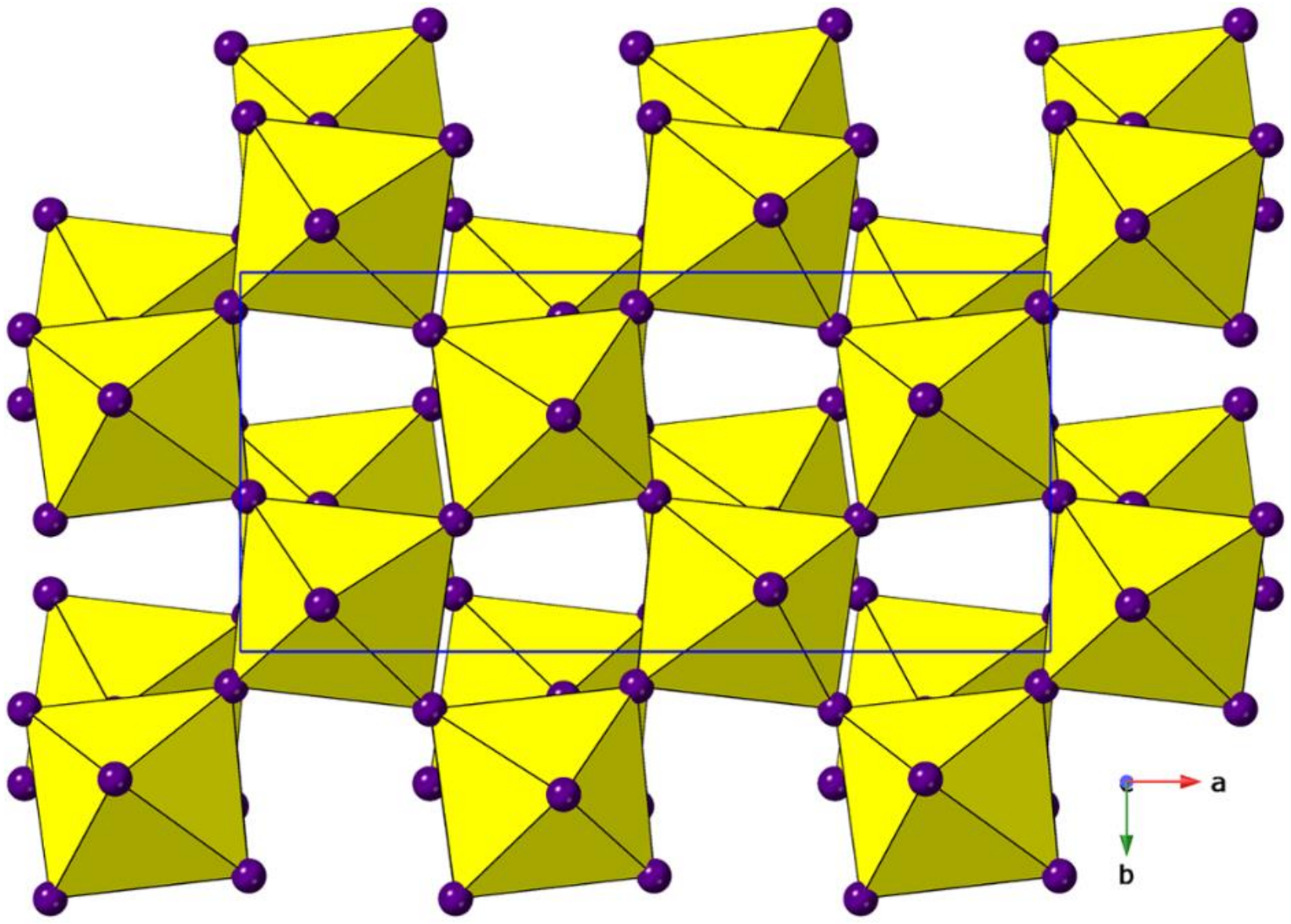

**Figure 13.** $[C_5H_{16}N_2]PbI_4$ (995699) viewed along the *c*-axis highlighting the nDJ arrangement with $a^0a^0c/a^0a^0c^-$ tilt system, corresponding to the $M_5^-$(b,0) shift mode and octahedral rotation around the *c*-axis attributable to $X_2^+$(0,a).

Looking down the $b$-axis a ‘sinusoidal’ rippling of the $[PbX_4]_\infty$ layers can be seen, with a repeat length of four octahedra (Figure 14a). Acting alone, the $X_2^+$ and $M_5^-$ modes would produce a unit cell of metrics type $\sqrt{2}\ a_{RP} \times \sqrt{2}\ a_{RP} \times c_{RP}$ and space group *Pbcm*: we note from section 3.1.2 that this type of distortion has not been seen, in isolation. The full mode label for this $X_2^+/M_5^-$ combination is (0,a|b,0). The additional supercell expansion is caused by the new layer rippling feature, which is described by modes with labels $\Delta_3$ and $Y_4$. This type of distortion was first noted in our recent example $(TzH)_2PbCl_4$[170] which has a ‘triple ripple’ rather than a ‘double ripple’ (Figure 14b). The following six *Pbca* examples adopt the same structure type. Naturally, the amplitudes of octahedral rotation, layer shift and ‘rippling’ are variable within this family with 1838616, for example showing almost zero octahedral rotation ($X_2^+$) and having a smaller $M_5^-$ shift, leading to nRP status. The next structure in Table 5 (724584) is a derivative of this structure type but with additional degrees of freedom. Although an $X_3^+$ rotation mode is permitted in this symmetry it has effectively zero amplitude; the resultant structure is nDJ, with tilt system close to $a^0a^0c/a^0a^0c^-$. 1995236 has a structure related to those above, but with an additional doubling of the $b$-axis. Taken together, the three structures in Figure 14 show a trend where the layer rippling feature produces axes of repeat lengths of four, six and eight octahedra, i.e. $2\sqrt{2}$, $3\sqrt{2}$ and $4\sqrt{2}$ times the parent). 1831521 has $\sqrt{2}\ a_{RP} \times 2\sqrt{2}\ a_{RP} \times c_{RP}$ metrics and space group $P2_1/c$. The structure is close to DJ, and has $X_3^+$ tilt, $M_5^-$ shift and more complex distortions.

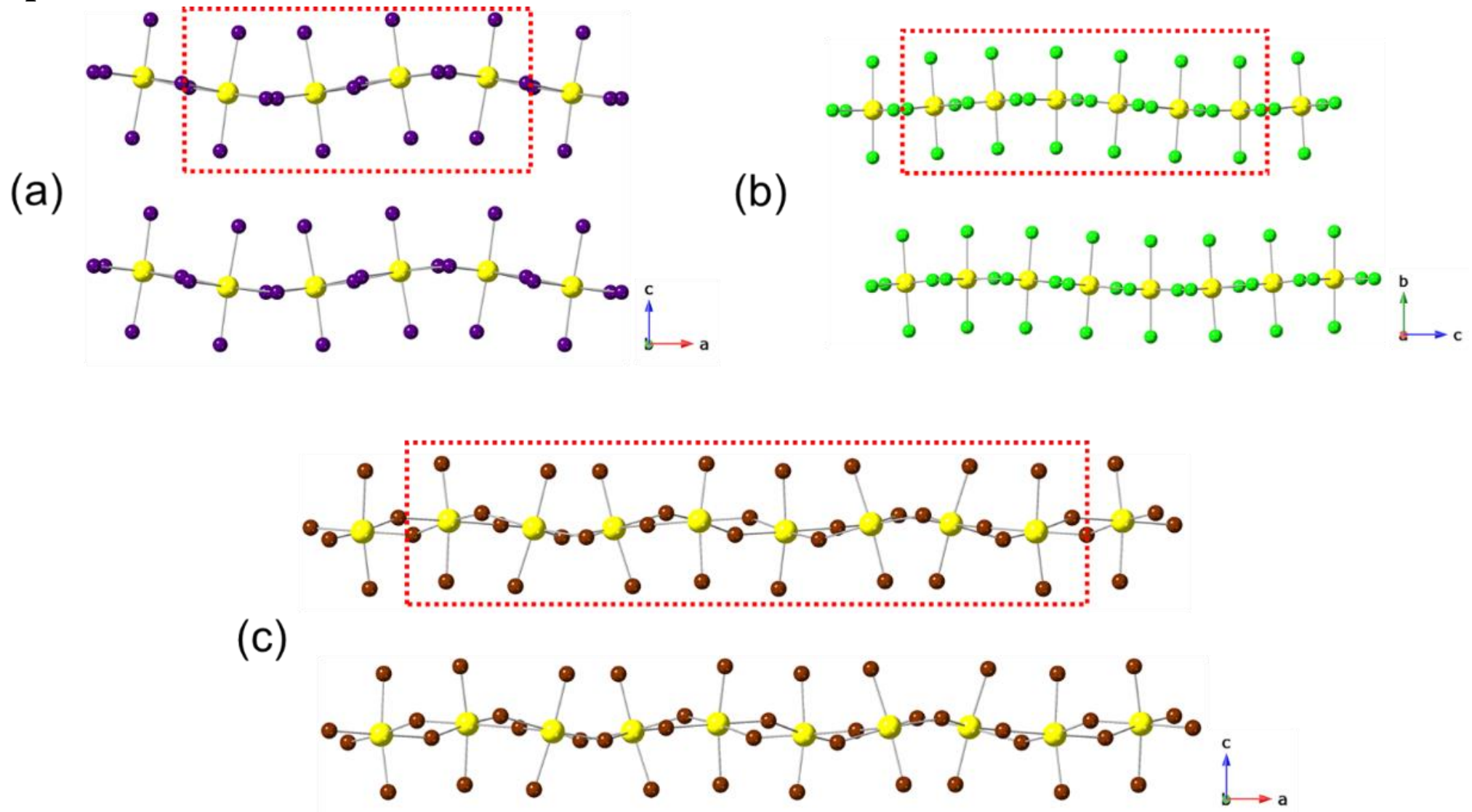

**Figure 14.** Ball and stick representations of (a) $[C_5H_{16}N_2]PbI_4$ (995699), (b) $[C_2H_4N_3]_2PbCl_4$ ($[TzH]_2PbCl_4$) and (c) $[C_9H_{14}N]_2PbBr_4$ (1995236). These compositions feature doubled, tripled and quadrupled ‘sinusoidal rippling’, respectively. This is highlighted by the dashed red box in each.

Three structures are reported to adopt metrics with six times the volume of the parent RP phase. Two polymorphs (1937299, 1937297) have a cell with metrics $c_{RP} \times 3\ a_{RP} \times 2\ a_{RP}$. These are part of a series of phases versus temperature: polymorphism and phase transitions are discussed further in section 3.5. Structure 1937299 has space group *Pccn*, with a very unusual tilt system. This is shown in Figure 15, where the two crystallographically distinct Pb-centred octahedra are shown in different colours. Viewed down the *c*-axis it can be seen that every 3rd octahedron (Pb1) along the *b*-axis is unique, giving rise to the unit cell tripling. Viewing along the *b*-axis, the underlying reason for this can be seen: The Pb1 octahedron is effectively untilted around *b*, but the two Pb2 octahedra are tilted out of phase relative to each other ($a^-$ type). The unusual tilt pattern is described by a C2 (⅓, ½, 0) mode, with the other key mode being the expected $M_5^-$ shift.

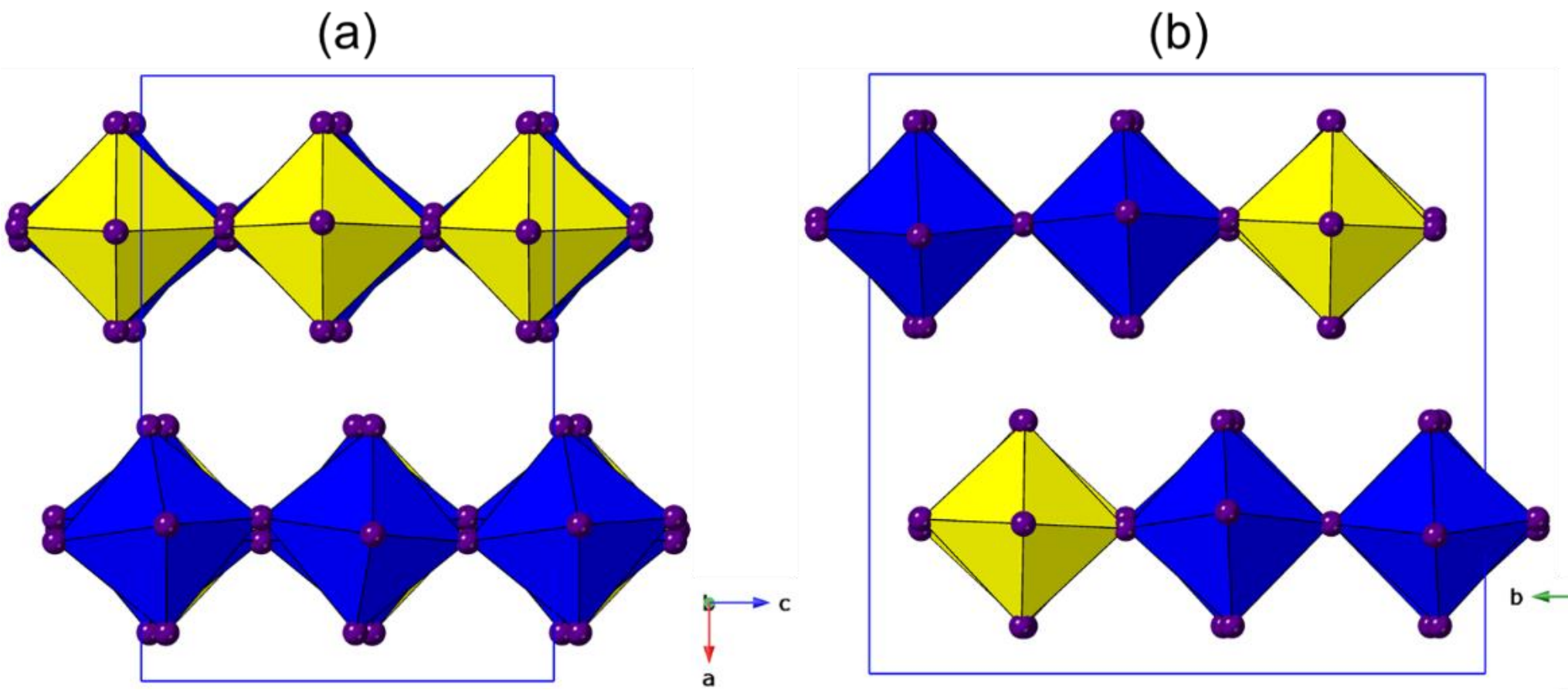

**Figure 15.** $[CH_7N_2]_2PbI_4$ (1937299) showing (a) no tilting of the Pb1 (yellow) and out-of-phase tilting of Pb2 (blue) octahedra around *b*. (b) Highlights that every third octahedron along the *b*-axis is unique, giving rise to unit cell tripling.

Structure 1937297 has essentially the same behaviour, with symmetry lowering caused by ordering of the organic moieties. Similar, unusual and complex tilt systems have been seen previously in traditional oxide perovskites such as $NaNbO_3$,[171] but we are unaware of any previous example of this type in layered perovskites. Structure 1963065 has metrics $3\ a_{RP} \times 2\ a_{RP} \times c_{RP}$; this has a distinct, but equally unusual combination of in-phase and out-of-phase tilts/ distortions (Figure 16). Three structures have unit cell volumes eight times the parent RP cell. 1887281 has metrics $2\sqrt{2}\ a_{RP} \times 2\sqrt{2}\ a_{RP} \times c_{RP}$, space group *Pbcn*. The structure is nRP, incorporating a minor sinusoidal undulation along the *a*-axis, with a repeat of four octahedra ($\Delta_3$ and $Y_4$ modes). There is also a slight offset of octahedra along *a*, in addition to the $M_5^-$ shift along *b*, which can be compared with the slightly simpler situation in the *Pbca* structures above. 1982717 and 1838611 appear to be polar analogues of this structure.

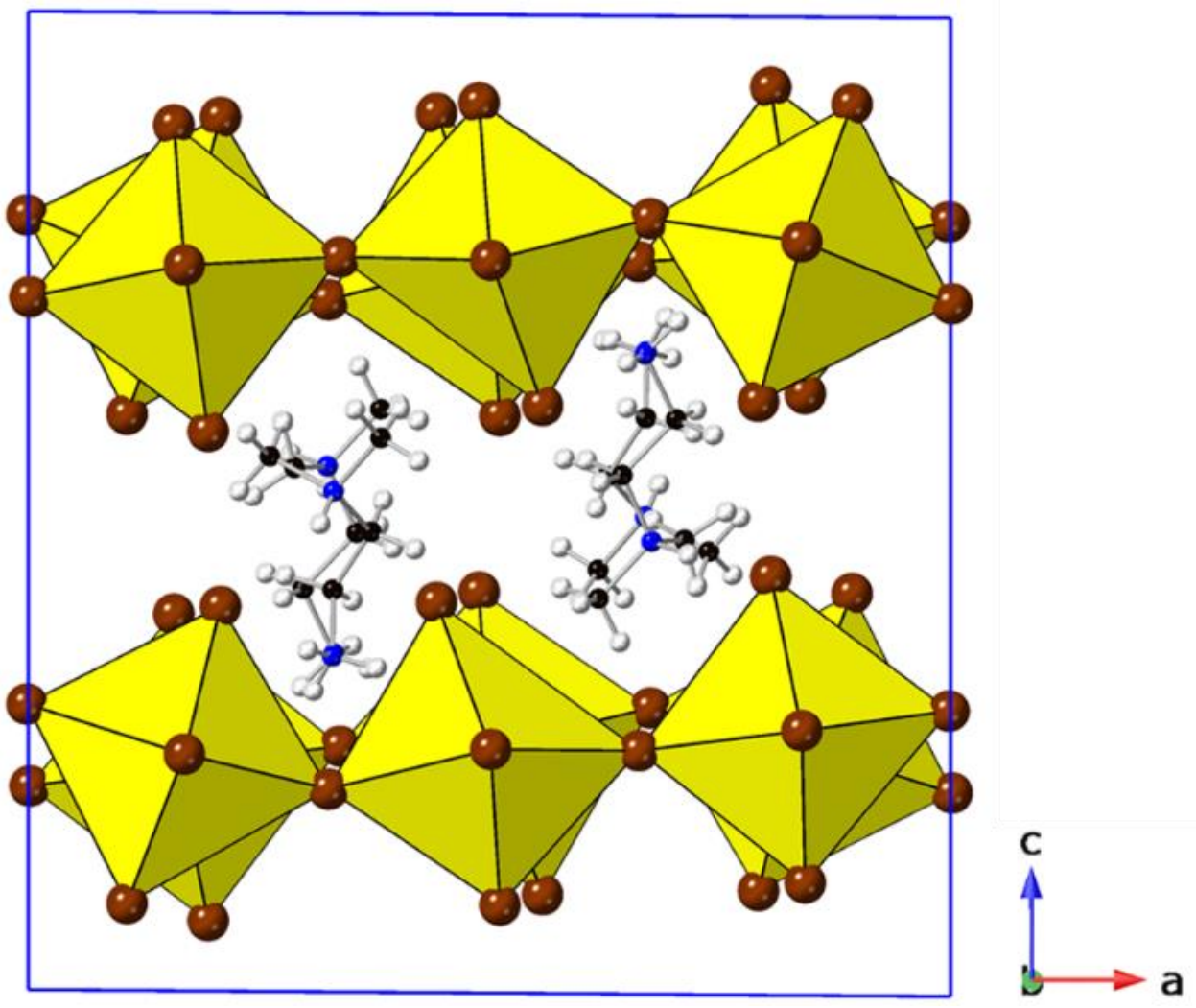

**Figure 16.** Unusual combination of in-phase and out-of-phase tilts/distortions observed in $[C_4H_{14}N_2]PbBr_4$ (1963065).

The two largest supercell derivatives have unit cell volumes of 12 and 16 times the RP parent, and both have four octahedral layers per unit cell repeat. The first is 1521060, having metrics $3\sqrt{2}\, a_{RP} \times \sqrt{2}\, a_{RP} \times 2\, c_{RP}$. This complex structure has seven independent Pb sites but does appear to have a relatively high pseudo-symmetry when viewed down the *c*-axis (Fig 17). The nDJ character is clear from this view, although there are slight shifts, consecutively from one layer to the next. Each layer also has the conventional $M_3^+$ rotation and $M_5^+$ tilt modes, when derived from the DJ parent, but there is an additional complex mode, which undulates the layers slightly.

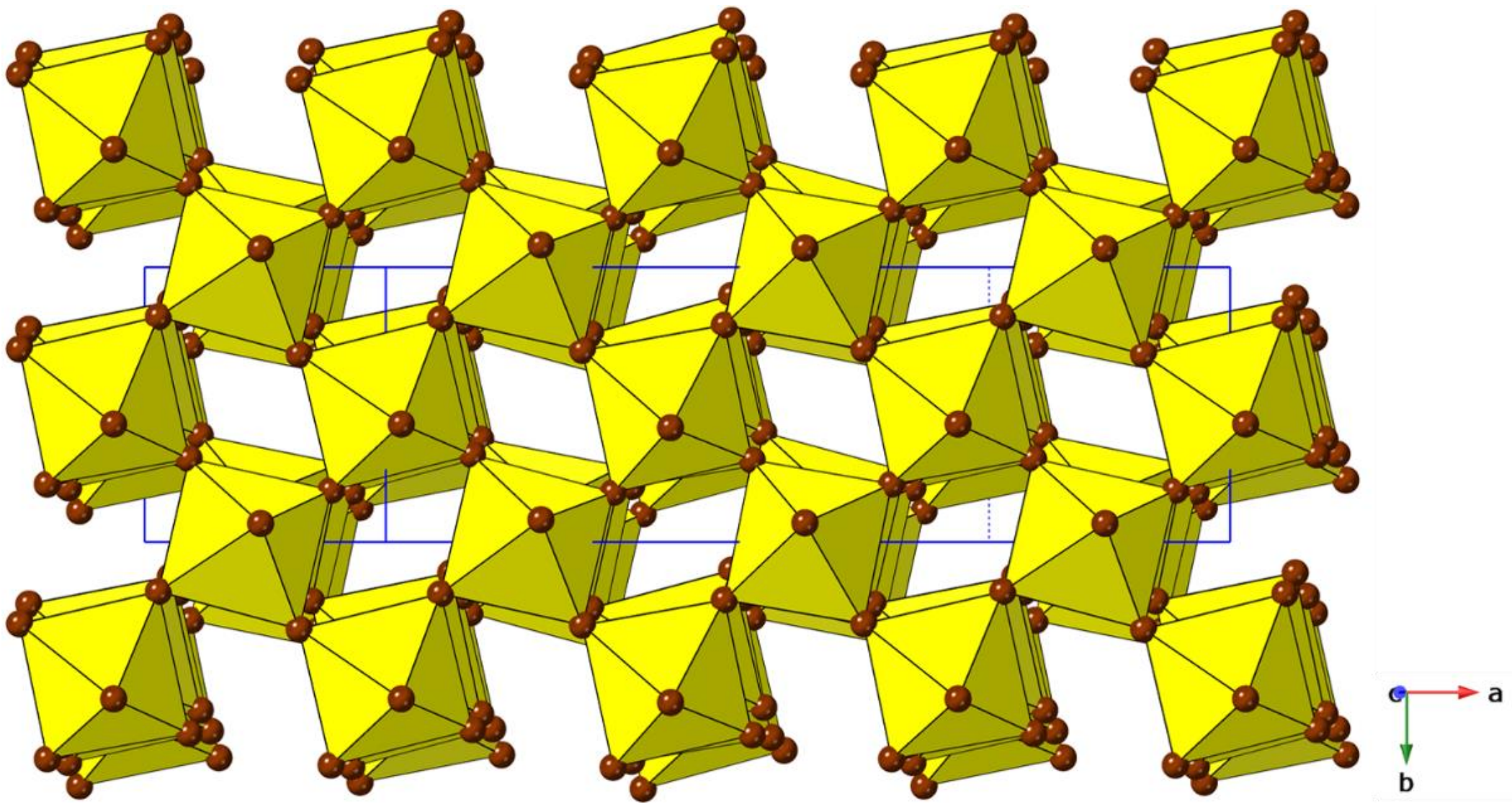

**Figure 17.** $[C_4H_{10}O_2N]_2PbBr_4$ (1521060) viewed in the *c** direction showing the large degree of pseudo-symmetry present.

The final structure, 1963067, has a unit cell volume 16 times the RP parent, metrics 2 $c_{RP}$ × 4 $a_{RP}$ × 2 $a_{RP}$, with polar orthorhombic space group *Aba2*. It is perhaps not helpful to describe the full mode details for such complex structures. In fact, the underlying key modes are much simpler, and they reveal an interesting result. The primary order parameters can be regarded as a C2 (¼, ½, 0) mode: a complex combination of tilts/distortions around *c*, and a complex layer shift, designated $\Lambda_5$, which acts along *b* and shifts only every alternate layer. Acting alone, the C2 mode would actually lead to directly to a polar space group, *Abm*2, and unit cell metrics $c_{RP}$ × 4 $a_{RP}$ × 2 $a_{RP}$ (Figure 18). This phase may therefore be regarded as a potential improper ferroelectric. The additional doubling of the *a*-axis arises from the combination C2 ⊕ $\Lambda_5$, which directly produces the observed unit cell metrics and space group. The familiar $M_5^-$ layer shift (acting along *c*) and the $\Gamma_5^-$ polar modes may be regarded as arising as secondary effects of C2. The C2 mode here is more complex than the C2 (⅓, ½, 0) mode observed in 1937299: in that case the C2 mode acting alone also leads to the observed unit cell metrics and space group, but centrosymmetricity is retained.

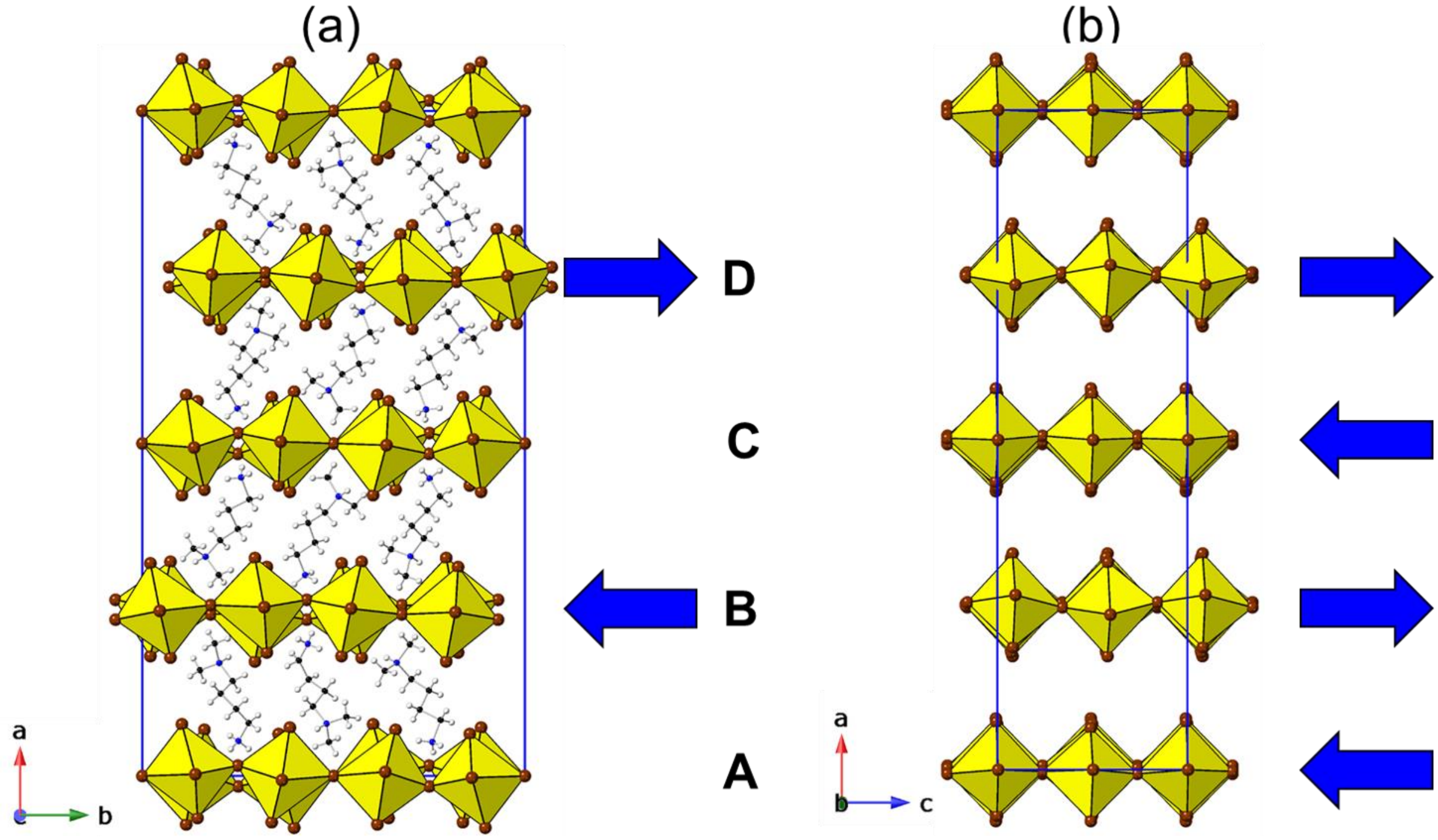

**Figure 18.** The complex structure of $[C_6H_{18}N_2]PbBr_4$ (1963067) (a) viewed down *c*, showing the complex C-mode tilt/distortion and effect of the unique $\Lambda_5$ shift mode (note only the B and D layers are affected by this) (b) viewed down *b*, showing the effect of the common $M_5^-$ mode. The result of these superposed shifts is that the A and C layers are perfectly eclipsed (DJ) relative to each other, but all directly neighbouring layers are nDJ2.

## 3.4 Influence of the interlayer species on octahedral layer architecture

All of the analysis in sections 3.1-3.3 is based on the architecture of the $[PbX_4]_\infty$ layers only, and makes no reference to the nature of the interlayer organic moieties; indeed, in most cases, it was carried out without knowledge of these! In the crystallography of purely inorganic perovskites and layered perovskites,[23–25,30] octahedral tilting is the primary influence on the resultant crystal symmetry, and space groups can be predicted and understood directly from the constituent tilts. In simple 'Glazer-like' tilt systems, these space groups are necessarily centrosymmetric (i.e. octahedral tilting retains inversion symmetry at the B-site). However, more complex tilt combinations or combinations of tilts plus cation ordering or other modes, can lead to non-centrosymmetric (NCS) space groups, especially in layered systems.[172] Indeed, this fact has been used as a design principle in recent work in the burgeoning field of hybrid improper ferroelectrics (note that 'hybrid' does not refer to inorganic-organic system here but to the inducement of polarity by cooperative action of two distinct modes[172,173]). From Tables 1-5 we can see that NCS, and polar, space groups are not uncommon in hybrid layered perovskites. However, any structure derived from the parent $n$ = 1 RP or DJ phase with combinations of only two tilt modes and no layer shift are necessarily centrosymmetric. The structures observed in these compounds are obviously dictated by the total energetics of the system, which will depend on a subtle interplay and co-operation between the molecular moieties and the inorganic framework. An interesting recent example,[155] which leads to unusual physical properties,[174] is that of $[C_6H_{16}N_2]PbI_4$ (1939809 and 1831525, Table 4), which exhibits an order-disorder transition of the 4-aminopiperidine, but this hardly changes the distortions within the $[PbI_4]_\infty$ layer itself. In this section, we will consider structural trends across this family of compounds, and we highlight a few interesting cases of the nature of the interlayer cations' influence on the overall structure. This is not intended to be an exhaustive survey or rationalisation; rather, we hope it encourages workers in the field to consider some of the structural principles we have introduced in this paper, in further understanding existing compounds, and possibly in designing-in particular features in future work.

### 3.4.1 The most common tilt systems

It can readily be seen from Tables 2 and 4 that some unit cell metrics and space groups occur very frequently, and our analysis in section 3.3 reveals clearly that these symmetries can be rationalised by the underlying octahedral tilt modes and resultant tilt systems. The two most common are based on $\sqrt{2} \times \sqrt{2}$ supercells within the layer plane, and space groups $P2_1/a$ (equivalently $P2_1/c$) in Table 4 (commencing no. 641641) and *Pbca* and *C*2/*c* (plus *Cc*) in Table 2 (commencing 1938881). Together, these examples account for 99 of the total of 259 structures in our review. Each of these types has a combination of two octahedral tilts, specifically out-of-phase tilts out of the $[PbX_4]_\infty$ plane and rotations of octahedra perpendicular to this plane, leading to symbols $a^-a^-c$, $a^-a^-c$/-($a^-a^-$)$c$ or $a^-a^-c$/$a^-a^-c$. On closer inspection of the types of amine that give rise to these structures, we see a marked tendency in Table 2 for the *Pbca* structure ($a^-a^-c$/-($a^-a^-$)$c$) to be directed by linear chain amines, $RNH_3^+$, resulting exclusively in stoichiometry $A_2PbX_4$. This corresponds to the 'traditional' idea of the RP phase in terms of both stoichiometry and (½, ½) layer staggering. Interestingly, however, the $a^-a^-c$/$a^-a^-c$ derivatives (*C*2/*c* and other lower symmetry variants such as 1826587, 1119707, 961380 etc.) sometimes arise from aliphatic diamines: i.e. these diamines apparently introduce a symmetry-lowering layer shift variable and prompt a tendency away from idealised RP towards intermediate layer shifts. For the $a^-a^-c$ systems in Table 4, a much wider variety of amines is accommodated, most commonly aliphatic amines, but also some aliphatic and other diamines. It seems that this type of tilt system (which is closely related to the most common $GdFeO_3$-type structure in conventional cubic perovskites) is intrinsically quite stable, and robust to many different types of interlayer species, especially given the additional degree of flexibility available via layer-shift modes.

### 3.4.2 Cl vs Br vs I

Several amines have been successfully incorporated within chloride, bromide and iodide systems. Some of these adopt the same structure type, some choose different structures, depending on the halide, and some have been shown to display temperature-dependent phase transitions. Obviously, care must always be taken in structural comparisons, to ensure that the temperature of structure determination is considered. The amines that form the same structure type for each halide are cyclopropylamine, cyclobutylamine and 4-methylbenzylamine (all of which adopt the most common $a^-a^-c$ structure, $P2_1/c$, Table 4) and *N*,*N*-Dimethyl-*p*-phenylenediamine which exhibit the more unusual $a \times 2a \times c$ ($P2_1/n$) structure in Table 3. The cyclopropylamine/cyclobutyamine series form part of a systematic series of studies by Billing and co-workers[43,74] which reveal several interesting trends in terms of the effect of ring size on the position and orientation of the amine between the layers (also revealing that ring sizes larger than six prefer to form structures containing chain-like architectures rather than layered perovskites).

Some examples of amines that form different structures for different halides are cyclopentylamine (numbers 708568, 708562 and 609994), cyclohexylamine (e.g. 708569, 708563, 609995), 1,6-diaminohexane (1914631 and 150501/150502) and cystamine (1841478, 628793 and 724583/4). The first two types here are part of Billing's studies,[43,74] which draws some interesting observations regarding layer staggering and interlayer distances, which we shall not repeat here. However, it should be noted that the analysis of Billing is incomplete in its interpretation of 'staggering' (reported only as either 'eclipsed' or 'staggered').

In the case of 1,6-diaminohexane, the chloride (1914631, Table 2) adopts a nDJ2 structure type. Although this symmetry permits both octahedral rotations and tilts (tilt system formally $a^-a^-c/$-($a^-a^-c^-$) the tilt mode amplitude is near zero, and the only significant modes are the $X_2^+$ rotation and the $M_5^-$ and $\Gamma_5^+$ shifts. In contrast, the bromide and iodide are isostructural, exhibiting nRP structures, with the common $a^-a^-c$ tilt system (150501/2, Table 4), differing only slightly in the degree of layer shift. In each case the tilt mode is significant. A key difference between the chloride and the bromide/iodide is that in the former the cation adopts a fully stretched (all *trans*) conformation, whereas in the latter pair the terminal C-C bond has a *gauche* kink. This leads to an interesting 'inverse' variation in the interlayer distances: Cl (12.32 Å) > Br (12.02 Å) > I (11.86 Å)

For the cystamine derivatives, the differences in crystal structure (triggered by the differing conformations of the cation) have been related in some detail to the optoelectronic properties by Krishnamurthy *et al.*[96]: the chloride displays broadband white luminescence, despite having a low level of structural distortion of the inorganic layers. In fact, the chloride (1841478) exhibits a structure very similar to that of the 1,6-diaminohexane analogue just discussed: the tilt system and degree of layer shift is almost the same. Although it might be expected that this similarity arises from the similar nature of the eight-atom linear diamine chain, in fact the chain conformation are very different, with a much 'tighter' configuration here (all torsion angles in the range -73 to -78º). The bromide (628793) has already been discussed in section 3.1.2, and exhibits the unusual $P_4$ tilt mode. Despite the difference apparent in unit cell metrics and space group compared to the chloride, the resulting layer topology is actually very similar: this time the tilt mode is disallowed, rather than just very small. In contrast, the iodide exists in two polymorphs which co-exist over a wide temperature range (ref). The high temperature $\beta$-phase, has the common $a^{-}a^{-}c$ $P2_1/a$ structure (724583) with disordered cystamine moieties, whereas the ambient temperature $\alpha$-phase has a more complex supercell discussed in section 3.3. Note that there is a significant change in the inorganic layer from $\alpha$ to $\beta$, with a change of rotation mode from $c$ to ($-c$), in addition to the changes in the conformation and positioning of the cystamine. Although the tilt system and layer shift in the $\alpha$-phase result in a similar overall structure to that of the chloride and bromide, the expanded superlattice is caused by ordering of two distinct enantiomeric forms of the cystamine: all these interesting features are discussed in more detail by Louvain *et al.*[121]

For *n*-butylamine a more diverse range of phases has been reported, which are discussed in section 3.5.

### 3.4.3 Homologues and isomers

Differences in amine structure can have a significant impact on bonding motifs, spatial arrangements and subsequently structural distortions. Due to the large diversity in amines that have been utilised in these materials only some general observations regarding structurally-related amines will be discussed. The simplest of these are linear aliphatic amines which have been extensively studied by Billing and co-workers.[47,71,78] In their work they prepared and characterised compositions of the general formula $[C_nH_{2n+1}NH_3]_2PbI_4$ where $n$ = 4-18. The ambient temperature structures for all compositions adopt the RP type *Pbca* structure ($a^{-}a^{-}c/-(a^{-}a^{-})c$) with no real structural change other than an increase in the interlayer distance, consistent with the increase in chain length. While there are phase transitions observed for $n$ = 8, 10, 12, 14, 16 & 18 compositions at higher temperature, these are primarily related to changes in conformation of the amines. These structures seem systematic in nature suggesting that the primary effect of increasing chain length is to increase the interlayer distance. While straight chain alkyl amines can be considered as the simplest amines used in these materials, various other amine types have been utilised. One of the most common structural features of the amines featured here is the inclusion of an aromatic component (>100 examples). Due to the large variability of substitution and type of these, only examples based on, or closely-related to, the (2-aminoethyl)pyridine structure will be discussed here (see ESI for amines being compared). Recent work by Febriansyah and co-workers[112] compared the structural effect of positional isomers of the (2-aminoethyl) substituent to the pyridine ring. In their work they reported an increase in the interlayer I-I separations consistent with the movement of the (2-aminoethyl) substituent away from the N of the pyridine ring. However, comparison of the tilts and layer shift of these materials shows very little change with respect to the substitution position, with the 2-, 3- and 4-(2-aminoethyl)pyridine lead iodide compositions (1944786, 1944783 and 1944782, respectively) all adopting the $a \times 2a \times c$ ($P2_1/n$) structure (Table 3) with very similar layer shifts. There is slight variation in the *N*-(2-aminoethyl)pyridine composition (1841680, section 3.1.3) which adopts the $P2_1/c$ variant, likely as a result of the closely situated positively charged N atoms.

The effect of positional isomerism can be further explored by considering the different structures obtained *via o*-, *m*- or *p*- fluorine substitution in the closely related phenylethylamine structure (1893383, 1893384 and 1488195)[107,134,148]. Unlike the related (2-aminoethyl)pyridine compositions there are significant differences in the structures adopted depending on the position of the fluorine atom (Fig. 19). In both the *m*- and *p*- substituted compositions the amines are arranged with the aromatic group oriented in the middle of the intralayer region, with configurations that maximise intermolecular F---F interactions (F---F ~3.04 and 3.42 Å, respectively). Both structures appear to be nRP with a greater degree of layer shift in the *m*-substituted composition, consistent with the enhanced F---F interaction present. In the *o*-substituted composition it is no longer possible for the amines to orient to maximise F---F interactions, however the proximity of the electronegative F group to the $NH_3^+$ of the ethylammonium chain induces a conformational change to maximise the F---H-N interaction (~3.25 Å) resulting in a layer shift close to DJ2. This difference between closely-related amines highlights the importance of the amine in the structural behaviour observed.

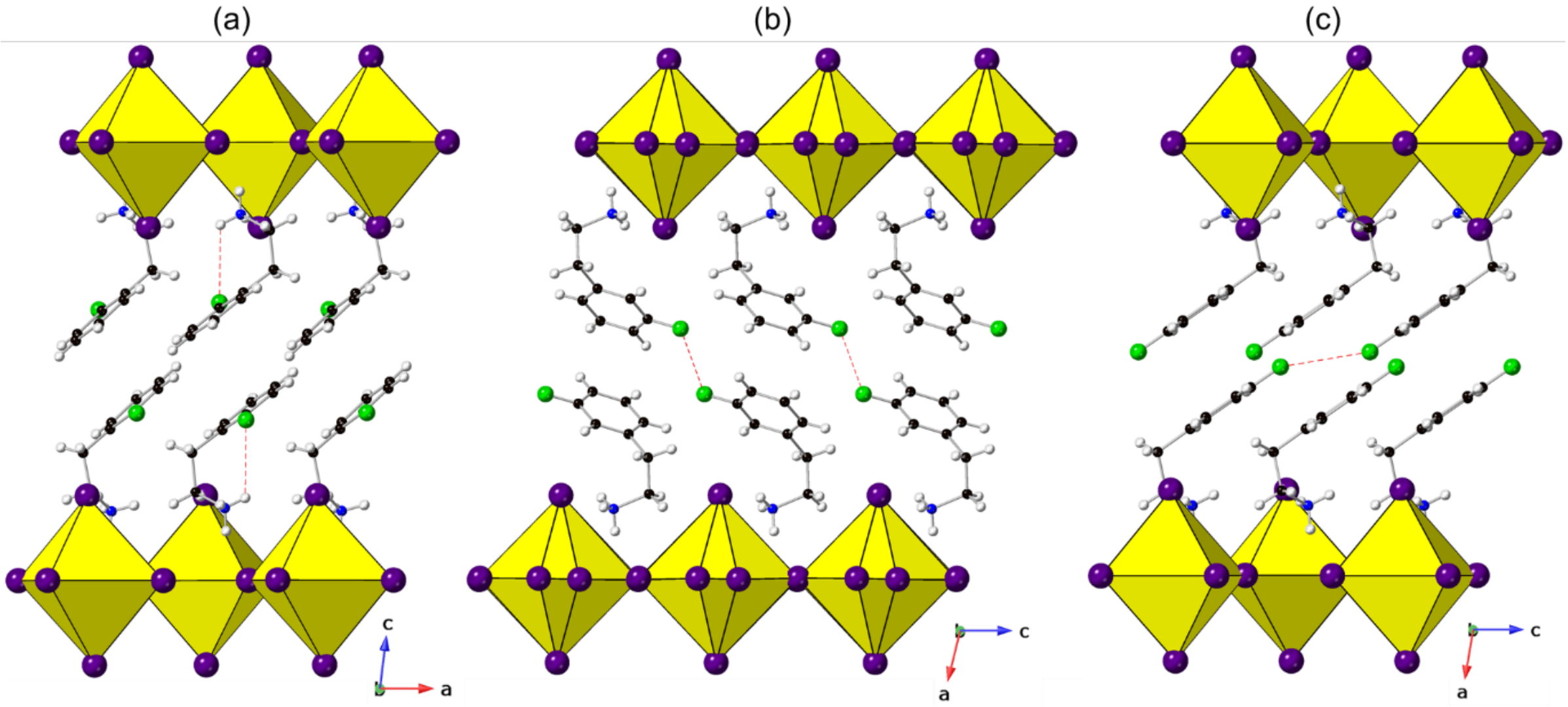

**Figure 19.** Structures of (a) *o*-, (b) *m*- and (c) *p*- positional isomers of $[C_8H_{11}FN]_2PbI_4$ (1893383, 1893384 and 1488195, respectively). Intramolecular F---H-N interactions in (a) and intermolecular F---F interactions in (b) and (c) are shown by the dashed red lines. Note the presence of disorder in the I positions in (b).

### 3.4.4 APbX4 and AA'PbX4 compositions

It can be noted that the vast majority of the structures presented here (190 out of 260) correspond to the $A_2PbX_4$ composition. A smaller number (67) correspond to $APbX_4$ types, where A is typically an aliphatic diamine or a cyclic amine containing a single N, with an amine side-chain. A much smaller number (3) correspond to mixed cation systems $AA'PbX_4$. There is clearly considerable scope for further, imaginative synthesis in targeting these less-represented stoichiometries. For the $APbX_4$ types templated by linear diamines, $[H_3NC_nH_{2n}NH_3]^{2+}$, early work[84] showed a trend whereby the parity of the carbon number (i.e. $n$ = odd or even) dictated the tilt system of the inorganic layers with compositions of $n$ = 8, 10 & 12 adopting the $a^-a^-c$ system. However, unlike the linear amine chains discussed in section 3.4.3 there is a significant change in the layer shift of the analogous diamines. As the chain length increases the degree of layer shift correspondingly decreases from nRP ($\Delta$ = 0.42, 0.42) for $n$ = 8 to almost perfectly DJ in $n$ = 12 ($\Delta$ = 0.01, 0.01). Due to the increased chain length, it appears that the degree of octahedral tilting required to optimise hydrogen bonding is reduced which subsequently reduces the need for layer shift in these materials. The absence of this effect in the related $A_2PbX_4$ compositions is likely due to the presence of the 'bilayer' of amines in the interlayer site that cannot optimise the hydrogen bonding in the same way.

The few examples of AA′PbX$_4$ structures all exhibit a common feature of alternating cations in the interlayer sites along the $[PbX_4]_\infty$ direction and seem to adopt the nDJ2 structure. Both $[CH_6N][CH_6N_3]PbI_4$ and $[CH_6N][Cs]PbBr_4$ (1588974 and 1552603, respectively) adopt the *Imma* structure corresponding to the $a^+b^0c^0$/–$(a^+)b^0c^0$ tilt system. Although the $[CH_6N_3][Cs]PbI_4$ structure (1552604) shares similarities with the two related compositions, both $M_5^-$ and $\Gamma_5^+$ layer shift modes are introduced and a more complex tilt system incorporating a rotation around the *c*-axis is observed resulting in the adoption of the *Pnnm* space group.

## 3.5 Polymorphs and phase transitions

We have already highlighted interesting polymorphic behaviour in the cystamine-based family. In addition, there are many more examples of competition between different phases (such as the layered versus chain options for Billing's cyclic amines) or polymorphs (such as 1963065, which adopts both (001) and (110)-oriented layered perovskite polymorphs[175]). Here we discuss a few further examples of temperature-phase transitions and some unusual examples of chemically-induced structural rearrangements. In some papers, structural phase transitions are reported from thermodynamic and spectroscopic behaviour, but full single crystal determinations are not available for both phases (e.g. 1938883). Likewise, there are some papers that do appear to report single crystal determinations at two temperatures but, unfortunately, they have not been deposited at CCDC.

The most common cause of temperature-induced structural phase transitions in these materials appears to be order-disorder behaviour of the organic moieties. The study of fluorinated benzylamines (BA) by Shi et al.[32] is a nice example. All four compounds, benzylamine, 2-fluorobenzylamine, 3-fluorobenzylamine and 4-fluorobenzylamine (Table 1) adopt the aristotype RP structure, *I4/mmm* in their highest temperature phase (i.e. above ambient) with disordered organic moieties. However, on cooling both BA itself and 2-FBA transform to polar space group $Cmc2_1$ via introduction of the $X_2^+$ rotation and a polar mode (Table 2, 120685 and 1944743). On the contrary, despite 3-FBA and 4-FBA also undergoing tilt transitions on cooling (1944741, $a^-a^-c$ and 1944739, $a^0a^0c/a^0a^0c$, respectively) they retain centrosymmetricity. In each case the low temperature phase has ordered organic moieties, but it is suggested that the lower steric hindrance in BA and 2-FBA permits the differences in crystal packing. Ferroelectricity has been confirmed in 2-FBA, and Xiong's group have also observed related behaviour in several other fluorinated amine systems.[32,45,50] 3-Fluoro-*N*-methylbenzylamine also displays an order-disorder transitions, but retains centrosymmetricity in both phases, this time from an unusual HT phase with a layer shift but no tilts (1852626, Table 2) to a RT phase (1845548) with the $X_3^+$ tilt mode added. There are several other examples of order-disorder transitions, involving changes of tilt system, for example 1417497 (HT, *Cmca*, Table 2) to 1417496 (LT, $P2_1/c$, Table 4). Structure 1962913 (section 3.1.2) is an unusual case, which shows a transition from a high symmetry phase with no octahedral tilting or layer shifts, but displaying antiferrodistortive displacements of the Pb atoms and disordered piperidinium cations at high temperature (352 K) to a layered phase with five-coordinate Pb at low temperature.[105]

Billing has described several phase transitions in the series of alkylammonium-templated materials $(RNH_3)_2PbI_4$, with *n*-butyl, *n*-pentyl or *n*-hexyl chains.[71] These do not involve disorder of the alkyl chain, but instead exhibit changes in packing of these chains, which induce shift/tilt transitions of the $[PbI_4]_\infty$ layers. For example, 665693 (Table 2) transforms to 665691 (Table 4) on cooling, with a change from RP type, *Pbca*, $a^-a^-c/-(a^-a^-)c$ to nRP type, $P2_1/c$, $a^-a^-c$ (Fig. 20). Several independent studies have been carried out on $(n\text{-BA})_2PbX_4$ (n-BA=n-butylamine). In the case of the iodide two phases exist (665689), both adopting the common RP-related *Pbca* structure, but differing in the orientation and H-bonding of the n-BA moiety.[71] The bromide (1455948) also adopts the *Pbca* structure at 100 K, but has been reported in a lower symmetry version of this structure (1903531) at room temperature. The chloride has three reported phases, from different studies: 1952028, 2016195 and 2003637, all of which adopt RP-like structure. Ji *et al.*[41] report ferroelectric behaviour and a transition from *Cmca* to $Cmc2_1$ at $T_C$ = 328 K, with order-disorder of the n-BA, but no change in tilt system. Tu *et al.*[48] also confirm $Cmc2_1$ at RT, whereas McClure *et al.*[69] suggest the lower symmetry space group *Pbca*, but this study is based on powder diffraction only.

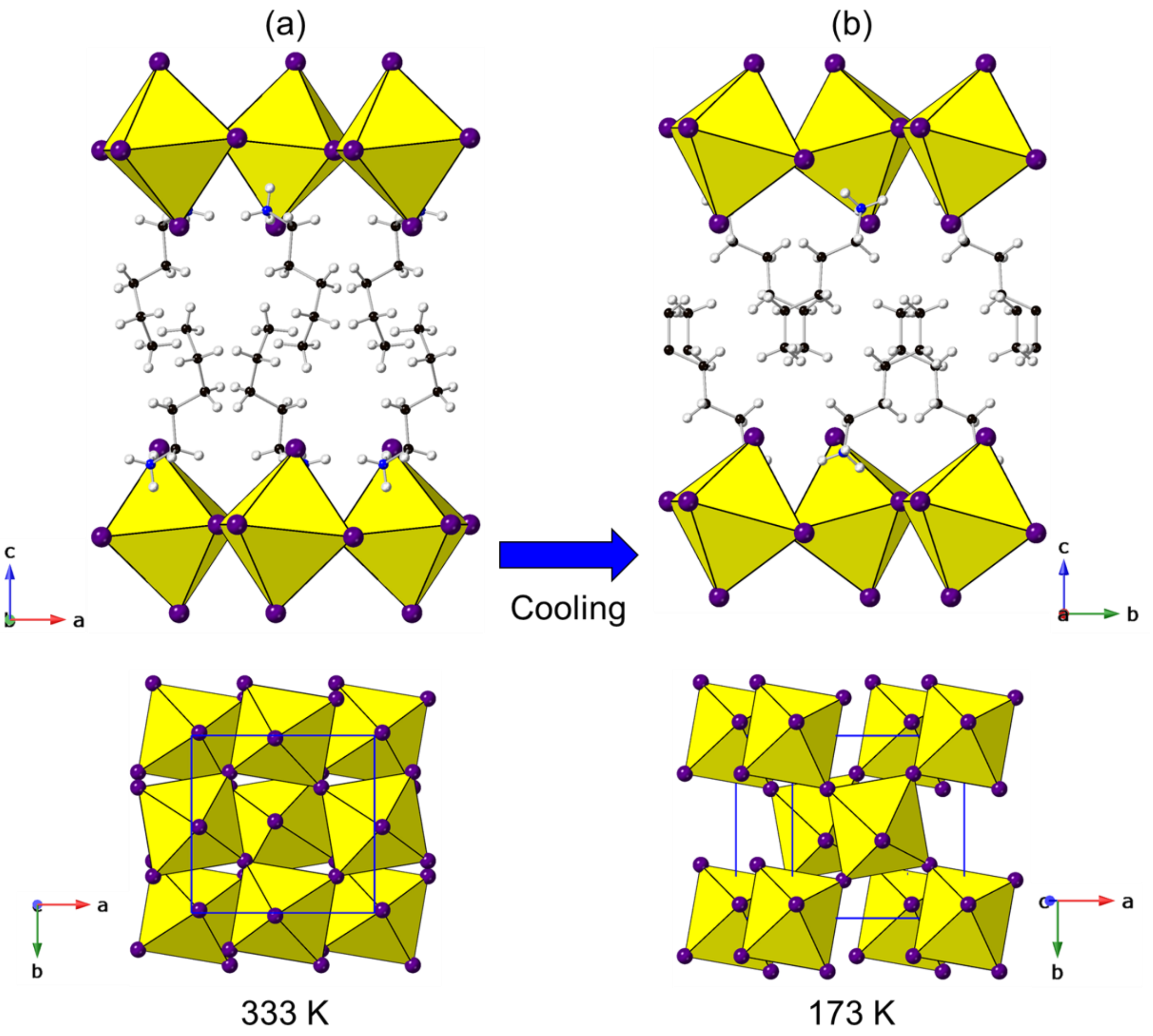

**Figure 20.** High (665693) and low (665691) temperature phases of $[C_5H_{14}N]_2PbI_4$ adopting *Pbca* and $P2_1/a$ structures, respectively. Note the change in degree of layer shift corresponding to the change in amine packing.

$(MHy)_2PbI_4$ (MHy = methylhydrazinium) is of interest as MHy is the smallest organic cation to be incorporated into any structure in this review. It undergoes three phase transitions versus temperature, mediated by order-disorder of the MHy, which also leads to some interesting physical properties.[34] The phases are 1937296 (Table 2) which displays only the $M_5^-$ shift mode (note there is reported to be an isostructural phase transition between two phases with this symmetry), 1937299 and 1937297 (Table 5) both of which display the unique 'triple tilt' C-mode (see section 3.3 and Fig. 15).

In addition to the differences in conformation of linear chain amines which can affect Cl vs Br vs I analogues (section 3.4.2), these changes may also occur within the same compound, as a function of temperature. An example is $(DAB)_2PbCl_4$ (DAB = 1,4-diaminobutane). Courseille *et al.*[94] report a complex structure at ambient temperature (1305732), and suggest a simpler structure (DJ-like, $a^-a^-c$) above RT; a full structural analysis of this phase is required, however.

Finally, there are intriguing cases where an *in situ* chemical reaction takes place, for example, reaction of alykynyl or alkenyl amines with $Br_2$ or $I_2$[126,144] leads to addition of $Br_2$ across the unsaturated C-C bond. In both cases (955778 to 1048947 and 955776 to 955777) the tilt system (the apparently very robust $a^-a^-c$, Table 4) remains unchanged, but the chemical changes are manifest both in changes of the $[PbX_4]_\infty$ interlayer distances (approximately an increase of around 3-5 Å in these cases) and in changes in layer shift from close to nDJ type towards DJ2 type. There are also reported examples of intercalation of intact $I_2$ molecules in similar reactions; unfortunately, full single crystal details are not deposited.

## 3.6 Summary and conclusions

The main aim of this comprehensive review is to provide a more systematic approach to understanding the crystallography of layered lead halide perovskites by introducing the concepts of symmetry mode analysis. Octahedral tilting and layer shifts are shown to be the key distortion modes underlying the nature of the inorganic perovskite-like layers in this family of materials. However, the complexity of these systems goes far beyond the octahedral tilt systems traditionally seen in purely inorganic layered perovskites. First and foremost, we use our analysis to show the relationships between apparently diverse groups of structures and compositions, by classifying in terms of unit cell metrics and key distortion modes, relative to idealised parent phases. We hope this approach, which may be unfamiliar to many current workers in the field, will be of use in informing and directing future work. Although we have tried to be exhaustive in the analysis of each individual structure type, we have by no means analysed the underlying reasons for the adoption of the particular distorted variants for specific compositions. We have merely highlighted a few trends and interesting cases relating the nature of the interlayer species to the behaviour of the inorganic layer. For example, we have shown that certain tilt systems, such as $a^-a^-c$, are very robust and resilient to changes in the nature of the interlayer species, being able to accommodate different species by adjustments to tilt amplitudes and layer shift factors (from DJ towards RP and even DJ2) whist still retaining essentially the same structure type. We have also highlighted some less common distortion modes, such an antiferrodistortive Pb displacements, tilt systems (such as triple and quadruple repeats) that go beyond the standard Glazer-like systems, and undulations (‘ripples’) of the perovskite-like layers themselves. Of particular note are complex tilt modes not generally found in conventional perovskites (but see, for example refs [171] and [176]), such as the C2 modes in 1937299 and 1963067, which may naturally lead to non-centrosymmetric or polar space groups.

It is clear that much more structural analysis can be done on the vast array of existing structures, for example, we have not referred at all to the various octahedral distortion indices or interlayer penetration effects that are typically quoted in relation to physical properties, and which are undoubtedly an important factor in determining those properties. We hope our novel view of the structural chemistry of this important family of materials will enhance understanding of both structure-composition and structure-property relationships, thus adding to the design and development of materials with enhanced features of interest.

**Acknowledgements**

We thank the Leverhulme Trust for a Research Fellowship to J.A.M. (RPG-2018-065).

## References

1 M. A. Green, A. Ho-Baillie and H. J. Snaith, *Nat. Photonics*, 2014, **8**, 506–514.

2 M. D. Smith, B. A. Connor and H. I. Karunadasa, *Chem. Rev.*, 2019, **119**, 3104–3139.

3 H.-Y. Zhang, X.-J. Song, H. Cheng, Y.-L. Zeng, Y. Zhang, P.-F. Li, W. Q. Liao and R.-G. Xiong, *J. Am. Chem. Soc.*, 2020, **142**, 4604–4608.

4 B. Saparov and D. B. Mitzi, *Chem. Rev.*, 2016, **116**, 4558–4596.

5 M. D. Smith, E. J. Crace, A. Jaffe and H. I. Karunadasa, *Annu. Rev. Mater. Res.*, 2018, **48**, 111–136.

6 L. Mao, C. C. Stoumpos and M. G. Kanatzidis, *J. Am. Chem. Soc.*, 2019, **141**, 1171–1190.

7 B. Raveau, *Prog. Solid State Chem.*, 2007, **35**, 171–173.

8 N. Mercier, *Angew. Chem. Int. Ed.*, 2019, **58**, 17912–17917.

9 J. Breternitz and S. Schorr, *Adv. Energy Mater.*, 2018, **8**, 1–2.

10 G. Kieslich and A. L. Goodwin, *Mater. Horizons*, 2017, **4**, 362–366.

11 A. M. Glazer, *Acta Crystallogr. Sect. B Struct. Crystallogr. Cryst. Chem.*, 1972, **28**, 3384–3392.

12 C. J. Howard and H. T. Stokes, *Acta Crystallogr.*, 1998, **B54**, 782–789.

13 B. J. Campbell, H. T. Stokes, D. E. Tanner and D. M. Hatch, *J. Appl. Crystallogr.*, 2006, **39**, 607–614.

14 D. Orobengoa, C. Capillas, M. I. Aroyo and J. M. Perez-Mato, *J. Appl. Crystallogr.*, 2009, **42**, 820–833.

15 S. C. Miller and W. F. Love, *Tables of irreducible representations of space groups and co-representations of magnetic space groups*, Pruett, Boulder, 1967.

16 M. V. Talanov, V. B. Shirokov and V. M. Talanov, *Acta Crystallogr.*, 2016, **A72**, 222–235.

17 H. L. B. Boström, M. S. Senn and A. L. Goodwin, *Nat. Commun.*, 2018, **9**, 1–7.

18 S. N. Ruddlesden and P. Popper, *Acta Crystallogr.*, 1957, **10**, 538–539.

19 M. Dion, M. Ganne and M. Tournoux, *Mater. Res. Bull.*, 1981, **16**, 1429–1435.

20 D. Balz and K. Plieth, *Zeitschrift fur Elektrochemie*, 1955, **59**, 545–551.

21 C. Brosset, *Z. Anorg. Allg. Chem.*, 1935, **8**, 139–147.

22 I. Abrahams, J. L. Nowinski, P. G. Bruce and V. C. Gibson, *J. Sol*, 1991, **94**, 254–259.

23 N. A. Benedek, J. M. Rondinelli, H. Djani, P. Ghosez and P. Lightfoot, *Dalt. Trans.*, 2015, **44**, 10543–10558.

24 E. E. Mccabe, E. Bousquet, C. P. J. Stockdale, C. A. Deacon, T. T. Tran, P. S. Halasyamani, M. C. Stennett and N. C. Hyatt, *Chem. Mater.*, 2015, **27**, 8298–8309.

25 K. S. Aleksandrov and J. Bartolomé, *Phase Transitions*, 2001, **74**, 255–335.

26 P. M. Woodward, *Acta Crystallogr.*, 1997, **B53**, 32–66.

27 W. H. Baur, *Acta Crystallogr.*, 1974, **B30**, 1195–1215.

28 M.-H. Tremblay, J. Bacsa, B. Zhao, F. Pulvirenti, S. Barlow and S. R. Marder, *Chem. Mater.*, 2019, **31**, 6145–6153.

29 C. R. Groom, I. J. Bruno, M. P. Lightfoot and S. C. Ward, *Acta Crystallogr.*, 2016, **B72**, 171-179.

30 P. V Balachandran, D. Puggioni and J. M. Rondinelli, *Inorg. Chem.*, 2014, **53**, 336–348.

31 H. Y. Zhang, X. J. Song, X. G. Chen, Z. X. Zhang, Y. M. You, Y. Y. Tang and R. G. Xiong, *J. Am. Chem. Soc.*, 2020, **142**, 4925–4931.

32 P. P. Shi, S. Q. Lu, X. J. Song, X. G. Chen, W. Q. Liao, P. F. Li, Y. Y. Tang and R. G. Xiong, *J. Am. Chem. Soc.*, 2019, **141**, 18334–18340.

33 X. G. Chen, X. J. Song, Z. X. Zhang, H. Y. Zhang, Q. Pan, J. Yao, Y. M. You and R. G. Xiong, *J. Am. Chem. Soc.*, 2020, **142**, 10212–10218.

34 M. Mączka, M. Ptak, A. Gągor, D. Stefańska and A. Sieradzki, *Chem. Mater.*, 2019, **31**, 8563–8575.

35 A. Bonamartini Corradi, A. M. Ferrari, L. Righi and P. Sgarabotto, *Inorg. Chem.*, 2001, **40**, 218–223.

36 J. Calabrese, N. L. Jones, R. L. Harlow, N. Herron, D. L. Thorn and Y. Wang, *J. Am. Chem. Soc.*, 1991, **113**, 2328–2330.

37 Y. Hao, S. Wen, J. Yao, Z. Wei, X. Zhang, Z. Jiang, Y. Mei and H. Cai, *J. Solid State Chem.*, 2019, **270**, 226–230.

38 S. Sourisseau, N. Louvain, W. Bi, N. Mercier, D. Rondeau, F. Boucher, J. Y. Buzaré and C. Legein, *Chem. Mater.*, 2007, **19**, 600–607.

39 Z. Tang, J. Guan and A. M. Guloy, *J. Mater. Chem.*, 2001, **11**, 479–482.

40 T. Li, W. A. Dunlap-Shohl, E. W. Reinheimer, P. Le Magueres and D. B. Mitzi, *Chem. Sci.*, 2019, **10**, 1168–1175.

41 C. Ji, S. Wang, L. Li, Z. Sun, M. Hong and J. Luo, *Adv. Funct. Mater.*, 2019, **29**, 2–7.

42 Z. X. Wang, W. Q. Liao, H. Y. Ye and Y. Zhang, *Dalt. Trans.*, 2015, **44**, 20406–20412.

43 D. G. Billing and A. Lemmerer, *CrystEngComm*, 2009, **11**, 1549–1562.

44 H. Y. Ye, W. Q. Liao, C. L. Hu, Y. Zhang, Y. M. You, J. G. Mao, P. F. Li and R. G. Xiong, *Adv. Mater.*, 2016, **28**, 2579–2586.

45 W. Q. Liao, Y. Zhang, C. L. Hu, J. G. Mao, H. Y. Ye, P. F. Li, S. D. Huang and R. G. Xiong, *Nat. Commun.*, 2015, **6**, 1–7.

46 X. N. Li, P. F. Li, W. Q. Liao, J. Z. Ge, D. H. Wu and H. Y. Ye, *Eur. J. Inorg. Chem.*, 2017, **2017**, 938–942.

47 A. Lemmerer and D. G. Billing, *Dalt. Trans.*, 2012, **41**, 1146–1157.

48 Q. Tu, I. Spanopoulos, E. S. Vasileiadou, X. Li, M. G. Kanatzidis, G. S. Shekhawat and V. P. Dravid, *ACS Appl. Mater. Interfaces*, 2020, **12**, 20440–20447.

49 X. G. Chen, X. J. Song, Z. X. Zhang, P. F. Li, J. Z. Ge, Y. Y. Tang, J. X. Gao, W. Y. Zhang, D. W. Fu, Y. M. You and R. G. Xiong, *J. Am. Chem. Soc.*, 2020, **142**, 1077–1082.

50 T. T. Sha, Y. A. Xiong, Q. Pan, X. G. Chen, X. J. Song, J. Yao, S. R. Miao, Z. Y. Jing, Z. J. Feng, Y. M. You and R. G. Xiong, *Adv. Mater.*, 2019, **31**, 1–7.

51 M. H. Jung, *J. Mater. Chem. A*, 2019, **7**, 14689–14704.

52 K. Z. Du, Q. Tu, X. Zhang, Q. Han, J. Liu, S. Zauscher and D. B. Mitzi, *Inorg. Chem.*, 2017, **56**, 9291–9302.

53 M. Braun and W. Frey, *Zeitschrift fur Krist. - New Cryst. Struct.*, 1999, **214**, 331–332.

54 N. R. Venkatesan, A. Mahdi, B. Barraza, G. Wu, M. L. Chabinyc and R. Seshadri, *Dalt. Trans.*, 2019, **48**, 14019–14026.

55 N. Mercier, S. Poiroux, A. Riou and P. Batail, *Inorg. Chem.*, 2004, **43**, 8361–8366.

56 S.-H. Jang and W. Kaminsky, *CCDC 2016665: Experimental Crystal Structure Determination*, 2020.

57 J.-J. Yao, *CCDC 1883324: Experimental Crystal Structure Determination*, 2018.

58 J.-J. Yao, *CCDC 1975113: Experimental Crystal Structure Determination*, 2020.

59 J.-J. Yao, *CCDC 1975109: Experimental Crystal Structure Determination*, 2020.

60 B. Luo, Y. Guo, Y. Xiao, X. Lian, T. Tan, D. Liang, X. Li and X. Huang, *J. Phys. Chem. Lett.*, 2019, **10**, 5271–5276.

61 C. Lermer, S. T. Birkhold, I. L. Moudrakovski, P. Mayer, L. M. Schoop, L. Schmidt-Mende and B. V. Lotsch, *Chem. Mater.*, 2016, **28**, 6560–6566.

62 N. Mercier, N. Louvain and W. Bi, *CrystEngComm*, 2009, **11**, 720–734.

63 A. Meresse and A. Daoud, *Acta Crystallogr.*, 1989, **C45**, 194–196.

64 V. Gómez, O. Fuhr and M. Ruben, *CrystEngComm*, 2016, **18**, 8207–8219.

65 A. Lemmerer and D. G. Billing, *South African J. Chem.*, 2013, **66**, 263–272.

66 M. K. Rayner and D. G. Billing, *Acta Crystallogr.*, 2010, **E66**, m659.

67 M. Fazayeli, M. Khatamian and G. Cruciani, *CrystEngComm*, 2020, *Advance Article*, DOI:10.1039/d0ce00184h.

68 N. Mercier, *CrystEngComm*, 2005, **7**, 429–432.

69 E. T. McClure, A. P. McCormick and P. M. Woodward, *Inorg. Chem.*, 2020, **59**, 6010–6017.

70 L. Dou, A. B. Wong, Y. Yu, M. Lai, N. Kornienko, S. W. Eaton, A. Fu, C. G. Bischak, J. Ma, T. Ding, N. S. Ginsberg, L. W. Wang, A. P. Alivisatos and P. Yang, *Science (80-. ).*, 2015, **349**, 1518–1521.

71 D. G. Billing and A. Lemmerer, *Acta Crystallogr.*, 2007, **B63**, 735–747.

72 W. J. Wei, C. Li, L. S. Li, Y. Z. Tang, X. X. Jiang and Z. S. Lin, *J. Mater. Chem. C*, 2019, **7**, 11964–11971.

73 X. H. Zhu, N. Mercier, A. Riou, P. Blanchard and P. Frère, *Chem. Commun.*, 2002, **2**, 2160–2161.

74 D. G. Billing and A. Lemmerer, *CrystEngComm*, 2007, **9**, 236–244.

75 A. Lemmerer and D. G. Billing, *CrystEngComm*, 2010, **12**, 1290–1301.

76 M. E. Kamminga, H. H. Fang, M. R. Filip, F. Giustino, J. Baas, G. R. Blake, M. A. Loi and T. T. M. Palstra, *Chem. Mater.*, 2016, **28**, 4554–4562.

77 C. Ortiz-Cervantes, P. I. Román-Román, J. Vazquez-Chavez, M. Hernández-Rodríguez and D. Solis-Ibarra, *Angew. Chemie - Int. Ed.*, 2018, **57**, 13882–13886.

78 D. G. Billing and A. Lemmerer, *New J. Chem.*, 2008, **32**, 1736–1746.

79 J. V. Passarelli, C. M. Mauck, S. W. Winslow, C. F. Perkinson, J. C. Bard, H. Sai, K. W. Williams, A. Narayanan, D. J. Fairfield, M. P. Hendricks, W. A. Tisdale and S. I. Stupp, *Nat. Chem.*, 2020, **12**, 672–682.

80 S. Wang, Y. Yao, Z. Wu, Y. Peng, L. Li and J. Luo, *J. Mater. Chem. C*, 2018, **6**, 12267–12272.

81 A. B. Corradi, A. M. Ferrari, G. C. Pellacani, A. Saccani, F. Sandrolini and P. Sgarabotto, *Inorg. Chem.*, 1999, **38**, 716–721.

82 L. Li, X. Shang, S. Wang, N. Dong, C. Ji, X. Chen, S. Zhao, J. Wang, Z. Sun, M. Hong and J. Luo, *J. Am. Chem. Soc.*, 2018, **140**, 6806–6809.

83 J. V. Passarelli, D. J. Fairfield, N. A. Sather, M. P. Hendricks, H. Sai, C. L. Stern and S. I. Stupp, *J. Am. Chem. Soc.*, 2018, **140**, 7313–7323.

84 A. Lemmerer and D. G. Billing, *CrystEngComm*, 2012, **14**, 1954–1966.

85 M. D. Smith, A. Jaffe, E. R. Dohner, A. M. Lindenberg and H. I. Karunadasa, *Chem. Sci.*, 2017, **8**, 4497–4504.

86 D. B. Straus, N. Iotov, M. R. Gau, Q. Zhao, P. J. Carroll and C. R. Kagan, *J. Phys. Chem. Lett.*, 2019, **10**, 1198–1205.

87 Z. Hong, W. K. Chong, A. Y. R. Ng, M. Li, R. Ganguly, T. C. Sum and H. Sen Soo, *Angew. Chemie - Int. Ed.*, 2019, **58**, 3456–3460.

88 J. V. Passarelli, C. M. Mauck, S. W. Winslow, C. F. Perkinson, J. C. Bard, H. Sai, K. W. Williams, A. Narayanan, D. J. Fairfield, M. P. Hendricks, W. A. Tisdale and S. I. Stupp, *Nat. Chem.*, 2020, **12**, 672–682.

89 X. Gong, O. Voznyy, A. Jain, W. Liu, R. Sabatini, Z. Piontkowski, G. Walters, G. Bappi, S. Nokhrin, O. Bushuyev, M. Yuan, R. Comin, D. McCamant, S. O. Kelley and E. H. Sargent, *Nat. Mater.*, 2018, **17**, 550–556.

90 D. G. Billing and A. Lemmerer, *CrystEngComm*, 2006, **8**, 686–695.

91 X. Li, W. Ke, B. Traoré, P. Guo, I. Hadar, M. Kepenekian, J. Even, C. Katan, C. C. Stoumpos, R. D. Schaller and M. G. Kanatzidis, *J. Am. Chem. Soc.*, 2019, **141**, 12880–12890.

92 S.-H. Jang and W. Kaminsky, *CCDC 2016669: Experimental Crystal Structure Determination*, 2020.

93 L. Dobrzycki and K. Woźniak, *CrystEngComm*, 2008, **10**, 577–589.

94 C. Courseiile, N. B. Chanh, T. Maris, A. Daoud, Y. Abid and M. Laguerre, *Phys. Stat. Sol. A*, 1994, **143**, 203–214.

95 E. R. Dohner, E. T. Hoke and H. I. Karunadasa, *J. Am. Chem. Soc.*, 2014, **136**, 1718–1721.

96 S. Krishnamurthy, P. Kour, A. Katre, S. Gosavi, S. Chakraborty and S. Ogale, *APL Mater.*, 2018, **6**, 114204.

97 L.-C. Pei, Z.-H. Wei, X.-X. Zhang and J.-J. Yao, *Jiegou Huaxue*, 2019, **38**, 1494.

98 T. Maris, *Phd Thesis*, 1996.

99 L. Mao, H. Tsai, W. Nie, L. Ma, J. Im, C. C. Stoumpos, C. D. Malliakas, F. Hao, M. R. Wasielewski, A. D. Mohite and M. G. Kanatzidis, *Chem. Mater.*, 2016, **28**, 7781–7792.

100 H. Hillebrecht, *CCDC 1999302: Experimental Crystal Structure Determination*, 2020.

101 Y. Hao, Z. Qiu, X. Zhang, Z. Wei, J. Yao and H. Cai, *Inorg. Chem. Commun.*, 2018, **97**, 134–138.

102 D. C. Palmer, *CrystalMaker*, 2014, version 10.5.2.

103 R. Zhang, B. M. Abbett, G. Read, F. Lang, T. Lancaster, T. T. Tran, P. S. Halasyamani, S. J. Blundell, N. A. Benedek and M. A. Hayward, *Inorg. Chem.*, 2016, **55**, 8951–8960.

104 T. Li, R. Clulow, A. J. Bradford, S. L. Lee, A. M. Z. Slawin and P. Lightfoot, *Dalt. Trans.*, 2019, **48**, 4784–4787.

105 S. Chai, J. Xiong, Y. Zheng, R. Shi and J. Xu, *Dalt. Trans.*, 2020, **49**, 2218–2224.

106 C. Lermer, S. P. Harm, S. T. Birkhold, J. A. Jaser, C. M. Kutz, P. Mayer, L. Schmidt-Mende and B. V. Lotsch, *Z. Anorg. Allg. Chem.*, 2016, **642**, 1369–1376.

107 J. Hu, I. W. H. Oswald, S. J. Stuard, M. M. Nahid, N. Zhou, O. F. Williams, Z. Guo, L. Yan, H. Huamin, Z. Chen, X. Xiao, Y. Lin, J. Huang, A. M. Moran, H. Ade, J. R. Neilson and W. You, *CCDC 1893384: Experimental Crystal Structure Determination*, 2019.

108 Y. Gao, E. Shi, S. Deng, S. B. Shiring, J. M. Snaider, C. Liang, B. Yuan, R. Song, S. M. Janke, A. Liebman-Peláez, P. Yoo, M. Zeller, B. W. Boudouris, P. Liao, C. Zhu, V. Blum, Y. Yu, B. M. Savoie, L. Huang and L. Dou, *Nat. Chem.*, 2019, **11**, 1151–1157.

109 C. M. M. Soe, C. C. Stoumpos, M. Kepenekian, B. Traoré, H. Tsai, W. Nie, B. Wang, C. Katan, R. Seshadri, A. D. Mohite, J. Even, T. J. Marks and M. G. Kanatzidis, *J. Am. Chem. Soc.*, 2017, **139**, 16297–16309.

110 O. Nazarenko, M. R. Kotyrba, M. Wörle, E. Cuervo-Reyes, S. Yakunin and M. V. Kovalenko, *Inorg. Chem.*, 2017, **56**, 11552–11564.

111 B. Febriansyah, D. Giovanni, S. Ramesh, T. M. Koh, Y. Li, T. C. Sum, N. Mathews and J. England, *J. Mater. Chem. C*, 2020, **8**, 889–893.

112 B. Febriansyah, Y. Lekina, B. Ghosh, P. C. Harikesh, T. M. Koh, Y. Li, Z. Shen, N. Mathews and J. England, *ChemSusChem*, 2020, **13**, 2693–2701.

113 M. P. Hautzinger, J. Dai, Y. Ji, Y. Fu, J. Chen, I. A. Guzei, J. C. Wright, Y. Li and S. Jin, *Inorg. Chem.*, 2017, **56**, 14991–14998.

114 B. Febriansyah, T. M. Koh, Y. Lekina, N. F. Jamaludin, A. Bruno, R. Ganguly, Z. X. Shen, S. G. Mhaisalkar and J. England, *Chem. Mater.*, 2019, **31**, 890–898.

115 N. Louvain, W. Bi, N. Mercier, J. Y. Buzaré, C. Legein and G. Corbel, *Dalt. Trans.*, 2007, **2**, 965–970.

116 Y. Guo, L. Yang, S. Biberger, J. A. McNulty, T. Li, K. Schötz, F. Panzer and P. Lightfoot, *Inorg. Chem.*, 2020, **59**, 12858-12866.

117 H. Dammak, S. Elleuch, H. Feki and Y. Abid, *Solid State Sci.*, 2016, **61**, 1–8.

118 S.-H. Jang and W. Kaminsky, *CCDC 2016668: Experimental Crystal Structure Determination*, 2020.

119 E. R. Dohner, A. Jaffe, L. R. Bradshaw and H. I. Karunadasa, *J. Am. Chem. Soc.*, 2014, **136**, 13154–13157.

120 I. Zimmermann, S. Aghazada and M. K. Nazeeruddin, *Angew. Chemie - Int. Ed.*, 2019, **58**, 1072–1076.

121 N. Louvain, G. Frison, J. Dittmer, C. Legein and N. Mercier, *Eur. J. Inorg. Chem.*, 2014, 364–376.

122 G. C. Papavassiliou, G. A. Mousdis, C. P. Raptopoulou and A. Terzis, *Zeitschrift fur Naturforsch. - Sect. B J. Chem. Sci.*, 2000, **55**, 536–540.

123 D. G. Billing, *Acta Crystallogr.*, 2002, **E58**, m669–m671.

124 S.-H. Jang and W. Kaminsky, *CCDC 2016667: Experimental Crystal Structure Determination*, 2020.

125 B. Luo, Y. Guo, X. Li, Y. Xiao, X. Huang and J. Z. Zhang, *J. Phys. Chem. C*, 2019, **123**, 14239–14245.

126 D. Solis-Ibarra and H. I. Karunadasa, *Angew. Chemie - Int. Ed.*, 2014, **53**, 1039–1042.

127 I. W. H. Oswald, A. A. Koegel and J. R. Neilson, *Chem. Mater.*, 2018, **30**, 8606–8614.

128 J. M. Hoffman, C. D. Malliakas, S. Sidhik, I. Hadar, R. McClain, A. D. Mohite and M. G. Kanatzidis, 2020, *Chem. Sci.*, **44**, 12139-12148.

129 Z. Liu, W.-T. Yu, X.-T. Tao, M.-H. Jiang, J.-X. Yang and L. Wang, *Z. Krist. NCS*, 2004, **219**, 457–458.

130 H. Dai, Z. Liu, J. Fan, J. Wang, X. Tao and M. Jiang, *Zeitschrift fur Krist. - New Cryst. Struct.*, 2009, **224**, 149–150.

131 L. Li, *CCDC 1939732: Experimental Crystal Structure Determination*, 2020.

132 G. A. Mousdis, G. C. Papavassiliou, C. P. Raptopoulou and A. Terzis, *J. Mater. Chem.*, 2000, **10**, 515–518.

133 M. H. Tremblay, J. Bacsa, S. Barlow and S. R. Marder, *Mater. Chem. Front.*, 2020, **4**, 2023–2028.

134 A. H. Slavney, R. W. Smaha, I. C. Smith, A. Jaffe, D. Umeyama and H. I. Karunadasa, *Inorg. Chem.*, 2017, **56**, 46–55.

135 C. K. Yang, W. N. Chen, Y. T. Ding, J. Wang, Y. Rao, W. Q. Liao, Y. Y. Tang, P. F. Li, Z. X. Wang and R. G. Xiong, *Adv. Mater.*, 2019, **31**, 1–7.

136 M. K. Rayner and D. G. Billing, *Acta Crystallogr.*, 2010, **E66**, m658.

137 M. K. Rayner and D. G. Billing, *Acta Crystallogr.*, 2010, **E66**, m660.

138 M. K. Jana, R. Song, H. Liu, D. R. Khanal, S. M. Janke, R. Zhao, C. Liu, Z. Valy Vardeny, V. Blum and D. B. Mitzi, *Nat. Commun.*, 2020, **11**, 1–10.

139 W. T. M. Van Gompel, R. Herckens, K. Van Hecke, B. Ruttens, J. D'Haen, L. Lutsen and D. Vanderzande, *Chem. Commun.*, 2019, **55**, 2481–2484.

140 O. J. Weber, K. L. Marshall, L. M. Dyson and M. T. Weller, *Acta Crystallogr.*, 2015, **B71**, 668–678.

141 T. Schmitt, S. Bourelle, N. Tye, G. Soavi, A. D. Bond, S. Feldmann, B. Traore, C. Katan, J. Even, S. E. Dutton and F. Deschler, *J. Am. Chem. Soc.*, 2020, **142**, 5060–5067.

142 D. B. Straus, S. Hurtado Parra, N. Iotov, Q. Zhao, M. R. Gau, P. J. Carroll, J. M. Kikkawa and C. R. Kagan, *ACS Nano*, 2020, **14**, 3621–3629.

143 M. K. Jana, R. Song, H. Liu, D. R. Khanal, S. M. Janke, R. Zhao, C. Liu, Z. Valy Vardeny, V. Blum and D. B. Mitzi, *Nat. Commun.*, 2020, **11**, 1–10.

144 D. Solis-Ibarra, I. C. Smith and H. I. Karunadasa, *Chem. Sci.*, 2015, **6**, 4054–4059.

145 K. Xiong, W. Liu, S. J. Teat, L. An, H. Wang, T. J. Emge and J. Li, *J. Solid State Chem.*, 2015, **230**, 143–148.

146 H. Hillebrecht, *CCDC 1999300: Experimental Crystal Structure Determination*, 2020.

147 L. Li, *CCDC 1939731: Experimental Crystal Structure Determination*, 2020.

148 J. Hu, I. W. H. Oswald, S. J. Stuard, M. M. Nahid, N. Zhou, O. F. Williams, Z. Guo, L. Yan, H. Huamin, Z. Chen, X. Xiao, Y. Lin, J. Huang, A. M. Moran, H. Ade, J. R. Neilson and W. You, *CCDC 1893383: Experimental Crystal Structure Determination*, 2019.

149 D. Ma, Y. Fu, L. Dang, J. Zhai, I. A. Guzei and S. Jin, *Nano Res.*, 2017, **10**, 2117–2129.

150 X. H. Zhu, N. Mercier, P. Frère, P. Blanchard, J. Roncali, M. Allain, C. Pasquier and A. Riou, *Inorg. Chem.*, 2003, **42**, 5330–5339.

151 L. Cheng and Y. Cao, *Acta Crystallogr. Sect. C Struct. Chem.*, 2019, **75**, 354–358.

152 Y. Li, C. Lin, G. Zheng and J. Lin, *J. Solid State Chem.*, 2007, **180**, 173–179.

153 C. Q. Jing, J. Wang, H. F. Zhao, W. X. Chu, Y. Yuan, Z. Wang, M. F. Han, T. Xu, J. Q. Zhao and X. W. Lei, *Chem. - A Eur. J.*, 2020, **26**, 10307–10313.

154 L. Mao, Y. Wu, C. C. Stoumpos, M. R. Wasielewski and M. G. Kanatzidis, *J. Am. Chem. Soc.*, 2017, **139**, 5210–5215.

155 I. H. Park, Q. Zhang, K. C. Kwon, Z. Zhu, W. Yu, K. Leng, D. Giovanni, H. S. Choi, I. Abdelwahab, Q. H. Xu, T. C. Sum and K. P. Loh, *J. Am. Chem. Soc.*, 2019, **141**, 15972–15976.

156 L. Mao, W. Ke, L. Pedesseau, Y. Wu, C. Katan, J. Even, M. R. Wasielewski, C. C. Stoumpos and M. G. Kanatzidis, *J. Am. Chem. Soc.*, 2018, **140**, 3775–3783.

157 X. Zhang, Z. Wei, Y. Cao, M. Li, J. Zhang and H. Cai, *J. Coord. Chem.*, 2020, **73**, 417–428.

158 K. Thirumal, W. K. Chong, W. Xie, R. Ganguly, S. K. Muduli, M. Sherburne, M. Asta, S. Mhaisalkar, T. C. Sum, H. Sen Soo and N. Mathews, *Chem. Mater.*, 2017, **29**, 3947–3953.

159 K. Shibuya, M. Koshimizu, F. Nishikido, H. Saito and S. Kishimoto, *Acta Crystallogr.*, 2009, **E65**, m1323-m1324.

160 D. G. Billing and A. Lemmerer, *Acta Crystallogr.*, 2006, **C62**, 269–271.

161 Jiahui, *CCDC 2011085: Experimental Crystal Structure Determination*, 2020.

162 L. Mao, P. Guo, M. Kepenekian, I. Hadar, C. Katan, J. Even, R. D. Schaller, C. C. Stoumpos and M. G. Kanatzidis, *J. Am. Chem. Soc.*, 2018, **140**, 13078–13088.

163 T. Yu, L. Zhang, J. Shen, Y. Fu and Y. Fu, *Dalt. Trans.*, 2014, **43**, 13115–13121.

164 Y. Li, G. Zheng, C. Lin and J. Lin, *Cryst. Growth Des.*, 2008, **8**, 1990–1996.

165 C. Lermer, A. Senocrate, I. Moudrakovski, T. Seewald, A. K. Hatz, P. Mayer, F. Pielnhofer, J. A. Jaser, L. Schmidt-Mende, J. Maier and B. V. Lotsch, *Chem. Mater.*, 2018, **30**, 6289–6297.

166 Y. Li, G. Zheng, C. Lin and J. Lin, *Solid State Sci.*, 2007, **9**, 855–861.

167 G. N. Liu, J. R. Shi, X. J. Han, X. Zhang, K. Li, J. Li, T. Zhang, Q. S. Liu, Z. W. Zhang and C. Li, *Dalt. Trans.*, 2015, **44**, 12561–12575.

168 K. Trujillo-Hernández, G. Rodríguez-López, A. Espinosa-Roa, J. González-Roque, A. P. Gómora-Figueroa, W. Zhang, P. S. Halasyamani, V. Jancik, M. Gembicky, G. Pirruccio and D. Solis-Ibarra, *J. Mater. Chem. C*, 2020, **8**, 9602–9607.

169 L. Y. Rong, C. Hu, Z. N. Xu, G. E. Wang and G. C. Guo, *Inorg. Chem. Commun.*, 2019, **102**, 90–94.

170 Y.-Y. Guo, L.-J. Yang, J. A. McNulty, A. M. Z. Slawin and P. Lightfoot, *Dalt. Trans.*, , DOI:10.1039/d0dt02936j.

171 M. D. Peel, S. P. Thompson, A. Daoud-Aladine, S. E. Ashbrook and P. Lightfoot, *Inorg. Chem.*, 2012, **51**, 6876–6889.

172 N. A. Benedek, A. T. Mulder and C. J. Fennie, *J. Solid State Chem.*, 2012, **195**, 11–20.

173 N. A. Benedek and C. J. Fennie, *Phys. Rev. Lett.*, 2011, **106**, 3–6.

174 F. Wang, H. Gao, C. de Graaf, J. M. Poblet, B. J. Campbell and A. Stroppa, *npj Comput. Mater.*, 2020, **6**, 183.

175 L. Mao, Y. Wu, C. C. Stoumpos, M. R. Wasielewski and M. G. Kanatzidis, *J. Am. Chem. Soc.*, 2017, **139**, 5210–5215.

176 C. A. L. Dixon and P. Lightfoot, *Phys. Rev. B*, 2018, **97**, 1–9.

**Table 2.** Summary of experimentally known structures with two octahedral layers per unit cell (derived from RP parent) and $\sqrt{2}a_{RP} \times \sqrt{2}a_{RP}$ cell metrics in the layer plane.

| CCDC Number | Amine | Type | Formula | Space group | Layer direction | Key modes | | Tilt system | Layer shift factor, Δ | Structure type | Ref. |
|---|---|---|---|---|---|---|---|---|---|---|---|
| | | | | | | Tilt | Layer shift | | | | |
| 1852626 | 3-Fluoro-*N*-methylbenzylamine | | $[C_8H_{11}FN]_2PbBr_4$ | *Cmcm* | *c* | – | $M_5^-$ | $a^0a^0c^0$ | 0.15, 0.15 | nDJ | 37 |
| 641642 | 2-Bromoethylamine | | $[C_2H_7BrN]_2PbI_4$ 293 K | *C*2/*c* | *c* | – | $M_5^-$, $\Gamma_5^+$ | $a^0a^0c^0$ | 0.5, 0.15 | nDJ2 | 38 |
| 167103 | 2,2′-Biimidazole | [A] | $[C_6H_8N_4]PbI_4$ | *C*2/*c* | *c* | – | $M_5^-$, $\Gamma_5^+$ | $a^0a^0c^0$ | 0.47, 0.12 | nDJ2 | 39 |
| 1863837 | Butan-2-amine | | $[C_4H_{12}N]_2PbBr_4$ | *P*4$_2$/*ncm* | *c* | $X_3^+$ | – | $a^-b^0c^0/b^0a^-c^0$ | 0.5, 0.5 | RP | 40 |
| 1934896 | 4,4-Difluoropiperidine | | $[C_5H_{10}F_2N]_2PbI_4$ 300 K | *Aba*2 | *b* | $X_3^+$ | – | $a^-a^-c^0/-(a^-a^-)c^0$ | 0.5, 0.5 | RP | 31 |
| 1992692 | 4,4-Difluorohexahydroazepine | | $[C_6H_{12}F_2N]_2PbI_4$ 343 K | *Cmc*2$_1$ | *b* | $X_3^+$ | – | $a^-a^-c^0/-(a^-a^-)c^0$ | 0.5, 0.5 | RP | 33 |
| 2016195 | Butylamine | | $[C_4H_{12}N]_2PbCl_4$ 330 K | *Cmca* | *a* | $X_2^+$ | – | $a^0a^0c/a^0a^0c$ | 0.5, 0.5 | RP | 41 |
| 1417497 | Isobutylamine | | $[C_4H_{12}N]_2PbBr_4$ 393 K | *Cmca* | *a* | $X_2^+$ | – | $a^0a^0c/a^0a^0c$ | 0.5, 0.5 | RP | 42 |
| 708568 | Cyclopentylamine | | $[C_5H_{12}N]_2PbCl_4$ | *Cmca* | *a* | $X_2^+$ | – | $a^0a^0c/a^0a^0c$ | 0.5, 0.5 | RP | 43 |
| 1409219 | Cyclohexylamine | | $[C_6H_{14}N]_2PbBr_4$ 383 K | *Cmca* | *a* | $X_2^+$ | – | $a^0a^0c/a^0a^0c$ | 0.5, 0.5 | RP | 44 |
| 1409221 | Cyclohexylamine | | $[C_6H_{14}N]_2PbI_4$ | *Cmca* | *a* | $X_2^+$ | – | $a^0a^0c/a^0a^0c$ | 0.5, 0.5 | RP | 44 |
| 1042749 | Benzylamine | | $[C_7H_{10}N]_2PbBr_4$ | *Cmca* | *a* | $X_2^+$ | – | $a^0a^0c/a^0a^0c$ | 0.5, 0.5 | RP | 45 |
| 1482272 | (Cyclohexylmethyl)amine | | $[C_7H_{16}N]_2PbBr_4$ 433 K | *Cmca* | *a* | $X_2^+$ | – | $a^0a^0c/a^0a^0c$ | 0.5, 0.5 | RP | 46 |
| 805432 | Octylamine | | $[C_8H_{20}N]_2PbI_4$ 314 K | *Acam* | *c* | $X_2^+$ | – | $a^0a^0c/a^0a^0c$ | 0.5, 0.5 | RP | 47 |
| 805440 | Decylamine | | $[C_{10}H_{24}N]_2PbI_4$ 343 K | *Acam* | *c* | $X_2^+$ | – | $a^0a^0c/a^0a^0c$ | 0.5, 0.5 | RP | 47 |
| 2003637 | Butylamine | | $[C_4H_{12}N]_2PbCl_4$ 293 K | *Cmc*2$_1$ | *a* | $X_2^+$ | – | $a^0a^0c/a^0a^0c$ | 0.5, 0.5 | RP | 48 |
| 1965897 | 4-Aminotetrahydropyran | | $[C_5H_{12}ON]_2PbBr_4$ | *Cmc*2$_1$ | *a* | $X_2^+$ | – | $a^0a^0c/a^0a^0c$ | 0.5, 0.5 | RP | 49 |
| 1904977 | 4,4-Difluorocyclohexylamine | | $[C_6H_{12}F_2N]_2PbI_4$ RT | $Cmc2_1$ | *a* | $X_2^+$ | – | $a^0a^0c/a^0a^0c$ | 0.5, 0.5 | RP | 50 |
| 708563 | Cyclohexylamine | | $[C_6H_{14}N]_2PbBr_4$ 173 K | $Cmc2_1$ | *a* | $X_2^+$ | – | $a^0a^0c/a^0a^0c$ | 0.5, 0.5 | RP | 43 |
| 1894433 | Hexylamine | | $[C_6H_{16}N]_2PbI_4$ | $Cmc2_1$ | *a* | $X_2^+$ | – | $a^0a^0c/a^0a^0c$ | 0.5, 0.5 | RP | 51 |
| 1944743 | 2-Fluorobenzylamine | | $[C_7H_9FN]_2PbCl_4$ 300 K | $Cmc2_1$ | *a* | $X_2^+$ | – | $a^0a^0c/a^0a^0c$ | 0.5, 0.5 | RP | 32 |
| 1542460 | Benzylamine | | $[C_7H_{10}N]_2PbBr_4$ RT | $Cmc2_1$ | *a* | $X_2^+$ | – | $a^0a^0c/a^0a^0c$ | 0.5, 0.5 | RP | 52 |
| 120685 | Benzylamine | | $[C_7H_{10}N]_2PbCl_4$ 293 K | $Cmc2_1$ | *a* | $X_2^+$ | – | $a^0a^0c/a^0a^0c$ | 0.5, 0.5 | RP | 53 |
| 1542462 | 1-(2-Naphthyl)-methylamine | | $[C_{11}H_{12}N]_2PbBr_4$ | $Cmc2_1$ | *a* | $X_2^+$ | – | $a^0a^0c/a^0a^0c$ | 0.5, 0.5 | RP | 52 |
| 1940831 | 2-(4-Biphenyl)ethylamine | | $[C_{14}H_{16}N]_2PbI_4$ | $Cmc2_1$ | *a* | $X_2^+$ | – | $a^0a^0c/a^0a^0c$ | 0.5, 0.5 | RP | 54 |
| 237190 | 2-Hydroxyethylamine | | $[C_2H_8ON]_2PbBr_4$ RT | *Pbcn* | *c* | $X_3^+$ | $M_5^-$ | $a^-a^-c^0/-(a^-a^-)c^0$ | 0.35, 0.35 | nRP | 55 |
| 2016665 | 4-Fluoro-*N*-methylaniline | | $[C_7H_9FN]_2PbI_4$ | *Pbcn* | *c* | $X_3^+$ | $M_5^-$ | $a^-a^-c^0/-(a^-a^-)c^0$ | 0.38, 0.38 | nRP | 56 |
| 1845548 | 3-Fluoro-*N*-methylbenzylamine | | $[C_8H_{11}FN]_2PbBr_4$ RT | *Pbcn* | *c* | $X_3^+$ | $M_5^-$ | $a^-a^-c^0/-(a^-a^-)c^0$ | 0.17, 0.17 | nDJ | 37 |
| 1883324 | 3-Chloro-*N*-methylbenzylamine | | $[C_8H_{11}ClN]_2PbI_4$ | *Pbcn* | *c* | $X_3^+$ | $M_5^-$ | $a^-a^-c^0/-(a^-a^-)c^0$ | 0.21, 0.21 | nDJ | 57 |
| 1975113 | 3-Bromo-*N*-methylbenzylamine | | $[C_8H_{11}BrN]_2PbCl_4$ | *Pbcn* | *c* | $X_3^+$ | $M_5^-$ | $a^-a^-c^0/-(a^-a^-)c^0$ | 0.13, 0.13 | nDJ | 58 |
| 1975109 | 3-Bromo-*N*-methylbenzylamine | | $[C_8H_{11}BrN]_2PbBr_4$ | *Pbcn* | *c* | $X_3^+$ | $M_5^-$ | $a^-a^-c^0/-(a^-a^-)c^0$ | 0.17, 0.17 | nDJ | 59 |
| 1938882 | 2,2,2-Trifluoroethylamine | | $[C_2H_5F_3N]_2PbBr_4$ | *Pnma* | *b* | $X_2^+$ | $M_5^-$ | $a^0a^0c/a^0a^0c$ | 0.48, 0.48 | nRP | 60 |
| 1938883 | 2-Fluoroethylamine | | $[C_2H_7FN]_2PbCl_4$ | *Pnma* | *b* | $X_2^+$ | $M_5^-$ | $a^0a^0c/a^0a^0c$ | 0.30, 0.30 | nRP | 61 |
| 705087 | 2-Cyanoethylamine | | $[C_3H_7N_2]_2PbI_4$ | *Pnma* | *b* | $X_2^+$ | $M_5^-$ | $a^0a^0c/a^0a^0c$ | 0.35, 0.35 | nRP | 62 |
| 1181686 | Propylamine | | $[C_3H_{10}N]_2PbCl_4$ | *Pnma* | *b* | $X_2^+$ | $M_5^-$ | $a^0a^0c/a^0a^0c$ | 0.32, 0.32 | nRP | 63 |
| 1495877 | 3-Bromopyridine | | $[C_5H_5BrN]_2PbBr_4$ | *Pnma* | *b* | $X_2^+$ | $M_5^-$ | $a^0a^0c/a^0a^0(-c)$ | 0.02, 0.02 | nDJ | 64 |
| 1944739 | 4-Fluorobenzylamine | | $[C_7H_9FN]_2PbCl_4$ 300 K | *Pnma* | *b* | $X_2^+$ | $M_5^-$ | $a^0a^0c/a^0a^0c$ | 0.33, 0.33 | nRP | 32 |
| 956549 | 1-Cyclohexylethylamine | | $[C_8H_{18}N]_2PbBr_4$ | *Pnma* | *b* | $X_2^+$ | $M_5^-$ | $a^0a^0c/a^0a^0c$ | 0.40, 0.40 | nRP | 65 |
| 781211 | 1,4-bis-(Aminomethyl)-cyclohexane | [A] | $[C_8H_{20}N_2]PbCl_4$ | *Pnma* | *b* | $X_2^+$ | $M_5^-$ | $a^0a^0c/a^0a^0(-c)$ | 0.17, 0.17 | nDJ | 66 |
| 1914148 | Ethyl-1,2-diamine | [A] | $[C_2H_{10}N_2]PbI_4$ | *Pbcm* | *c* | $X_2^+$ | $M_5^-$ | $a^0a^0c/a^0a^0c$ | 0.35, 0.35 | nRP | 67 |
| 1938881 | 2,2-Difluoroethylamine | | $[C_2H_6F_2N]_2PbBr_4$ | *Pbca* | *c* | $X_2^+$, $X_3^+$ | – | $a^-a^-c/-(a^-a^-)c$ | 0.5, 0.5 | RP | [60] |
| 267398 | 4-Aminobutanoic acid | | $[C_4H_{10}O_2N]_2PbI_4$ | *Pbca* | *c* | $X_2^+$, $X_3^+$ | – | $a^-a^-c/-(a^-a^-)c$ | 0.5, 0.5 | RP | [68] |
| 1952028 | Butylamine | | $[C_4H_{12}N]_2PbCl_4$ | *Pbca* | *c* | $X_2^+$, $X_3^+$ | – | $a^-a^-c/-(a^-a^-)c$ | 0.5, 0.5 | RP | [69] |
| 1455948 | Butylamine | | $[C_4H_{12}N]_2PbBr_4$<br>100 K | *Pbca* | *c* | $X_2^+$, $X_3^+$ | – | $a^-a^-c/-(a^-a^-)c$ | 0.5, 0.5 | RP | [70] |
| 665689 | Butylamine | | $[C_4H_{12}N]_2PbI_4$ | *Pbca* | *c* | $X_2^+$, $X_3^+$ | – | $a^-a^-c/-(a^-a^-)c$ | 0.5, 0.5 | RP | [71] |
| 1935994 | 2-Thiophenemethylamine | | $[C_5H_8SN]_2PbCl_4$ | *Pbca* | *c* | $X_2^+$, $X_3^+$ | – | $a^-a^-c/-(a^-a^-)c$ | 0.5, 0.5 | RP | [72] |
| 1935993 | 2-Thiophenemethylamine | | $[C_5H_8SN]_2PbBr_4$ | *Pbca* | *c* | $X_2^+$, $X_3^+$ | – | $a^-a^-c/-(a^-a^-)c$ | 0.5, 0.5 | RP | [72] |
| 187952 | 2-Thiophenemethylamine | | $[C_5H_8SN]_2PbI_4$ | *Pbca* | *c* | $X_2^+$, $X_3^+$ | – | $a^-a^-c/-(a^-a^-)c$ | 0.5, 0.5 | RP | [73] |
| 665693 | Pentylamine | | $[C_5H_{14}N]_2PbI_4$<br>333 K | *Pbca* | *c* | $X_2^+$, $X_3^+$ | – | $a^-a^-c/-(a^-a^-)c$ | 0.5, 0.5 | RP | [71] |
| 1904976 | 4,4-<br>Difluorocyclohexylamine | | $[C_6H_{12}F_2N]_2PbI_4$<br>398 K | *Pbca* | *c* | $X_2^+$, $X_3^+$ | – | $a^-a^-c/-(a^-a^-)c$ | 0.5, 0.5 | RP | [50] |
| 609995 | Cyclohexylamine | | $[C_6H_{14}N]_2PbI_4$ | *Pbca* | *c* | $X_2^+$, $X_3^+$ | – | $a^-a^-c/-(a^-a^-)c$ | 0.5, 0.5 | RP | [74] |
| 746130 | 6-Iodohexylamine | | $[C_6H_{15}IN]_2PbI_4$ | *Pbca* | *c* | $X_2^+$, $X_3^+$ | – | $a^-a^-c/-(a^-a^-)c$ | 0.5, 0.5 | RP | [75] |
| 665695 | Hexylamine | | $[C_6H_{16}N]_2PbI_4$<br>293 K | *Pbca* | *c* | $X_2^+$, $X_3^+$ | – | $a^-a^-c/-(a^-a^-)c$ | 0.5, 0.5 | RP | [71] |
| 1493135 | Benzylamine | | $[C_7H_{10}N]_2PbI_4$ | *Pbca* | *c* | $X_2^+$, $X_3^+$ | – | $a^-a^-c/-(a^-a^-)c$ | 0.5, 0.5 | RP | [76] |
| 805428 | Heptylamine | | $[C_7H_{18}N]_2PbI_4$<br>278 K | *Pbca* | *c* | $X_2^+$, $X_3^+$ | – | $a^-a^-c/-(a^-a^-)c$ | 0.5, 0.5 | RP | [47] |
| 805431 | Octylamine | | $[C_8H_{20}N]_2PbI_4$<br>293 K | *Pbca* | *c* | $X_2^+$, $X_3^+$ | – | $a^-a^-c/-(a^-a^-)c$ | 0.5, 0.5 | RP | [47] |
| 805434 | Nonylamine | | $[C_9H_{22}N]_2PbI_4$<br>293 K | *Pbca* | *c* | $X_2^+$, $X_3^+$ | – | $a^-a^-c/-(a^-a^-)c$ | 0.5, 0.5 | RP | [47] |
| 1859258 | Deca-3,5-diyn-1-amine | | $[C_{10}H_{16}N]_2PbBr_4$ | *Pbca* | *c* | $X_2^+$, $X_3^+$ | – | $a^-a^-c/-(a^-a^-)c$ | 0.5, 0.5 | RP | [77] |
| 805436 | Decylamine | | $[C_{10}H_{24}N]_2PbI_4$<br>268 K | *Pbca* | *c* | $X_2^+$, $X_3^+$ | – | $a^-a^-c/-(a^-a^-)c$ | 0.5, 0.5 | RP | [47] |
| 692951 | Dodecylamine | | $[C_{12}H_{28}N]_2PbI_4$<br>293 K | *Pbca* | *c* | $X_2^+$, $X_3^+$ | – | $a^-a^-c/-(a^-a^-)c$ | 0.5, 0.5 | RP | [78] |
| 1934872 | 2-((5-Methoxynaphthalen-1-<br>yl)oxy)ethan-1-amine | | $[C_{13}H_{16}O_2N]_2PbI_4$ | *Pbca* | *c* | $X_2^+$, $X_3^+$ | – | $a^-a^-c/-(a^-a^-)c$ | 0.5, 0.5 | RP | [79] |
| 692953 | Tetradecylamine | | $[C_{14}H_{32}N]_2PbI_4$<br>293 K | *Pbca* | *c* | $X_2^+$, $X_3^+$ | – | $a^-a^-c/-(a^-a^-)c$ | 0.5, 0.5 | RP | [78] |
| 692955 | Hexadecylamine | | $[C_{16}H_{36}N]_2PbI_4$<br>293 K | *Pbca* | *c* | $X_2^+$, $X_3^+$ | – | $a^-a^-c/-(a^-a^-)c$ | 0.5, 0.5 | RP | [78] |
| 692957 | Octadecylamine | | $[C_{18}H_{40}N]_2PbI_4$ 293K | *Pbca* | *c* | $X_2^+, X_3^+$ | – | $a^-a^-c/-(a^-a^-)c$ | 0.5, 0.5 | RP | 78 |
| 1826587 | 2-Methylpentane-1,5-diamine | [A] | $[C_6H_{18}N_2]PbCl_4$ | *Cc* | *a* | $X_2^+, X_4^+$ | $\Gamma_5^+$ | $a^-a^-c/a^-a^-c$ | 0.30, 0.30 | nRP | 80 |
| 1119686 | 2-Methylpentane-1,5-diamine | [A] | $[C_6H_{18}N_2]PbBr_4$ | *Cc* | *a* | $X_2^+, X_4^+$ | $\Gamma_5^+$ | $a^-a^-c/a^-a^-c$ | 0.26, 0.26 | nRP | 81 |
| 1869673 | 1,9-diaminononane | [A] | $[C_9H_{24}N_2]PbI_4$ | *Cc* | *a* | $X_2^+, X_4^+$ | $\Gamma_5^+$ | $a^-a^-c/a^-a^-c$ | 0.42, 0.42 | nRP | 82 |
| 1840806 | Pyrene-*O*-propylamine | | $[C_{19}H_{18}ON]_2PbI_4$ | *Cc* | *a* | $X_2^+, X_4^+$ | $\Gamma_5^+$ | $a^-a^-c/a^-a^-c$ | 0.27, 0.27 | nRP | 83 |
| 853207 | 1,4-Diaminobutane | [A] | $[C_4H_{14}N_2]PbI_4$ | $C2/c$ | *a* | $X_2^+, X_4^+$ | $\Gamma_5^+$ | $a^-a^-c/a^-a^-c$ | 0.22, 0.22 | nDJ | 84 |
| 1521067 | 2-Methylpentane-1,5-diamine | [A] | $[C_6H_{18}N_2]PbBr_4$ | $C2/c$ | *a* | $X_2^+, X_4^+$ | $\Gamma_5^+$ | $a^-a^-c/a^-a^-c$ | 0.24, 0.24 | nDJ | 85 |
| 1977186 | 4-Chlorophenethylamine | | $[C_8H_{11}ClN]_2PbI_4$ | $C2/c$ | *a* | $X_2^+, X_4^+$ | $\Gamma_5^+$ | $a^-a^-c/a^-a^-c$ | 0.40, 0.40 | nRP | 86 |
| 1977187 | 4-Bromophenethylamine | | $[C_8H_{11}BrN]_2PbI_4$ | $C2/c$ | *a* | $X_2^+, X_4^+$ | $\Gamma_5^+$ | $a^-a^-c/a^-a^-c$ | 0.44, 0.44 | nRP | 86 |
| 956552 | (RS)-1-Cyclohexylethylamine | | $[C_8H_{18}N]_2PbCl_4$ | $C2/c$ | *a* | $X_2^+, X_4^+$ | $\Gamma_5^+$ | $a^-a^-c/a^-a^-c$ | 0.49, 0.49 | nRP | 65 |
| 1856671 | Hexadecylamine | | $[C_{16}H_{36}N]_2PbI_4$ | $Pca2_1$ | *b* | $X_2^+, X_3^+$ | $M_5^-$ | $a^-a^-c/-(a^-a^-)c$ | 0.5, 0.5 | RP | 87 |
| 1934873 | 4-[(Naphthalen-1-yl)oxy]butyl-1-amine | | $[C_{14}H_{18}ON]_2PbI_4$ | $Pca2_1$ | *c* | $X_2^+, X_3^+$ | $M_5^-$ | $a^-a^-c/-(a^-a^-)c$ | 0.08, 0.08 | nDJ | 88 |
| 1903531 | Butylamine | | $[C_4H_{12}N]_2PbBr_4$ RT | $P2_1/c$ | *a* | $X_2^+, X_3^+$ | $M_5^-$ | $a^-a^-b/-(a^-a^-)c$ | | nRP | 89 |
| 1119707 | Propane-1,3-diamine | [A] | $[C_3H_{12}N_2]PbCl_4$ | $P2_12_12_1$ | *a* | $X_2^+, X_3^+$ | $M_5^-$ | $a^-a^-c/-(a^-a^-)c$ | 0.13, 0.13 | nDJ | 81 |
| 607740 & 607741 | (R)- or (S)-1-Phenylethylamine | | $[C_8H_{12}N]_2PbI_4$ | $P2_12_12_1$ | *c* | $X_2^+, X_3^+$ | $M_5^-$ | $a^-a^-c/-(a^-a^-)c$ | 0.29, 0.29 | nRP | 90 |
| 1942543 | 3-(Aminoethyl)pyridine | [A] | $[C_6H_{10}N_2]PbI_4$ | *Pn* | *c* | $X_2^+, X_4^+$ | $M_5^-, \Gamma_5^+$ | | | nDJ | 91 |
| 2016669 | 3-Fluoro-*N*-methylaniline | | $[C_7H_9FN]_2PbI_4$ | *Pn* | *b* | $X_3^+, X_4^+$ | $M_5^-, \Gamma_5^+$ | | | nRP | 92 |
| 659021 | *m*-Phenylenediamine | [A] | $[C_6H_{10}N_2]PbCl_4$ | $P2/c$ | *a* | $X_2^+, X_4^+$ | $M_5^-, \Gamma_5^+$ | | | nDJ | 93 |
| 1305732 | 1,4-Diaminobutane | [A] | $[C_4H_{14}N_2]PbCl_4$ | $P2_1/c$ | *c* | $X_2^+, X_3^+$ | $M_5^-, \Gamma_5^+$ | | | nDJ | 94 |
| 961380 | *N*-methylpropane-1,3-diamine | [A] | $[C_4H_{14}N_2]PbBr_4$ | $P2_1/c$ | *c* | $X_2^+, X_3^+$ | $M_5^-, \Gamma_5^+$ | | | nDJ2 | 95 |
| 1841478 | 2,2′-Dithiobis(ethylammonium) | [A] | $[C_4H_{14}S_2N_2]PbCl_4$ | $P2_1/c$ | *c* | $X_2^+, X_3^+$ | $M_5^-, \Gamma_5^+$ | | | nDJ2 | 96 |
| 1877264 | 1-Methylpiperidin-4-amine | [A] | $[C_6H_{16}N_2]PbI_4$ | $P2_1/c$ | *c* | $X_2^+, X_3^+$ | $M_5^-, \Gamma_5^+$ | | | nDJ | 97 |
| 1914631 | 1,6-Diaminohexane | [A] | $[C_6H_{18}N_2]PbCl_4$ | $P2_1/c$ | *c* | $X_2^+, X_3^+$ | $M_5^-, \Gamma_5^+$ | | | nDJ2 | 98 |
| 1521055 | 3-(2-Aminoethyl)aniline | [A] | $[C_8H_{14}N_2]PbBr_4$ | $P2_1/c$ | *c* | $X_2^+$, $X_3^+$ | $M_5^-$, $\Gamma_5^+$ | | | nDJ | 85 |
| 1525376 | Histamine | [A] | $[C_5H_{11}N_3]PbI_4$ | $P2_1/n$ | *b* | $X_2^+$, $X_4^+$ | $M_5^-$, $\Gamma_5^+$ | | | nDJ2 | 99 |
| 1999302 | 5-Aminopentanoic acid | | $[C_5H_{12}O_2N]_2PbBr_4$ | $P2_1/n$ | *c* | $X_2^+$, $X_4^+$ | $M_5^-$, $\Gamma_5^+$ | | | nDJ2 | 100 |
| 1819854 | 4-Fluorobenzylamine | | $[C_7H_9FN]_2PbI_4$ | $P2_1/n$ | *c* | $X_3^+$, $X_4^+$ | $M_5^-$, $\Gamma_5^+$ | | | nDJ | 101 |
| 1934876 | 3-(Pentachlorophenoxy)propyl-1-amine | | $[C_9H_9Cl_5ON]_2PbI_4$ | $P2_1/n$ | *c* | $X_3^+$, $X_4^+$ | $M_5^-$, $\Gamma_5^+$ | | | nRP | 88 |

**Table 3.** Summary of experimentally known structures with two octahedral layers per unit cell (derived from RP parent) and at least one axis in the layer plane doubled.

| CCDC Number | Amine | Type | Formula | Space group | Metrics | Key modes | | Tilt system | Layer shift factor, Δ | Structure type | Ref. |
|---|---|---|---|---|---|---|---|---|---|---|---|
| | | | | | | Tilt | Layer shift | | | | |
| 1962913 | Piperidine | | $[C_5H_{12}N]_2PbCl_4$ | $C2/c$ | $c \times a \times 2a$ | – | $\Gamma_5^+$ | none (see text) | | nRP | 105 |
| 1507154 | Benzimidazole | | $[C_7H_7N_2]_2PbI_4$ | $C2/c$ | $c \times a \times 2a$ | – | $\Gamma_5^+$ | – | | nDJ2 | 106 |
| 1893384 | 3-Fluorophenethylamine | | $[C_8H_{11}FN]_2PbI_4$ | $C2/c$ | $c \times a \times 2a$ | – | $\Gamma_5^+$ | – | | nRP | 107 |
| 1840805 | Pyrene-*O*-butylamine | | $[C_{20}H_{20}ON]_2PbI_4$ | $C2/c$ | $c \times a \times 2a$ | – | $\Gamma_5^+$ | – | | nDJ2 | 83 |
| 1846391 | 2-(3‴,4′-dimethyl-[2,2′:5′,2″:5″,2‴-quaterthiophen]-5-yl)ethan-1-amine | | $[C_{20}H_{20}S_4N]_2PbI_4$ | $C2/c$ | $c \times a \times 2a$ | – | $\Gamma_5^+$ | – | | nRP | 108 |
| 1938883 | 2-Fluoroethylamine | | $[C_2H_7FN]_2PbBr_4$ | *Pnma* | $2a \times c \times a$ | – | $M_5^-$ | – | 0.17, 0.43 | nDJ2 | 60 |
| 746131 | 2-Bromoethylamine | | $[C_2H_7BrN]_2PbI_4$ 73 K | *Pnma* | $2a \times c \times a$ | – | $M_5^-$ | – | 0.16, 0.42 | nDJ2 | 75 |
| 1962914 | Piperidine | | $[C_5H_{12}N]_2PbBr_4$ | *Pnma* | $2a \times c \times a$ | – | $M_5^-$ | – | 0.48, 0.49 | nRP | 105 |
| 641644 | 2-Chloroethylamine | | $[C_2H_7ClN]_2PbI_4$ | *Pbnm* | $a \times 2a \times c$ | – | $M_5^-$ | – | 0.44, 0.17 | nDJ2 | 38 |
| 641643 | 2-Bromoethylamine | | $[C_2H_7BrN]_2PbI_4$ 293 K | *Pbnm* | $a \times 2a \times c$ | – | $M_5^-$ | – | 0.46, 0.17 | nDJ2 | 38 |
| 1588974 | Methylamine & Guanidine | [A][A′] | $[CH_6N][CH_6N_3]PbI_4$ | *Imma* | $a \times 2a \times c$ | $T_3^+$ | – | $a^+b^0c^0/–(a^+)b^0c^0$ | 0.5, 0 | nDJ2 | 109 |
| 1552603 | Guanidine & Cs | [A][A′] | $[CH_6N_3]CsPbBr_4$ | *Imma* | $a \times 2a \times c$ | $T_3^+$ | – | $a^+b^0c^0/–(a^+)b^0c^0$ | 0.5, 0 | nDJ2 | 110 |
| 1915486 | 4-(2-Aminoethyl)pyridine | [A] | $[C_7H_{12}N_2]PbBr_4$ | $P2_1/n$ | $a \times 2a \times c$ | Σ | $M_5^-$, $\Gamma_5^+$ | complex: see text | | nRP | 111 |
| 1944786 | 2-(2-Aminoethyl)pyridine | [A] | $[C_7H_{12}N_2]PbI_4$ | $P2_1/n$ | $a \times 2a \times c$ | Σ | $M_5^-$, $\Gamma_5^+$ | | | nDJ2 | 112 |
| 1944783 | 3-(2-Aminoethyl)pyridine | [A] | $[C_7H_{12}N_2]PbI_4$ | $P2_1/n$ | $a \times 2a \times c$ | Σ | $M_5^-$, $\Gamma_5^+$ | | | nDJ2 | 112 |
| 1944782 | 4-(2-Aminoethyl)pyridine | [A] | $[C_7H_{12}N_2]PbI_4$ | $P2_1/n$ | $a \times 2a \times c$ | Σ | $M_5^-$, $\Gamma_5^+$ | | | nDJ2 | 112 |
| 659016 | *N*,*N*-Dimethyl-*p*-phenylenediamine | [A] | $[C_8H_{14}N_2]PbCl_4$ | $P2_1/n$ | $a \times 2a \times c$ | Σ | $M_5^-$, $\Gamma_5^+$ | | | nDJ2 | 93 |
| 1572154 | *N*,*N*-Dimethyl-*p*-phenylenediamine | [A] | $[C_8H_{14}N_2]PbBr_4$ | $P2_1/n$ | $a \times 2a \times c$ | Σ | $M_5^-$, $\Gamma_5^+$ | | | nDJ | 113 |
| 1572156 | *N*,*N*-Dimethyl-*p*-phenylenediamine | [A] | $[C_8H_{14}N_2]PbI_4$ | $P2_1/n$ | $a \times 2a \times c$ | Σ | $M_5^-$, $\Gamma_5^+$ | | | nDJ | 113 |
| 1841680 | *N*-(2-Aminoethyl)pyridine | [A] | $[C_7H_{12}N_2]PbI_4$ | $P2_1/c$ | $a \times c \times 2a$ | | | | | nDJ2 | 114 |
| 628793 | Cystamine | [A] | $[C_4H_{14}S_2N_2]PbBr_4$ | $C2/c$ | $2a \times 2a \times c$ | $P_4$ | $M_5^-$, $\Gamma_5^+$ | $a^0a^0c/a^0a^0(-)c$ | | nDJ2 | 115 |
| 1985833 | 1,2,4-Triazole | | $[C_2H_4N_3]_2PbBr_4$ | $C2/c$ | $2a \times 2a \times c$ | $P_4$, $P_5$ | $M_5^-$, $\Gamma_5^+$ | complex: see text | | nDJ | 116 |
| 1043214 | (2-Thiophene)ethylamine | | $[C_6H_{10}SN]_2PbI_4$ | $Cc$ | $2a \times 2a \times c$ | $P_4$ | $M_5^-$, $\Gamma_5^+$ | | | nDJ2 | 117 |
| 1841681 | Phenethylamine | | $[C_8H_{12}N]_2PbI_4$ | $Cc$ | $2a \times 2a \times c$ | $P_4$ | $M_5^-$, $\Gamma_5^+$ | | | nDJ2 | 114 |
| 1840808 | Naphthalene-*O*-ethylamine | | $[C_{12}H_{14}ON]_2PbI_4$ | $Cc$ | $2a \times 2a \times c$ | $P_4$ | $M_5^-$, $\Gamma_5^+$ | | | nRP | 83 |
| 1840802 | Pyrene-*O*-ethylamine | | $[C_{18}H_{16}ON]_2PbI_4$ | $Cc$ | $2a \times 2a \times c$ | $P_4$ | $M_5^-$, $\Gamma_5^+$ | | | nRP | 83 |
| 2016668 | *N*-Methylaniline | | $[C_7H_{10}N]_2PbI_4$ | $Cc$ | $2a \times 2a \times c$ | $P_5$ | $M_5^-$, $\Gamma_5^+$ | | | nRP | 118 |
| 1542463 | 2-(2-Naphthyl)ethylamine | | $[C_{12}H_{14}N]_2PbI_4$ | $Pn$ | $2a \times 2a \times c$ | $P_4$ | $M_5^-$, $\Gamma_5^+$ | | | nRP | 52 |
| 1552604 | Guanidine & Cs | [A][A′] | $[CH_6N_3]CsPbI_4$ | $Pnnm$ | $2a \times c \times 2a$ | $X_2^+$, $X_3^+$ | $M_5^-$, $\Gamma_5^+$ | | | nDJ2 | 110 |
| 708569 | Cyclohexylamine | | $[C_6H_{14}N]_2PbCl_4$ | $P2_1/m$ | $2a \times c \times 2a$ | $X_2^+$ | $M_5^-$ | | | nRP | 43 |
| 1841683 | *N*-(2-Aminoethyl)piperidine | [A] | $[C_7H_{18}N_2]PbI_4$ | $P2_1/n$ | $2a \times c \times 2a$ | $\Sigma_3$ | $M_5^-$ | | | nDJ | 114 |

**Table 4.** Summary of experimentally known structures with one octahedral layer per unit cell (derived from DJ parent).

| CCDC Number | Amine | Type | Formula | Space group | Metrics | Key modes | | Tilt system | Layer shift factor, Δ | Structure type | Ref. |
|---|---|---|---|---|---|---|---|---|---|---|---|
| | | | | | | Tilt | Layer shift | | | | |
| 993479 | 2,2’-(Ethylenedioxy)bis(ethylamine) | [A] | $[C_6H_{18}O_2N_2]PbCl_4$ | *C*2 | $\sqrt{2}a \times \sqrt{2}a \times c$ | – | $\Gamma_5^+$ | $a^0a^0c^0$ | 0.37, 0.37 | nRP | 119 |
| 1871404 | 3,5-Dihydroimidazo[4,5-f]benzimidazole | [A] | $[C_8H_8N_4]PbI_4$ | *C*2/*m* | $\sqrt{2}a \times \sqrt{2}a \times c$ | – | $\Gamma_5^+$ | $a^0a^0c^0$ | 0.31, 0.31 | nRP | 120 |
| 120686 | 1-(2-Naphthyl)-methylamine | | $[C_{11}H_{12}N]_2PbCl_4$ | *Pbam* | $\sqrt{2}a \times \sqrt{2}a \times c$ | $M_3^+$ | – | $a^0a^0c$ | 0, 0 | DJ | 53 |
| 641641 | 2-Iodoethylamine | | $[C_2H_7IN]_2PbI_4$ 293 K | $P2_1/a$ | $\sqrt{2}a \times \sqrt{2}a \times c$ | $M_3^+$, $M_5^+$ | $\Gamma_5^+$ | $a^-a^-c$ | 0.19, 0.19 | nDJ | 38 |
| 237189 | 2-Hydroxyethylamine | | $[C_2H_8ON]_2PbI_4$ 293 K | $P2_1/a$ | $\sqrt{2}a \times \sqrt{2}a \times c$ | $M_3^+$, $M_5^+$ | $\Gamma_5^+$ | $a^-a^-c$ | 0.20, 0.20 | nDJ | 55 |
| 724583 | 2,2’-Disulfanediyldiethanamine | [A] | $[C_4H_{14}S_2N_2]PbI_4$ *β*-phase | $P2_1/a$ | $\sqrt{2}a \times \sqrt{2}a \times c$ | $M_3^+$, $M_5^+$ | $\Gamma_5^+$ | $a^-a^-c$ | 0.33, 0.33 | nRP | 121 |
| 665691 | Pentylamine | | $[C_5H_{14}N]_2PbI_4$ 173 K | $P2_1/a$ | $\sqrt{2}a \times \sqrt{2}a \times c$ | $M_3^+$, $M_5^+$ | $\Gamma_5^+$ | $a^-a^-c$ | 0.33, 0.33 | nRP | 71 |
| 665694 | Hexylamine | | $[C_6H_{16}N]_2PbI_4$ 173 K | $P2_1/a$ | $\sqrt{2}a \times \sqrt{2}a \times c$ | $M_3^+$, $M_5^+$ | $\Gamma_5^+$ | $a^-a^-c$ | 0.06, 0.06 | nDJ | 71 |
| 805427 | Heptylamine | | $[C_7H_{18}N]_2PbI_4$ 253 K | $P2_1/a$ | $\sqrt{2}a \times \sqrt{2}a \times c$ | $M_3^+$, $M_5^+$ | $\Gamma_5^+$ | $a^-a^-c$ | 0.24, 0.24 | nDJ | 47 |
| 141193 | 4-Methylbenzylamine | | $[C_8H_{12}N]_2PbCl_4$ | $P2_1/a$ | $\sqrt{2}a \times \sqrt{2}a \times c$ | $M_3^+$, $M_5^+$ | $\Gamma_5^+$ | $a^-a^-c$ | 0.21, 0.21 | nDJ | 122 |
| 141192 | 4-Methylbenzylamine | | $[C_8H_{12}N]_2PbBr_4$ | $P2_1/a$ | $\sqrt{2}a \times \sqrt{2}a \times c$ | $M_3^+$, $M_5^+$ | $\Gamma_5^+$ | $a^-a^-c$ | 0.32, 0.32 | nRP | 122 |
| 141190 | 4-Methylbenzylamine | | $[C_8H_{12}N]_2PbI_4$ | $P2_1/a$ | $\sqrt{2}a \times \sqrt{2}a \times c$ | $M_3^+$, $M_5^+$ | $\Gamma_5^+$ | $a^-a^-c$ | 0.44, 0.44 | nRP | 122 |
| 200737 | (RS)-1-Phenylethylamine | | $[C_8H_{12}N]_2PbI_4$ | $P2_1/a$ | $\sqrt{2}a \times \sqrt{2}a \times c$ | $M_3^+$, $M_5^+$ | $\Gamma_5^+$ | $a^-a^-c$ | 0.18, 0.18 | nDJ | 123 |
| 2016667 | 4-Methoxy-*N*-methylaniline | | $[C_8H_{12}ON]_2PbI_4$ | $P2_1/a$ | $\sqrt{2}a \times \sqrt{2}a \times c$ | $M_3^+$, $M_5^+$ | $\Gamma_5^+$ | $a^-a^-c$ | 0.50, 0.50 | RP | 124 |
| 805430 | Octylamine | | $[C_8H_{20}N]_2PbI_4$ 173 K | $P2_1/a$ | $\sqrt{2}a \times \sqrt{2}a \times c$ | $M_3^+$, $M_5^+$ | $\Gamma_5^+$ | $a^-a^-c$ | 0.24, 0.24 | nDJ | 47 |
| 805433 | Nonylamine | | $[C_9H_{22}N]_2PbI_4$ 223 K | $P2_1/a$ | $\sqrt{2}a \times \sqrt{2}a \times c$ | $M_3^+$, $M_5^+$ | $\Gamma_5^+$ | $a^-a^-c$ | 0.25, 0.25 | nDJ/nRP | 47 |
| 805435 | Decylamine | | $[C_{10}H_{24}N]_2PbI_4$ 243 K | $P2_1/a$ | $\sqrt{2}a \times \sqrt{2}a \times c$ | $M_3^+$, $M_5^+$ | $\Gamma_5^+$ | $a^-a^-c$ | 0.11, 0.11 | nDJ | 47 |
| 692952 | Dodecylamine | | $[C_{12}H_{28}N]_2PbI_4$ 319 K | $P2_1/a$ | $\sqrt{2}a \times \sqrt{2}a \times c$ | $M_3^+$, $M_5^+$ | $\Gamma_5^+$ | $a^-a^-c$ | 0.14, 0.14 | nDJ | 78 |
| 692954 | Tetradecylamine | | $[C_{14}H_{32}N]_2PbI_4$ 335 K | $P2_1/a$ | $\sqrt{2}a \times \sqrt{2}a \times c$ | $M_3^+$, $M_5^+$ | $\Gamma_5^+$ | $a^-a^-c$ | 0.13, 0.13 | nDJ | 78 |
| 692956 | Hexadecylamine | | $[C_{16}H_{36}N]_2PbI_4$ 341 K | $P2_1/a$ | $\sqrt{2}a \times \sqrt{2}a \times c$ | $M_3^+$, $M_5^+$ | $\Gamma_5^+$ | $a^-a^-c$ | 0.11, 0.11 | nDJ | 78 |
| 692958 | Octadecylamine | | $[C_{18}H_{40}N]_2PbI_4$ 348 K | $P2_1/a$ | $\sqrt{2}a \times \sqrt{2}a \times c$ | $M_3^+$, $M_5^+$ | $\Gamma_5^+$ | $a^-a^-c$ | 0.10, 0.10 | nDJ | 78 |
| 746126 | 2-Iodoethylamine | | $[C_2H_7IN]_2PbI_4$ | $P2_1/c$ | $c \times \sqrt{2}a \times \sqrt{2}a$ | $M_3^+$, $M_5^+$ | $\Gamma_5^+$ | $a^-a^-c$ | 0.20, 0.20 | nDJ | 75 |
| 1855044 | Ethylamine | | $[C_2H_8N]_2PbBr_4$ | $P2_1/c$ | $c \times \sqrt{2}a \times \sqrt{2}a$ | $M_3^+$, $M_5^+$ | $\Gamma_5^+$ | $a^-a^-c$ | 0.46, 0.46 | nRP | 125 |
| 746124 | 2-Hydroxyethylamine | | $[C_2H_8ON]_2PbI_4$ 173 K | $P2_1/c$ | $c \times \sqrt{2}a \times \sqrt{2}a$ | $M_3^+$, $M_5^+$ | $\Gamma_5^+$ | $a^-a^-c$ | 0.21, 0.21 | nDJ | 75 |
| 708566 | Cyclopropylamine | | $[C_3H_8N]_2PbCl_4$ | $P2_1/c$ | $c \times \sqrt{2}a \times \sqrt{2}a$ | $M_3^+$, $M_5^+$ | $\Gamma_5^+$ | $a^-a^-c$ | 0.50, 0.50 | RP | 43 |
| 708560 | Cyclopropylamine | | $[C_3H_8N]_2PbBr_4$ | $P2_1/c$ | $c \times \sqrt{2}a \times \sqrt{2}a$ | $M_3^+$, $M_5^+$ | $\Gamma_5^+$ | $a^-a^-c$ | 0.46, 0.46 | nRP | 43 |
| 609992 | Cyclopropylamine | | $[C_3H_8N]_2PbI_4$ | $P2_1/c$ | $c \times \sqrt{2}a \times \sqrt{2}a$ | $M_3^+$, $M_5^+$ | $\Gamma_5^+$ | $a^-a^-c$ | 0.49, 0.49 | nRP | 74 |
| 746127 | 3-Iodopropylamine | | $[C_3H_9IN]_2PbI_4$ | $P2_1/c$ | $c \times \sqrt{2}a \times \sqrt{2}a$ | $M_3^+$, $M_5^+$ | $\Gamma_5^+$ | $a^-a^-c$ | 0.27, 0.27 | nRP | 75 |
| 746125 | 3-Hydroxypropylamine | | $[C_3H_{10}ON]_2PbI_4$ | $P2_1/c$ | $c \times \sqrt{2}a \times \sqrt{2}a$ | $M_3^+$, $M_5^+$ | $\Gamma_5^+$ | $a^-a^-c$ | 0.06, 0.06 | nDJ | 75 |
| 955776 | 3-Butyn-1-amine | | $[C_4H_8N]_2PbBr_4$ | $P2_1/c$ | $c \times \sqrt{2}a \times \sqrt{2}a$ | $M_3^+$, $M_5^+$ | $\Gamma_5^+$ | $a^-a^-c$ | 0.09, 0.09 | nDJ | 126 |
| 955778 | But-3-en-1-amine | | $[C_4H_8I_2N]_2PbBr_4$ | $P2_1/c$ | $c \times \sqrt{2}a \times \sqrt{2}a$ | $M_3^+$, $M_5^+$ | $\Gamma_5^+$ | $a^-a^-c$ | 0.18, 0.18 | nDJ | 126 |
| 708567 | Cyclobutylamine | | $[C_4H_{10}N]_2PbCl_4$ | $P2_1/c$ | $c \times \sqrt{2}a \times \sqrt{2}a$ | $M_3^+$, $M_5^+$ | $\Gamma_5^+$ | $a^-a^-c$ | 0.47, 0.47 | nRP | 43 |
| 708561 | Cyclobutylamine | | $[C_4H_{10}N]_2PbBr_4$ | $P2_1/c$ | $c \times \sqrt{2}a \times \sqrt{2}a$ | $M_3^+$, $M_5^+$ | $\Gamma_5^+$ | $a^-a^-c$ | 0.50, 0.50 | RP | 43 |
| 609993 | Cyclobutylamine | | $[C_4H_{10}N]_2PbI_4$ | $P2_1/c$ | $c \times \sqrt{2}a \times \sqrt{2}a$ | $M_3^+$, $M_5^+$ | $\Gamma_5^+$ | $a^-a^-c$ | 0.45, 0.45 | nRP | 74 |
| 746128 | 4-Iodobutylamine | | $[C_4H_{11}IN]_2PbI_4$ | $P2_1/c$ | $c \times \sqrt{2}a \times \sqrt{2}a$ | $M_3^+$, $M_5^+$ | $\Gamma_5^+$ | $a^-a^-c$ | 0.06, 0.06 | nDJ | 75 |
| 1417496 | Isobutylamine | | $[C_4H_{12}N]_2PbBr_4$ | $P2_1/c$ | $c \times \sqrt{2}a \times \sqrt{2}a$ | $M_3^+$, $M_5^+$ | $\Gamma_5^+$ | $a^-a^-c$ | 0.46, 0.46 | nRP | 42 |
| 1876240 | Isobutylamine | | $[C_4H_{12}N]_2PbI_4$ | $P2_1/c$ | $c \times \sqrt{2}a \times \sqrt{2}a$ | $M_3^+$, $M_5^+$ | $\Gamma_5^+$ | $a^-a^-c$ | 0.49, 0.49 | nRP | 127 |
| 1495869 | 3-Bromopyridine | | $[C_5H_5BrN]_2PbI_4$ | $P2_1/c$ | $c \times \sqrt{2}a \times \sqrt{2}a$ | $M_3^+$, $M_5^+$ | $\Gamma_5^+$ | $a^-a^-c$ | 0.22, 0.22 | nDJ | 64 |
| 708562 | Cyclopentylamine | | $[C_5H_{12}N]_2PbBr_4$ | $P2_1/c$ | $c \times \sqrt{2}a \times \sqrt{2}a$ | $M_3^+$, $M_5^+$ | $\Gamma_5^+$ | $a^-a^-c$ | 0.43, 0.43 | nRP | 43 |
| 609994 | Cyclopentylamine | | $[C_5H_{12}N]_2PbI_4$ | $P2_1/c$ | $c \times \sqrt{2}a \times \sqrt{2}a$ | $M_3^+$, $M_5^+$ | $\Gamma_5^+$ | $a^-a^-c$ | 0.50, 0.50 | RP | 74 |
| 746129 | 5-Iodopentylamine | | $[C_5H_{13}IN]_2PbI_4$ | $P2_1/c$ | $c \times \sqrt{2}a \times \sqrt{2}a$ | $M_3^+$, $M_5^+$ | $\Gamma_5^+$ | $a^-a^-c$ | 0.00, 0.00 | DJ | 75 |
| 1863836 | 2-Aminopentane | | $[C_5H_{14}N]_2PbI_4$ | $P2_1/c$ | $c \times \sqrt{2}a \times \sqrt{2}a$ | $M_3^+$, $M_5^+$ | $\Gamma_5^+$ | $a^-a^-c$ | 0.49, 0.49 | nRP | 40 |
| 2018083 | 3-Methylbutyl-1-amine | | $[C_5H_{14}N]_2PbI_4$ | $P2_1/c$ | $c \times \sqrt{2}a \times \sqrt{2}a$ | $M_3^+$, $M_5^+$ | $\Gamma_5^+$ | $a^-a^-c$ | 0.50, 0.50 | RP | 128 |
| 249243 | 4-Chloroaniline | | $[C_6H_7ClN]_2PbI_4$ | $P2_1/c$ | $c \times \sqrt{2}a \times \sqrt{2}a$ | $M_3^+$, $M_5^+$ | $\Gamma_5^+$ | $a^-a^-c$ | 0.30, 0.30 | nRP | 129 |
| 723486 | 4-Bromoaniline | | $[C_6H_7BrN]_2PbI_4$ | $P2_1/c$ | $c \times \sqrt{2}a \times \sqrt{2}a$ | $M_3^+$, $M_5^+$ | $\Gamma_5^+$ | $a^-a^-c$ | 0.30, 0.30 | nRP | 130 |
| 1939732 | 6-Aminohexanoic acid | | $[C_6H_{14}O_2N]_2PbBr_4$ | $P2_1/c$ | $c \times \sqrt{2}a \times \sqrt{2}a$ | $M_3^+$, $M_5^+$ | $\Gamma_5^+$ | $a^-a^-c$ | 0.39, 0.39 | nRP | 131 |
| 150502 | 1,6-Diaminohexane | [A] | $[C_6H_{18}N_2]PbBr_4$ RT | $P2_1/c$ | $c \times \sqrt{2}a \times \sqrt{2}a$ | $M_3^+$, $M_5^+$ | $\Gamma_5^+$ | $a^-a^-c$ | 0.39, 0.39 | nRP | 132 |
| 150501 | 1,6-Diaminohexane | [A] | $[C_6H_{18}N_2]PbI_4$ | $P2_1/c$ | $c \times \sqrt{2}a \times \sqrt{2}a$ | $M_3^+$, $M_5^+$ | $\Gamma_5^+$ | $a^-a^-c$ | 0.38, 0.38 | nRP | 132 |
| 1944741 | 3-Fluorobenzylamine | | $[C_7H_9FN]_2PbCl_4$ 300 K | $P2_1/c$ | $c \times \sqrt{2}a \times \sqrt{2}a$ | $M_3^+$, $M_5^+$ | $\Gamma_5^+$ | $a^-a^-c$ | 0.15, 0.15 | nDJ | 32 |
| 1950233 | 4-Iodobenzylamine | | $[C_7H_9IN]_2PbI_4$ | $P2_1/c$ | $c \times \sqrt{2}a \times \sqrt{2}a$ | $M_3^+$, $M_5^+$ | $\Gamma_5^+$ | $a^-a^-c$ | 0.45, 0.45 | nRP | 28 |
| 1482271 | (Cyclohexylmethyl)amine | | $[C_7H_{16}N]_2PbBr_4$ RT | $P2_1/c$ | $c \times \sqrt{2}a \times \sqrt{2}a$ | $M_3^+$, $M_5^+$ | $\Gamma_5^+$ | $a^-a^-c$ | 0.33, 0.33 | nRP | 46 |
| 1863839 | 2-Aminoheptane | | $[C_7H_{18}N]_2PbI_4$ | $P2_1/c$ | $c \times \sqrt{2}a \times \sqrt{2}a$ | $M_3^+$, $M_5^+$ | $\Gamma_5^+$ | $a^-a^-c$ | 0.47, 0.47 | nRP | 40 |
| 1986789 | 2-(3,5-Dichlorophenyl)ethyl-1-amine | | $[C_8H_{10}Cl_2N]_2PbI_4$ | $P2_1/c$ | $c \times \sqrt{2}a \times \sqrt{2}a$ | $M_3^+$, $M_5^+$ | $\Gamma_5^+$ | $a^-a^-c$ | 0.36, 0.36 | nRP | 133 |
| 1986786 | 2-(3,5-Dibromophenyl)ethyl-1-amine | | $[C_8H_{10}Br_2N]_2PbI_4$ | $P2_1/c$ | $c \times \sqrt{2}a \times \sqrt{2}a$ | $M_3^+$, $M_5^+$ | $\Gamma_5^+$ | $a^-a^-c$ | 0.19, 0.19 | nDJ | 133 |
| 1488195 | 4-Fluorophenethylamine | | $[C_8H_{11}FN]_2PbI_4$ | $P2_1/c$ | $c \times \sqrt{2}a \times \sqrt{2}a$ | $M_3^+$, $M_5^+$ | $\Gamma_5^+$ | $a^-a^-c$ | 0.32, 0.32 | nRP | 134 |
| 1885084 | (RS)-1-(4-Chlorophenyl)ethylamine | | $[C_8H_{11}ClN]_2PbI_4$ | $P2_1/c$ | $c \times \sqrt{2}a \times \sqrt{2}a$ | $M_3^+$, $M_5^+$ | $\Gamma_5^+$ | $a^-a^-c$ | 0.22,0.22 | nDJ | 135 |
| 1863838 | 2-Ethylhexylamine | | $[C_8H_{20}N]_2PbI_4$ | $P2_1/c$ | $c \times \sqrt{2}a \times \sqrt{2}a$ | $M_3^+$, $M_5^+$ | $\Gamma_5^+$ | $a^-a^-c$ | 0.20, 0.20 | nDJ | 40 |
| 781210 | 1,4-Bis(aminomethyl)cyclohexane | [A] | $[C_8H_{20}N_2]PbBr_4$ | $P2_1/c$ | $c \times \sqrt{2}a \times \sqrt{2}a$ | $M_3^+$, $M_5^+$ | $\Gamma_5^+$ | $a^-a^-c$ | 0.15, 0.15 | nDJ | 136 |
| 781212 | 1,4-Bis(aminomethyl)cyclohexane | [A] | $[C_8H_{20}N_2]PbI_4$ | $P2_1/c$ | $c \times \sqrt{2}a \times \sqrt{2}a$ | $M_3^+$, $M_5^+$ | $\Gamma_5^+$ | $a^-a^-c$ | 0.14, 0.14 | nDJ | 137 |
| 1521059 | 1,8-Diaminooctane | [A] | $[C_8H_{22}N_2]PbBr_4$ | $P2_1/c$ | $c \times \sqrt{2}a \times \sqrt{2}a$ | $M_3^+$, $M_5^+$ | $\Gamma_5^+$ | $a^-a^-c$ | 0.41, 0.41 | nRP | 85 |
| 853209 | 1,8-Diaminooctane | [A] | $[C_8H_{22}N_2]PbI_4$ | $P2_1/c$ | $c \times \sqrt{2}a \times \sqrt{2}a$ | $M_3^+$, $M_5^+$ | $\Gamma_5^+$ | $a^-a^-c$ | 0.44, 0.44 | nRP | 84 |
| 853212 | 1,5-Diaminonaphthalene | [A] | $[C_{10}H_{12}N_2]PbI_4$ | $P2_1/c$ | $c \times \sqrt{2}a \times \sqrt{2}a$ | $M_3^+$, $M_5^+$ | $\Gamma_5^+$ | $a^-a^-c$ | 0.01, 0.01 | nDJ | 84 |
| 1986788 | 2-(3,5-Dimethylphenyl)ethyl-1-amine | | $[C_{10}H_{16}N]_2PbI_4$ | $P2_1/c$ | $c \times \sqrt{2}a \times \sqrt{2}a$ | $M_3^+$, $M_5^+$ | $\Gamma_5^+$ | $a^-a^-c$ | 0.31, 0.31 | nRP | 133 |
| 853210 | 1,10-Diaminodecane | [A] | $[C_{10}H_{26}N_2]PbBr_4$ | $P2_1/c$ | $c \times \sqrt{2}a \times \sqrt{2}a$ | $M_3^+$, $M_5^+$ | $\Gamma_5^+$ | $a^-a^-c$ | 0.20, 0.20 | nDJ | 84 |
| 2015614 | (RS)-1-(Naphthalen-1-yl)ethan-1-aminium | | $[C_{12}H_{14}N]_2PbBr_4$ | $P2_1/c$ | $c \times \sqrt{2}a \times \sqrt{2}a$ | $M_3^+$, $M_5^+$ | $\Gamma_5^+$ | $a^-a^-c$ | 0.01, 0.01 | nDJ | 138 |
| 853211 | 1,12-Diaminododecane | [A] | $[C_{12}H_{30}N_2]PbI_4$ | $P2_1/c$ | $c \times \sqrt{2}a \times \sqrt{2}a$ | $M_3^+$, $M_5^+$ | $\Gamma_5^+$ | $a^-a^-c$ | 0.01, 0.01 | nDJ | 84 |
| 1840803 | Naphthalene-*O*-propylamine | | $[C_{13}H_{16}ON]_2PbI_4$ | $P2_1/c$ | $c \times \sqrt{2}a \times \sqrt{2}a$ | $M_3^+$, $M_5^+$ | $\Gamma_5^+$ | $a^-a^-c$ | 0.26, 0.26 | nRP | 83 |
| 1876191 | 1-(4-Aminobutyl)pyrene | | $[C_{20}H_{20}N]_2PbI_4$ | $P2_1/c$ | $c \times \sqrt{2}a \times \sqrt{2}a$ | $M_3^+$, $M_5^+$ | $\Gamma_5^+$ | $a^-a^-c$ | 0.13, 0.13 | nDJ | 139 |
| 1840807 | Perylene-*O*-ethylamine | | $[C_{22}H_{18}ON]_2PbI_4$ | $P2_1/c$ | $c \times \sqrt{2}a \times \sqrt{2}a$ | $M_3^+$, $M_5^+$ | $\Gamma_5^+$ | $a^-a^-c$ | 0.23, 0.23 | nDJ | 83 |
| 1432453 | 4-Chlorobenzylamine | | $[C_7H_9ClN]_2PbI_4$ | $P2_1$ | $\sqrt{2}a \times \sqrt{2}a \times c$ | $M_3^+$, $M_5^+$ | $\Gamma_5^+$ | $a^-a^-c$ | 0.18, 0.18 | nDJ | 140 |
| 1947898 | 4-Bromobenzylamine | | $[C_7H_9BrN]_2PbI_4$ | $P2_1$ | $\sqrt{2}a \times \sqrt{2}a \times c$ | $M_3^+$, $M_5^+$ | $\Gamma_5^+$ | $a^-a^-c$ | 0.18, 0.18 | nDJ | 141 |
| 2000017 | 2-Bromophenethylamine | | $[C_8H_{11}BrN]_2PbI_4$ | $P2_1$ | $\sqrt{2}a \times \sqrt{2}a \times c$ | $M_3^+$, $M_5^+$ | $\Gamma_5^+$ | $a^-a^-c$ | 0.17, 0.17 | nDJ | 142 |
| 956553 | (R)-1-Cyclohexylethylamine | | $[C_8H_{18}N]_2PbCl_4$ | $P2_1$ | $\sqrt{2}a \times \sqrt{2}a \times c$ | $M_3^+$, $M_5^+$ | $\Gamma_5^+$ | $a^-a^-c$ | 0.41, 0.41 | nRP | 65 |
| 956554 | (S)-1-Cyclohexylethylamine | | $[C_8H_{18}N]_2PbCl_4$ | $P2_1$ | $\sqrt{2}a \times \sqrt{2}a \times c$ | $M_3^+$, $M_5^+$ | $\Gamma_5^+$ | $a^-a^-c$ | 0.37, 0.37 | nRP | 65 |
| 2015620 | (R)-1-(Naphthalen-1-yl)ethan-1-aminium | | $[C_{12}H_{14}N]_2PbBr_4$ | $P2_1$ | $\sqrt{2}a \times \sqrt{2}a \times c$ | $M_3^+$, $M_5^+$ | $\Gamma_5^+$ | $a^-a^-c$ | 0.15, 0.15 | nDJ | 143 |
| 2015618 | (S)-1-(Naphthalen-1-yl)ethan-1-aminium | | $[C_{12}H_{14}N]_2PbBr_4$ | $P2_1$ | $\sqrt{2}a \times \sqrt{2}a \times c$ | $M_3^+$, $M_5^+$ | $\Gamma_5^+$ | $a^-a^-c$ | 0.12, 0.12 | nDJ | 143 |
| 1992691 | 4,4-Difluorohexahydroazepine | | $[C_6H_{12}F_2N]_2PbI_4$ 293 K | $P2_1$ | $c \times \sqrt{2}a \times \sqrt{2}a$ | $M_3^+$, $M_5^+$ | $\Gamma_5^+$ | $a^-a^-c$ | 0.41, 0.41 | nRP | 33 |
| 955777 | 3,4-Diiodobut-3-en-1-amine | | $[C_4H_8I_2N]_2PbBr_4$ | $P$-1 | $\sqrt{2}a \times \sqrt{2}a \times c$ | $M_3^+$, $M_5^+$ | $\Gamma_5^+$ | $a^-a^-c$ | | nDJ | 126 |
| 1048947 | 3,4-Dibromobutan-1-amine | | $[C_4H_{10}Br_2N]_2PbBr_4$ | $P$-1 | $\sqrt{2}a \times \sqrt{2}a \times c$ | $M_3^+$, $M_5^+$ | $\Gamma_5^+$ | $a^-a^-c$ | | nDJ2 | 144 |
| 1048945 | 3,4-Dichlorobutan-1-amine | | $[C_4H_{10}Cl_2N]_2PbBr_4$ | $P$-1 | $\sqrt{2}a \times \sqrt{2}a \times c$ | $M_3^+$, $M_5^+$ | $\Gamma_5^+$ | $a^-a^-c$ | | nDJ2 | 144 |
| 853206 | 1,4-Diaminobutane | [A] | $[C_4H_{14}N_2]PbBr_4$ | $P$-1 | $\sqrt{2}a \times \sqrt{2}a \times c$ | $M_3^+$, $M_5^+$ | $\Gamma_5^+$ | $a^-a^-c$ | | nRP | 84 |
| 1053651 | 1,4-Diaminobutane | [A] | $[C_4H_{14}N_2]PbI_4$ | $P$-1 | $\sqrt{2}a \times \sqrt{2}a \times c$ | $M_3^+$, $M_5^+$ | $\Gamma_5^+$ | $a^-a^-c$ | | nDJ | 145 |
| 1999300 | 5-Aminopentanoic acid | | $[C_5H_{12}O_2N]_2PbCl_4$ | $P$-1 | $\sqrt{2}a \times \sqrt{2}a \times c$ | $M_3^+$, $M_5^+$ | $\Gamma_5^+$ | $a^-a^-c$ | | nRP | 146 |
| 1939731 | 5-Aminopentanoic acid | | $[C_5H_{12}O_2N]_2PbBr_4$ | $P$-1 | $\sqrt{2}a \times \sqrt{2}a \times c$ | $M_3^+$, $M_5^+$ | $\Gamma_5^+$ | $a^-a^-c$ | | nDJ2 | 147 |
| 1893383 | 2-Fluorophenethylamine | | $[C_8H_{11}FN]_2PbI_4$ | $P$-1 | $\sqrt{2}a \times \sqrt{2}a \times c$ | $M_3^+$, $M_5^+$ | $\Gamma_5^+$ | $a^-a^-c$ | | nDJ2 | 148 |
| 1515121 | 2-Phenethylamine | | $[C_8H_{12}N]_2PbI_4$ | $P$-1 | $\sqrt{2}a \times \sqrt{2}a \times c$ | $M_3^+$, $M_5^+$ | $\Gamma_5^+$ | $a^-a^-c$ | | nRP | 149 |
| 219791 | 5-ammoniumethylsulfanyl)-2,2'-bithiophene | [A] | $[C_{12}H_{18}S_4N_2]PbI_4$ | $P$-1 | $\sqrt{2}a \times \sqrt{2}a \times c$ | $M_3^+$, $M_5^+$ | $\Gamma_5^+$ | $a^-a^-c$ | | nRP | 150 |
| 1934875 | 6-[(5-Methoxynaphthalen-1-yl)oxy]hexyl-1-amine | | $[C_{17}H_{24}O_2N]_2PbI_4$ | $P$-1 | $\sqrt{2}a \times \sqrt{2}a \times c$ | $M_3^+$, $M_5^+$ | $\Gamma_5^+$ | $a^-a^-c$ | | nDJ | 79 |
| 1885085 | (R)-1-(4-Chlorophenyl)ethylamine | | $[C_8H_{11}ClN]_2PbI_4$ | $P1$ | $\sqrt{2}a \times \sqrt{2}a \times c$ | $M_3^+$, $M_5^+$ | $\Gamma_5^+$ | $a^-a^-c$ | | nRP | 135 |
| 1885086 | (S)-1-(4-Chlorophenyl)ethylamine | | $[C_8H_{11}ClN]_2PbI_4$ | $P1$ | $\sqrt{2}a \times \sqrt{2}a \times c$ | $M_3^+$, $M_5^+$ | $\Gamma_5^+$ | $a^-a^-c$ | | nRP | 135 |
| 1542464 | 2-(2-Naphthyl)ethylamine | | $[C_{12}H_{14}N]_2PbBr_4$ | $P1$ | $\sqrt{2}a \times \sqrt{2}a \times c$ | $M_3^+$, $M_5^+$ | $\Gamma_5^+$ | $a^-a^-c$ | | nRP | 52 |
| 1883687 | (2-Azaniumylethyl)trimethylphosphonium | [A] | $[C_5H_{16}PN]PbBr_4$ | $P2_1$ | $a \times 2a \times c$ | $X_3^+$ | $\Gamma_5^+$ | $a^+b^0c$ | 0.01, 0.01 | nDJ | 151 |
| 1942547 | 2-(Aminomethyl)pyridine | [A] | $[C_6H_{10}N_2]PbI_4$ | $Pn$ | $2a \times c \times 2a$ | $M_3^+$, $X_3^+$ | $\Gamma_5^+$ | $a^+b^+c$ | 0.00, 0.00 | DJ | 91 |
| 295291 | 4-(2-aminoethyl)imidazole | [A] | $[C_5H_{11}N_3]PbBr_4$ RT | $P2_1/c$ | $c \times 2a \times 2a$ | $M_3^+$, $M_5^+$ | $\Gamma_5^+$ | $a^-b^0c$ | | nDJ2 | 152 |
| 1915484 | 2-(2-aminoethyl)imidazole | [A] | $[C_5H_{11}N_3]PbBr_4$ | $P2_1/c$ | $c \times 2a \times 2a$ | $M_3^+$, $M_5^+$ | $\Gamma_5^+$ | $a^-b^0c$ | | nDJ2 | 111 |
| 1982716 | *N*$^1$,*N*$^1$-Dimethylpropane-1,3-diamine | [A] | $[C_5H_{16}N_2]PbCl_4$ | $P2_1/c$ | $c \times 2a \times 2a$ | $M_3^+$, $M_5^+$ | $\Gamma_5^+$ | $a^-b^0c$ | | nDJ2 | 153 |
| 1963066 | *N*$^1$,*N*$^1$-Dimethylpropane-1,3-diamine | [A] | $[C_5H_{16}N_2]PbBr_4$ | $P2_1/c$ | $c \times 2a \times 2a$ | $M_3^+$, $M_5^+$ | $\Gamma_5^+$ | $a^-b^0c$ | | nDJ2 | 154 |
| 1939809 | 4-(Aminomethyl)piperidine | [A] | $[C_6H_{16}N_2]PbI_4$ 373 K | $P2_1/c$ | $c \times 2a \times 2a$ | $M_3^+$, $M_5^+$ | $\Gamma_5^+$ | $a^0a^0c$ | 0, 0 | DJ | 155 |
| 1831525 | 4-(Aminomethyl)piperidine | [A] | $[C_6H_{16}N_2]PbI_4$ 293 K | $Pc$ | $c \times 2a \times 2a$ | $M_3^+$, $M_5^+$ | $\Gamma_5^+$ | $a^0a^0c$ | 0, 0 | DJ | 156 |
| 1816279 | Cyclohexane-1,2-diamine | [A] | $[C_6H_{16}N_2]PbI_4$ | $P2_12_12$ | $2a \times 2a \times c$ | $M_3^+$ | | $a^0a^0c$ | 0, 0 | DJ | 157 |
| 1498513 | 2-Phenylethylamine | | $[C_8H_{12}N]_2PbCl_4$ | $P$-1 | $2a \times 2a \times c$ | $A_3^+$ | $\Gamma_5^+$ | $a^0a^0c$/ $a^0a^0(-c)$ | | nDJ2 | 158 |
| 754084 | 2-Phenylethylamine | | $[C_8H_{12}N]_2PbBr_4$ | $P$-1 | $2a \times 2a \times c$ | $A_3^+$ | $\Gamma_5^+$ | $a^0a^0c$/ $a^0a^0(-c)$ | | nDJ2 | 159 |
| 616101 | 2-(1-Cyclohexenyl)ethylamine | | $[C_8H_{16}N]_2PbI_4$ | $P$-1 | $2a \times 2a \times c$ | $A_3^+$ | $\Gamma_5^+$ | $a^0a^0c$/ $a^0a^0(-c)$ | | nDJ2 | 160 |
| 2011085 | 2-(4-Methoxyphenyl)ethyl-1-amine | | $[C_9H_{14}ON]_2PbI_4$ | $P$-1 | $2a \times 2a \times c$ | $A_3^+$ | $\Gamma_5^+$ | $a^0a^0c$/ $a^0a^0(-c)$ | | nRP | 161 |
| 1861843 | 2-([2,2'-bithiophen]-5-yl)ethyl-1-amine | | $[C_{10}H_{12}S_2N]_2PbI_4$ | $P1$ | $2a \times 2a \times c$ | | | | | nRP | 108 |
| 1840804 | (Naphthalene-*O*-propylamine$^*$ | | $[C_{13}H_{16}ON]_2[C_4H_6O_2]PbI_4$ | $P$-1 | $2a \times 2a \times c$ | $A_3^+$, $A_5^+$ | | | | nDJ2 | 83 |
| 1846391 | 2-(2$^4$,4$^3$-dimethyl[1$^2$,2$^2$:2$^5$,3$^2$:3$^5$,4$^2$-quaterthiophen]-1$^5$-yl)ethylamine | | $[C_{24}H_{22}S_4N]_2PbI_4$ | $P$-1 | $2a \times 2a \times c$ | | | | | nRP | 108 |
| 1875165 | 4-(Aminomethyl)piperidine | [A] | $[C_6H_{16}N_2]PbBr_4$ | $Pca2_1$ | $2\sqrt{2}a \times c \times \sqrt{2}a$ | $M_3^+$ | | | | DJ | 162 |
| 1934874 | 6-[(Naphthalen-1-yl)oxy]hexyl-1-amine | | $[C_{16}H_{22}ON]_2PbI_4$ | $P$-1 | $c \times 3\sqrt{2}a \times \sqrt{2}a$ | $M_3^+$, $M_5^+$ | | | | nRP | 79 |

$^*[C_4H_6O_2]$ = dihydrofuran-2(3H)-one is solvated into the crystal structure

**Table 5.** Summary of experimentally known structures with at least one unit cell metric larger than $2a_{RP}$ or $2c_{RP}$.

| CCDC Number | Amine | Type | Formula | Space group | Metrics | Key modes | | Other modes | Structure type | Ref. |
|---|---|---|---|---|---|---|---|---|---|---|
| | | | | | | Tilt | Layer shift | | | |
| 995699 | $N^1$,$N^1$-dimethylpropane-1,3-diamine | [A] | $[C_5H_{16}N_2]PbI_4$ | *Pbca* | $2\sqrt{2}a \times \sqrt{2}a \times c$ | $X_2^+$ | $M_5^-$ | $\Delta_3$, $Y_4$ | nDJ | 163 |
| 702987 | 4-Amidinopyridine | [A] | $[C_6H_9N_3]PbBr_4$ | *Pbca* | $2\sqrt{2}a \times \sqrt{2}a \times c$ | $X_2^+$ | $M_5^-$ | $\Delta_3$, $Y_4$ | nDJ | 164 |
| 1838616 | 2-(Aminomethyl)pyridine | [A] | $[C_6H_{10}N_2]PbCl_4$ RT | *Pbca* | $2\sqrt{2}a \times \sqrt{2}a \times c$ | $X_2^+$ | $M_5^-$ | $\Delta_3$, $Y_4$ | nRP | 165 |
| 666178 | 2-(Aminomethyl)pyridine | [A] | $[C_6H_{10}N_2]PbBr_4$ | *Pbca* | $2\sqrt{2}a \times \sqrt{2}a \times c$ | $X_2^+$ | $M_5^-$ | $\Delta_3$, $Y_4$ | nRP | 166 |
| 1838617 | 2-(Aminomethyl)pyridine | [A] | $[C_6H_{10}N_2]PbI_4$ | *Pbca* | $2\sqrt{2}a \times \sqrt{2}a \times c$ | $X_2^+$ | $M_5^-$ | $\Delta_3$, $Y_4$ | nDJ | 165 |
| 1048275 | 1-(2-Aminoethyl)piperazine | [A] | $[C_6H_{17}N_3]PbI_4$ | *Pbca* | $2\sqrt{2}a \times \sqrt{2}a \times c$ | $X_2^+$ | $M_5^-$ | $\Delta_3$, $Y_4$ | nDJ | 167 |
| 1915485 | 3-(2-Aminoethyl)pyridine | [A] | $[C_7H_{12}N_2]PbBr_4$ | *Pbca* | $2\sqrt{2}a \times \sqrt{2}a \times c$ | $X_2^+$ | $M_5^-$ | $\Delta_3$, $Y_4$ | nRP | 111 |
| 724584 | Cystamine | [A] | $[C_4H_{14}S_2N_2]PbI_4$ | $P2_1/n$ | $2\sqrt{2}a \times \sqrt{2}a \times c$ | $X_2^+$ | $M_5^-$, $\Gamma_5^+$ | $Y_3$ | nDJ | 121 |
| 1995236 | (RS)-2-Phenylpropyl-1-amine | | $[C_9H_{14}N]_2PbBr_4$ | *Pbca* | $4\sqrt{2}a \times \sqrt{2}a \times c$ | $X_2^+$ | $M_5^-$ | $\Delta_3$, $Y_2$, $Y_4$ | nDJ | 168 |
| 1831521 | 3-(Aminomethyl)piperidine | [A] | $[C_6H_{16}N_2]PbI_4$ | $P2_1/c$ | $\sqrt{2}a \times 2\sqrt{2}a \times c$ | $X_2^+$ | $M_5^-$ | | nDJ | 156 |
| 1937299 | Methylhydrazine | | $[CH_7N_2]_2PbI_4$ 280 K | *Pccn* | $c \times 3a \times 2a$ | C | $M_5^-$ | $\Sigma_4$ | nRP | 34 |
| 1937297 | Methylhydrazine | | $[CH_7N_2]_2PbI_4$ 100 K | *P*-1 | $c \times 3a \times 2a$ | C | $M_5^-$ | $\Sigma_4$ | nRP | 34 |
| 1963065 | $N^1$,$N^1$-Dimethylethyl-1,2-diamine | [A] | $[C_4H_{14}N_2]PbBr_4$ | $P2_1/c$ | $3a \times 2a \times c$ | | | | nDJ | 154 |
| 1887281 | $N^1$,$N^1$-Dimethylethyl-1,2-diamine | [A] | $[C_4H_{14}N_2]PbCl_4$ | *Pbcn* | $2\sqrt{2}a \times 2\sqrt{2}a \times c$ | | $M_5^-$ | $\Sigma_1$, $\Sigma_4$, $\Delta_3$, $Y_4$ | nDJ | 169 |
| 1982717 | $N^1$,$N^1$-Dimethylethyl-1,2-diamine | [A] | $[C_4H_{14}N_2]PbCl_4$ | $Pca2_1$ | $2\sqrt{2}a \times 2\sqrt{2}a \times c$ | $X_3^+$ | $M_5^-$ | $\Sigma_1$, $\Sigma_4$, $\Delta_3$, $Y_4$ | nDJ | 153 |
| 1838611 | 2-(Aminomethyl)pyridine | [A] | $[C_6H_{10}N_2]PbCl_4$ 100 K | $Pna2_1$ | $c \times 2\sqrt{2}a \times 2\sqrt{2}a$ | | $M_5^-$ | $\Sigma_1$, $\Sigma_4$, $\Delta_3$, $Y_3$, $Y_4$ | nRP | 165 |
| 1521060 | 4-Aminobutanoic acid | | $[C_4H_{10}O_2N]_2PbBr_4$ | $P2_1/c$ | $3\sqrt{2}a \times \sqrt{2}a \times 2c$ | | | | nDJ | 85 |
| 1963067 | $N^1$,$N^1$-Dimethylbutyl-1,4-diamine | [A] | $[C_6H_{18}N_2]PbBr_4$ | *Aba*2 | $2c \times 4a \times 2a$ | C | $M_5^-$, $\Lambda_5$ | $\Gamma_5^-$ | nDJ2 | 154 |